\documentclass[11pt]{usthesis} 

\usepackage{epsfig,amssymb,multirow,amsmath, bm}
\usepackage[font={bf}]{caption}
\usepackage[sort&compress,numbers]{natbib}
\usepackage{eso-pic}
    \newcommand*{\WaterMark}[2][0.15\paperwidth]{%
        \AddToShipoutPicture*{\AtTextCenter{%
                \parbox[c]{0pt}{\makebox[0pt][c]{%
                    \includegraphics[height=#1]{#2}}}}}}
\usepackage{afterpage}
\newcommand\blankpage{%
    \null
    \thispagestyle{empty}%
    \newpage}
\usepackage[linktocpage]{hyperref} 

\newcommand{\ave}[1]{\left<#1\right>}

\def\dblone{\hbox{$1\hskip -1.2pt\vrule depth 0pt height 1.6ex width 0.7pt\vrule depth 0pt height 0.3pt width 0.12em$}}

\begin{document}
 \WaterMark{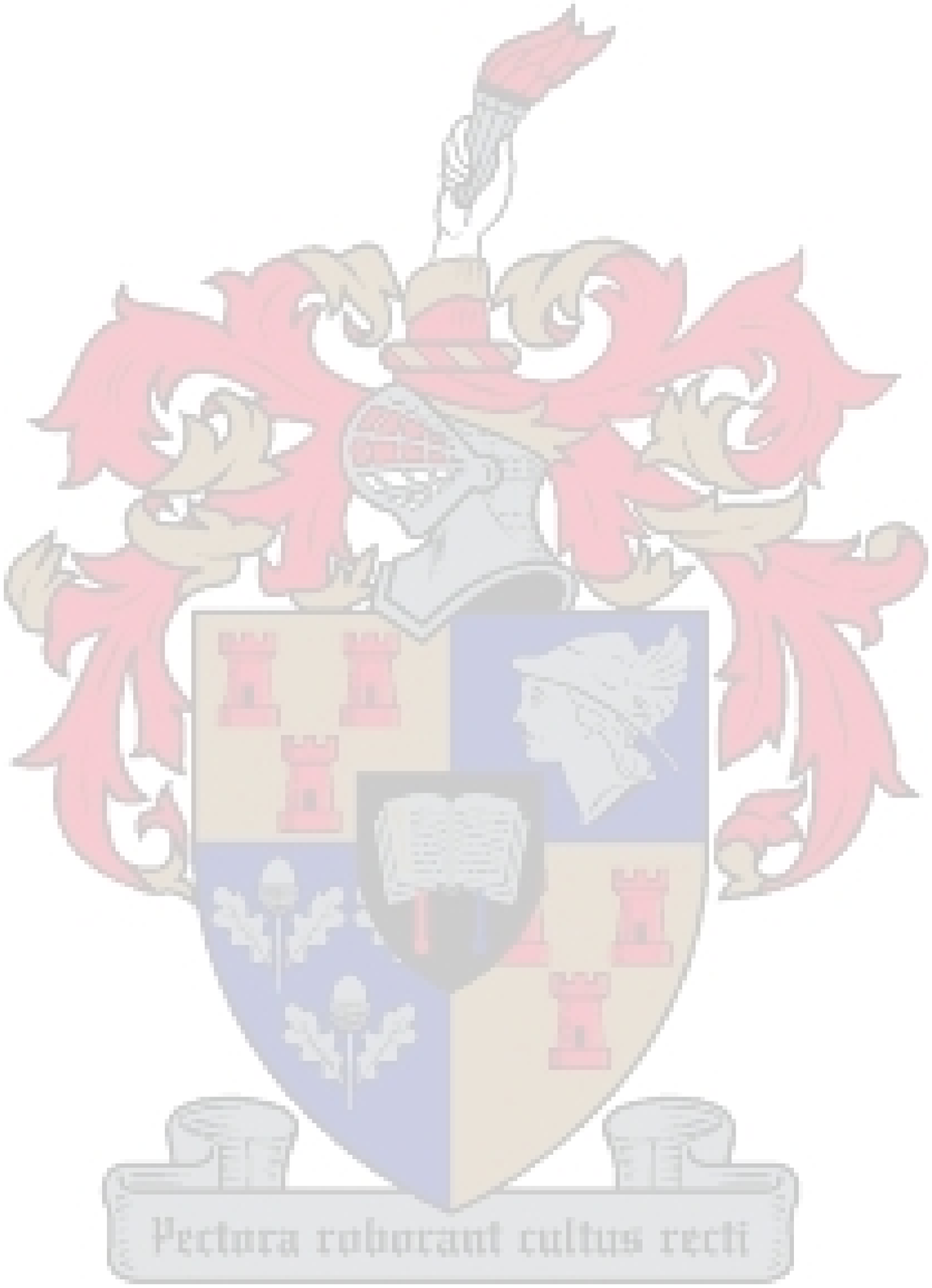}
\thesistitle{Ferromagnetic Phase Transitions in Neutron Stars}
\author{Jacobus Petrus Willem Diener}         
\degree{Doctor of Philosophy} 
\faculty{Science}
\supervisor{Prof. Frederik G. Scholtz} 
\cosupervisor{Prof. Hendrik B. Geyer and Dr Gregory C. Hillhouse}
\submitdate{December 2012}  
\declarationdate{December 2012}
\copyrightyear{2012}

\titlepage 
\blankpage
\declaration

\specialhead{ABSTRACT}
We consider the ferromagnetic phase in pure neutron matter as well as charge neutral, beta-equilibrated nuclear matter. We employ Quantum Hadrodynamics, a relativistic field theory description of nuclear matter with meson degrees of freedom, and include couplings between the baryon (proton and neutron) magnetic dipole moment as well as between their charge and the magnetic field in the Lagrangian density describing such a system. We vary the strength of the baryon magnetic dipole moment till a non-zero value of the magnetic field, for which the total energy density of the magnetised system is at a minimum, is found. The system is then assumed to be in the ferromagnetic state.\\
\\
The ferromagnetic equation of state is employed to study matter in the neutron star interior. We find that as the density increases the ferromagnetic field does not increase continuously, but exhibit sudden rapid increases. These sudden increases in the magnetic field correspond to shifts between different configurations of the charged particle's Landau levels and can have significant observational consequences for neutron stars. We also found that although the ferromagnetic phase softens the neutron star equation of state it does not significantly alter the star's mass-radius relationship.\\
\\
The properties of magnetised symmetric nuclear matter were also studied. We confirm that magnetised matter tends to be more proton-rich but become more weakly bound for stronger magnetic fields. We show that the behaviour of the compressibility of nuclear matter is influenced by the Landau quantisation and tends to have an oscillatory character as it increases with the magnetic field. The symmetry energy also exhibits similar behaviour.

\specialhead{OPSOMMING}
In hierdie studie het ons die ferromagnetiese fase in suiwer neutronmaterie, sowel as in ladingsneutrale, beta-ge\"ekwilibreerde neutronstermaterie, ondersoek. Vir die doeleindes het ons die Kwantum Hadrodinamika-model van kernmaterie gebruik. Dit is 'n relatiwistiese, veldteoretiese model wat mesone inspan om die interaksies tussen die protone en neutrone te bemiddel. Om die impak  van die magneetveld te bestudeer, sluit ons 'n koppeling tussen die barioonlading en die magneetveld, asook barioondipoolmoment en die magneetveld, in by die Lagrange digtheid wat ons sisteem beskryf. Om die ferromagnetiese fase te ondersoek, varieer ons die sterkte van die barioondipoolmoment om 'n nie-nul waarde van die magneetveld wat energie digtheid sal minimeer te vind. \\
\\
Die ferromagnetiese toestandsvergelyking word toegepas op materie aan die binnekant van die neutronster en die impak hiervan op die waarneembare eienskappe van die ster word ondersoek. Ons vind dat die ferromagnetiese magneetveld nie kontinu toeneem soos die digtheid verhoog nie. Die skielike toenames in die magneetveld is die gevolg van die sisteem wat die konfigurasie van die gelaaide deeltjies se Landau-vlakke skielik verander en dit kan beduidende waarneembare gevolge vir die ster inhou. Ons vind ook dat die ferromagnetiese fase die toestandsvergelyking versag, maar dat die versagting die massa-radius verhouding van die ster nie grootliks be\"invloed nie.\\
\\
Die eienskappe van gemagnetiseerde kernmaterie word ook ondersoek. Ons bevestig dat gemagnetiseerde materie meer proton-ryk, maar minder sterk gebind word. Ons wys dat die saampersbaarheid van kernmaterie deur die teenwoordigheid van Landau-vlakke be\"invloed word en ossilerend saam met die magneetveld toeneem. Die simmetrie-energie manifesteer ook soortgelyke gedrag.
\tableofcontents  
\listoftables       
\listoffigures
\specialhead{ACKNOWLEDGEMENTS}
Compiling this dissertation has been an arduous, but enjoyable journey.\\
\\
I would like to acknowledge the invaluable input and support from prof Frikkie Scholtz in this study. He inherited me after the first year of my PhD study and got it back on track. His intuition and knowledge never failed to amaze me. \\
\\
I would like to thank the rest of the village for also raising the child. Hannes, for your open door and patience with lesser mortals, as well as Lee, Nanna, Rohwer and Rikus for the all the friendly chats, jokes and occasional expletives.\\
\\
I gratefully acknowledge financial support from the following institutions:
\begin{itemize}
	\item the South African SKA project, and
	\item Stellenbosch University.
\end{itemize}
I would to also acknowledge the late prof Okkie de Jager for the idea which lead to this investigation.\\
\\
Without support from my parents, family and in particular Sanette, this work would not have been possible. Soli Deo gloria.

%

\blankpage
\chapter{Introduction}
This dissertation aims to present a relativistic covariant description of ferromagnetism in beta-equilibrated nuclear matter with special emphasis on the use of this description to study ferromagnetism in the neutron star interior. 
\section{Neutron stars}
In 1934 Walter Baade and Fritz Zwicky proposed that some supernovae are driven by the energy released in forming a dense compact stellar object out of the core of a massive star \cite{csg}. These objects have since been named {\em neutron stars} and are associated with the remnant cores of massive stars that exploded in core-collapse supernovae. Neutron stars are observed as pulsars, which are rapidly rotating neutron stars emitting radio waves from their magnetic poles. If the star's magnetic axis is not aligned with the rotation axis these emissions are observed as pulse trains.  These cosmic lighthouses were first observed by Jocelyn Bell in 1967 \cite{hewish}. \\
\\
Since that time neutron stars/pulsars have been the subject of intensive studies as they provide us with a laboratory to study matter under extreme conditions: neutron stars are inferred to have average densities of the order of nuclear matter $\approx 10^{14}$ g/cm$^3$ \cite{csg} and magnetic field strengths of between $10^8$ and $10^{13}$ G \cite{pnas}. Currently we cannot replicate these conditions in any other laboratory. For a review of possible nuclear and particle physics that can be studied with neutron stars, see \cite{webertxt}.\\
\\
Most of our information about (radio-) pulsars is gained by monitoring their radio emission. Since the emitted pulses are very stable and well defined, the rotation of the pulsar can be very precisely timed and monitored. Pulsars are observed to be spinning slower at very stable rates, but every now and again undergo rapid acceleration events known as {\em glitches}. After the sudden spin-up of the star, it relaxes again to its pre-glitch deceleration tempo \cite{csg}. The spin-up and relaxation timescales are of particular interest, since they relate information to us about the processes in the neutron star interior. For a review of neutron star properties and observations see \cite{Lattimer07}.
\subsection{Soft Gamma Repeaters}
With the launch of space telescopes capable of detecting high-energy radiation, a whole new class of neutron stars was discovered. The first of these were the Soft Gamma Repeaters (SGRs) which are sources of repeated low energy (soft) $\gamma$-rays {\em bursts} with peak luminosities reaching $10^{41}$ ergs$\cdot$s$^{-1}$ and photon energies above $20$ keV. These bursts have short timescales of around $0.1$s and a repetition rate that varies from seconds to years \cite{chap14}. \\
\\
SGRs also produce more rare (on the order of 50 - 100 years) {\em giant flares}: these are very luminous, $10^{44}$ ergs$\cdot$s$^{-1}$, flashes of hard $\gamma$-rays with photon energies of 50 - 500 keV. 
{\em Intermediate bursts} are also observed, with energies and luminosities between that of bursts and giant flares, sometimes with a frequency on the order of weeks \cite{chap14}.  \\
\\
These objects also have relative steady, persistent emission in the X-ray spectrum, $10^{35}$ ergs$\cdot$s$^{-1}$ of 0.5 to 10 keV photons, but X-ray pulses are also observed from these sources. X-ray pulses are also observed from a very similar class of objects, namely Anomalous X-ray Pulsars (AXPs). 
\subsection{Anomalous X-ray Pulsars}
Most AXPs have stable X-ray pulses and consequently their spin-down behaviour can be monitored in the same way as pulsars are monitored. 
However, they were termed {\em anomalous} since it was unclear what powers their radiation \cite{2gether}. In addition to their distinguishing pulsed and persistent X-ray emissions, these objects are also observed to glitch \cite{AXPprogress}. \\
\\
Initially AXPs and SGRs did not appear to have much in common but, as instruments and observational techniques improved, the emphasis has shifted to rather establishing how these objects differ 
\cite{2gether}.  AXPs also exhibits radiative events: {\em bursts} similar to that of SGRs \cite{chap14} as well as larger {\em outbursts} \cite{AXPprogress}. AXP outbursts share some properties with SGR giant flares, but \cite{AXPprogress} reports that the 
tail observed after an AXP outburst is much longer than observed for SGR giant flares. 
In June 2002 an outburst in AXP 1E 2259+586 was accompanied by a large glitch which seems to suggest that the glitches are accompanied by radiative events \cite{AXPprogress}. \\
\\
Both AXPs and SGRs have long spin periods, but large period derivatives and are believed to be powered by the decay of their superstrong magnetic fields of $10^{14} - 10^{15}$ G \cite{pnas}.
\subsection{Magnetars}
In general AXPs and SGRs are grouped together as {\em magnetars} or magnetar candidates \cite{chap14}. Magnetars are highly magnetised neutron stars whose emission is driven by the decay of the magnetic field. The strong magnetic fields are believed to be the result of dynamo action in the proto-neutron star \cite{T+D93} which results in the X- and $\gamma$-ray emission, that distinguishes them from radio-emitting pulsars \cite{D+T92}. However, radio-emissions has been detected during a SGR outburst, but no persistent emission is detected from magnetar candidates \cite{chap14}.\\
\\
In the current model for magnetars, the decay of the magnetic field powers the persistent X-ray emission through low-level seismic activity in the crust and heating of the stellar interior \cite{T+D96}. While the bursts are the result of large-scale crustal fractures caused by the evolving magnetic field \cite{T+D95}. 
\section{Ferromagnetism and response of dense matter in extreme magnetic fields}
Since the observation that pulsars have very strong magnetic fields, the origin of these magnetic fields, as well as its interaction with the matter in the interior of the star, has been a topic of discussion and research. Soon after the discovery of pulsars Brownell and Callaway \cite{BN}, as well as Silverstein \cite{S}, proposed that a ferromagnetic phase of interior nuclear matter of a neutron star can make a significant contribution to the magnetic field.\\
\\
Various authors built on this notion and investigated the magnetisation and/or ferromagnetic phase transition in various types of nuclear matter with varied results: most recently Bigdeli \cite{bigdeli} found, calculating the Helmholtz free energy of magnetised asymmetric nuclear matter, that an external magnetic field can induce an antiferromagnetic phase transition in said matter\footnote{From the data presented in the paper, we believe that the author might have come to the wrong conclusion: if the larger fraction of neutrons align their dipole moments antiparallel to a positive magnetic field, while the largest fraction of protons align parallel, then the resultant magnetic dipole will be positive and result in a ferromagnetic state. It would also appear that the Landau problem for charged protons was ignored in this paper, which may have also influenced the results.}.  A concise summary of other approaches to the question of the ferromagnetic phase in nuclear matter 
is also presented in \cite{bigdeli}.\\
\\
For this study of ferromagnetism in nucleonic matter we are primarily concerned with magnetised charge neutral, beta-equilibrated matter consisting of protons, neutrons and leptons. Various similar studies have already been conducted on this topic and we will give a short summary of the recent ones, applicable to this work.\\
\\
The first paper related to this study is that of Chakrabarty and collaborators \cite{chakPRL}. In this paper the authors investigated the effect of a strong magnetic field on the composition of nuclear matter within the context of Quantum Hadrodynamics. 
They found magnetised matter is more strongly bound than unmagnetised matter and that the proton fraction of beta-equilibrated charge neutral nuclear matter increases, as the magnetic field gets stronger. They also suggested that the maximum mass of neutron stars appears to be insensitive to the magnetic field, but the corresponding radii would be smaller, leading to more compactified objects. It was pointed out by Broderick et al. \cite{Brod00} that in \cite{chakPRL}, amongst others, the electromagnetic contribution to the energy density, thus also to the pressure, was not included in their calculations. \\
\\
Broderick et al. \cite{Brod00} also included a coupling between the magnetic dipole moment and the magnetic field and investigated the influence thereof on the equation of state of charge neutral, beta-equilibrated nuclear matter. This was to include the higher-order contributions to the dipole moments of the nucleons\footnote{In the paper these contributions are referred to as the {\em anomalous contribution}. We would rather not use that term when referring to the baryon dipole moment, see section \ref{sec:neumag}}. They found that the Landau quantisation softens the equation of state (the pressure of the matter increases less rapidly with density). However, they also found that this softening is overwhelmed by the stiffening induced by including the coupling between the dipole moments and the magnetic field.\\ 
\\
G. Mao and collaborators in \cite{mao} and \cite{mao2} also considered the inclusion of the anomalous contribution to the electron magnetic dipole moment in charge neutral, beta-equilibrated atomic matter (excluding muons). They concluded that the effect of including this coupling is negligible. \\
\\
However, in contrast to electrons, baryons have substructure from which they derive their dipole moment or contributions to it. Ryu et al. investigated the neutron star equation of state with density-dependent dipole moments for the baryon octet using the quark-meson coupling (QMC) models and extensions thereof \cite{Ryu}. They report that the baryon dipole moment is dependent on the magnetic field and the size of the MIT-bag in the QMC models. They do not report significant increases in the neutron star maximum masses, but that, since protons are the lightest baryon, as the proton fraction increases with the magnetic field strength, the formation of hyperons are suppressed. \\
\\
Our study, reported on here, shares similarities with all the studies mentioned above. We assume that, as the density in a charge neutral beta-equilibrated system increases, the strength of the coupling between the baryon magnetic dipole moments and the magnetic field will increase to the point at which a ferromagnetic state will be energetically favoured. A further assumption is that the equilibrium value of the ferromagnetic field will always be such that the energy density is at a minimum.\\
\\
Based on these assumptions we calculated the ferromagnetic phase diagram as a function of the adjusted dipole moment coupling strength. We investigated the behaviour of magnetised and ferromagnetised nuclear matter with adjusted baryon magnetic dipole moments and report on its implications for the neutron star equation of state. After this we will also speculate on possible observational consequences of a ferromagnetic state in the neutron star interior.
\section{Magnetised matter}
In order to clarify notation, and as a point of reference, the description of matter in a magnetic field given by Griffiths in the {\em Introduction to Electrodynamics} \cite{griff} will be summarised here.\\
\\
In any laboratory investigation of the electromagnetic responses of matter, the quantity that the experimenter is able to adjust can be called the free charge (in the case of an electric response) or free current (in the case of a magnetic response). However, what is measured is the matter's total response, which will include the response of any bound charges or currents in the matter. In the case of an electric field the bound charge is related to the alignments of the constituent particles' charges (polarisation). For a magnetic field, bound currents are induced by the magnetic dipole moments of the matter's constituent particles called the magnetisation. \\
\\
In order to include all effects and responses in the electromagnetic description of the matter the electric displacement, $\bm D$, and $\bm H$ are introduced\footnote{$\bm H$ is often referred to as the magnetic field but we will, as is done in \cite{griff}, simply refer to it as ``$\bm H$''.}. These quantities are vectors and are defined as
\begin{subequations}
	\begin{eqnarray}
		{\bm D}&\equiv&\epsilon_0 {\bm E}+{\bm P}\mbox{, and}\\
		{\bm H}&\equiv&\frac{1}{\mu_0} {\bm B}-{\bm M}\label{H},
	\end{eqnarray}
\end{subequations}
where
\begin{itemize}
	\item $\bm E$ is the electric field,
	\item $\bm P$ is the polarisation which is defined in terms of the bound charge density, $\rho_b$, as
	\begin{eqnarray}
		\rho_b\equiv-{\bm \nabla}\cdot\bm P,
	\end{eqnarray}
	\item $\bm B$ is the magnetic field, 
	\item $\bm M$ the magnetic dipole moment per unit volume of the matter, also known as the magnetisation, which can be
	defined in terms of a bound current ${\bf J}_b$ as 
	\begin{eqnarray}\label{jb}
		{\bm\nabla}\times{\bf M}\equiv{\bm J}_b,
	\end{eqnarray}	
	and furthermore
	\item $\epsilon_0$ and $\mu_0$ are the permittivity and permeability of free space respectively.
\end{itemize}
With these definitions Maxwell's equation, in particular Gauss's and Amp\'ere's laws can be written in terms of only the free charges, $\rho_f$, and currents, ${\bm J}_f$, as
\begin{subequations}\label{maxwell}
	\begin{eqnarray}
		\bm \nabla\cdot{\bm D}&=&\rho_f\label{maxD}\mbox{, and}\\
		\bm \nabla\times{\bm H}&=&\bm J_f+\frac{\partial \bm D}{\partial t}\label{maxH}
	\end{eqnarray}
	when the Maxwell correction is also included Amp\'ere's law \cite{gross}.
	Since we will only consider charge neutral matter, equation (\ref{maxD}) is not of particular importance to us. In contrast, equation (\ref{maxH}) will feature quite prominently so we rewrite it, using (\ref{H}), as
	\begin{eqnarray}
		\bm \nabla\times\left(\frac{\bm B}{\mu_0}-\bm M\right)&=&{\bm J}_f+\frac{\partial}{\partial t}\left(\epsilon_0 {\bm E}+{\bm P}\right).\label{maxH2}
	\end{eqnarray}
\end{subequations}
Note that the remaining two Maxwell's equations stay the same when magnetised matter is considered:
\begin{subequations}
	\begin{eqnarray}
		\bm \nabla\times{\bm E}&=&-\frac{\partial \bm B}{\partial t}\mbox{, and}\\
		\bm \nabla\cdot{\bm B}&=&0.
	\end{eqnarray}
\end{subequations}

\section{Units and conventions}\label{units}
In this work natural units will be used, i.e.
\begin{eqnarray}
	\hbar = c = 1,
\end{eqnarray}
where
\begin{itemize}
	\item $\hbar$ is Planck's constant divided by $2\pi$, and
	\item $c$ the speed of light in vacuum.
\end{itemize}
This implies that 
\begin{eqnarray}
	1 = 197.33\ \mbox{MeV}\cdot\mbox{fm}\label{MeV2fm},
\end{eqnarray}
which will serve as the conversion factor between energy (in mega-electronvolts) and length (in fermi). Additionally, since $c=1$, we have that
\begin{eqnarray}
	2.998\times10^{8}\mbox{ m}\cdot\mbox{s}^{-1} = 1
\end{eqnarray}
so that
\begin{eqnarray}
	1\mbox{ s}=2.998\times10^{23}\mbox{ fm}.
\end{eqnarray}
\subsection{Gaussian units}
In the context of nuclear and neutron star matter Gaussian units, instead of SI units, are used for the electromagnetic field and charges. To convert electrostatic equations from SI to Gaussian units, one sets \cite{griff}
\begin{eqnarray}
	\epsilon_0=\frac{1}{4\pi}.
\end{eqnarray}
In SI units \cite{griff}
\begin{eqnarray}
	c = \frac{1}{\sqrt{\epsilon_0\mu_0}}
\end{eqnarray}
and this implies that
\begin{eqnarray}
	\mu_0  = \frac{1}{c^2 \epsilon_0} = \frac{ 4\pi}{c^2}.\label{muG}
\end{eqnarray}
Therefore, when using Gaussian units, $\epsilon_0$ and $\mu_0$ 
are simply equal to $4\pi$ and $4\pi c^{-2}$ as their respective effect have been absorbed in the conversion factors between units. When combined with natural units these expressions simplify even further, since $c$ is defined to be equal to $1$.  In SI units the free electromagnetic Lagrangian is \cite{waleckaadv}
\begin{eqnarray}\label{Lem}
	{\cal L}_{em} =-\frac{\epsilon_0}{4}F^{\mu\nu}F_{\mu\nu}
\end{eqnarray}
and, irrespective of the choice of the gauge field $A^{\mu}$, contracts to \cite{Brouder}
\begin{eqnarray}\label{Lem2}
	{\cal L}_{em} =\frac{\epsilon_0}{2}\left(\left|{\bm E}^2\right|-c^2\left|{\bm B}^2\right|\right).
\end{eqnarray}
If we combine Gaussian and natural units, (\ref{Lem}) becomes
\begin{eqnarray}
	{\cal L}_{em} =-\frac{1}{4\times4\pi}F^{\mu\nu}F_{\mu\nu}=\frac{1}{8\pi}\left(\left|{\bm E}^2\right|-\left|{\bm B}^2\right|\right).
\end{eqnarray}
Combining the expression for $\mu_0$ in SI units \cite{griff} with (\ref{muG}) in natural units, we have that
\begin{eqnarray}\label{mu00}
	\mu_0 = 4 \pi\times10^{-7}\, \frac{\mbox{N}}{\mbox{ A}^2} =4 \pi\times10^{-7}\, \frac{\mbox{kg}\cdot\mbox{m}}{\mbox{C}^{2}} = 4\pi.
\end{eqnarray}
In natural units mass (kg) has the unit of (length)$^{-1}$. Therefore, in Gaussian units, the unit of charge becomes dimensionless and (\ref{mu00}) establishes a conversion factor for charge:
\begin{eqnarray}
	1\,\mbox{C}= 5.331\times10^{17}\, .\label{coulombG}
\end{eqnarray}
%
%
%
\subsection{Heaviside-Lorentz units}
The Heaviside-Lorentz system of units only differs by a factor of $4\pi$ from Gaussian units \cite{jackson} and can also be easily used in conjunction with natural units. Here $\epsilon_0$ is defined to be 
\begin{eqnarray}
	\epsilon_0=1
\end{eqnarray}
and thus
\begin{eqnarray}
	\mu_0=\frac{1}{c^2}\,.
\end{eqnarray}
Since we have already declared to be using natural units ($\hbar = c = 1$) this will mean that $\mu_0 = 1$. 
In these units the contribution of the free electromagnetic Lagrangian will be 
\begin{eqnarray}
	{\cal L}_{em} =-\frac{1}{4}F^{\mu\nu}F_{\mu\nu}=\frac{1}{2}\left(\left|{\bm E}^2\right|-\left|{\bm B}^2\right|\right).
\end{eqnarray}
Using the definition of $\mu_0$ in SI units, together with the choice of Heaviside-Lorentz units, we have that
\begin{eqnarray}
	\mu_0 = 4 \pi\times10^{-7}\, \frac{\mbox{kg}\cdot\mbox{ m}}{ \mbox{C}^{2}} = 1
\end{eqnarray}
in which case charge is again dimensionless and
\begin{eqnarray}
	1\,\mbox{C}= 1.890\times10^{18}.\label{coulombLH}
\end{eqnarray}
The conversion factors for the unit of charge, (\ref{coulombG} and \ref{coulombLH}), will be used to convert charges of particles into dimensionless quantities, see table \ref{Tab:charges}. Also see \cite{waleckaintro} for more on Heaviside-Lorentz units and dimensionless charges.\\
\\
In this work the Heaviside-Lorentz units will be used, since in conjunction with natural units the equations in our model appear the simplest. Conversion factors for Heaviside-Lorentz units are listed in table \ref{tab:constants}.
\begin{table*}[tbh]
	\centering
		\begin{tabular}{lcccc}
		\\
				\hline
				Particle&Charge [C] &Symbol&Value (Gaussian)&Value (Heaviside-Lorentz)\\
				\hline\hline
				Proton & $\ 1.602\times10^{-19}$&$q_p$ &0.0854&0.303\\
				Electron &$ -1.602\times10^{-19}$& $q_e$&-0.0854 & -0.303\\
				Muon & $-1.602\times10^{-19}$& $q_{\mu}$ &-0.0854& -0.303\\
				\hline	
		\end{tabular}
	\caption{Particle charges as dimensionless quantities in different unit systems.}
	\label{Tab:charges}
\end{table*}
\begin{table*}[bbh]
	\centering
		\begin{tabular}{lll}
		\\
				\hline
				Name&Symbol&Value\\
				\hline\hline
				Solar mass & \ $M_{\odot}$& $1.98892\times10^{30}$ kg or 
				\\&& $1.1155\times10^{60}$ MeV\\
				Gravitational constant &\  $\cal G$ & $6.6726\times10^{-11}$  m$^3 \cdot$kg$^{-1}\cdot$s$^{-2}$ or\\&&
				 $1.325\times10^{-42} \mbox{fm}/$MeV\\
				 \hline\hline
				 Conversion factors&&\\
				 \hline
				Length & $1\ \mbox{fm}$& $1\times10^{-15}$ m\\
				Energy & $1$ MeV& $6.2415\times10^{8}$ ergs \\
				Energy density & 1\ MeV/fm$^{3}$ & $6.2415\times10^{53}$ ergs/m$^{3}$\\
				Mass & 1 MeV/c$^2$& $1.783\times10^{30}$ kg \\
				Time & 1 s& $2.998\times10^{23}$ fm \\
				Magnetic field & 1 fm$^{-2}$&$1.993 \times10^{18}$ G\\
				\hline	
		\end{tabular}
	\caption{Constants and conversion factors used in this work.}
	\label{tab:constants}
\end{table*}
\subsection{Magnetic field}
The magnitude of the magnetic field, $B$, will thus have the units of $(\mbox{length})^{-2}$, since it is actually a flux\footnote{Therefore, when authors refers to $\bm H$ as the magnetic field, $\bm B$ is sometimes referred to as the ``magnetic flux density'' \cite{griff}.}. In particular, $B$ will be expressed in fm$^{-2}$ in most expressions and calculations (unless explicitly stated). However, magnetic fields are commonly expressed in units of gauss (G) where
\begin{align}
	\begin {split}
	1 \mbox{ G}&=10^{-4}\mbox{ kg}\cdot\mbox{C}^{-1}\cdot\mbox{s}^{-1}\\
	&=5.016\times10^{-19}\mbox{ fm}^{-2}
	\end{split}
\end{align}
in Heaviside-Lorentz units. Thus we will express magnetic fields in gauss although these fields were calculated in units of fm$^{-2}$. 
We used the conversion factor of
 \begin{eqnarray}
	B\,[\mbox{fm}^{-2}] = B\times 1.993\times10^{18}\,[\mbox{G}]\label{b2g}
\end{eqnarray}
to convert the magnitude of the calculated magnetic fields to quantities in gauss.
\subsection{Subscripts}
We will use the following subscripts to denote different quantities. 
\begin{table*}[hhh]
	\centering
		\begin{tabular}{cl}
		\\
				\hline
				Subscript&Quantity\\
				\hline\hline
				B&magnetic field\\
				b&baryons\\
				l&leptons\\
				n&neutrons\\
				p&protons\\
				e&electrons\\
				$\mu$&muons\\
				\hline	
		\end{tabular}
	\caption{Various subscripts used in this dissertation.}
	\label{tab:sub}
\end{table*}
For instance $\rho$ in general refers to a particle number density, while $\rho_n$ refers to the neutron particle density.
\subsection{Chemical potential and magnetic moment}
In addition to the subscript ``$_\mu$'' referring to muons, in the literature ``$\mu$'' can also refer to both the chemical potential and the dipole moment of a particle. Here we make the distinction that ``$\mu$'' refers to the chemical potential while ``$\mu^{(dip)}$'' refers to the magnetic dipole moment. Although the convention of referring to the nuclear magneton as ``$\mu_N$'' is well established, we will also refer to it as ``$\mu_N^{(dip)}$''.\\
\\
Using this convention the muon chemical potential will be ``$\mu_{\mu}$'', although such confusing expressions are avoided, where possible, in this work.
\subsection{Energy}
We include mesons in our description of nuclear matter, thus the total energy of a baryon (including the meson contributions) will be referred to as ``$e$'', where the free baryon contributions will be referred to as ``$E$''. As an example, in the case of unmagnetised neutrons where the single particle energy is given by (\ref{ekfsu})
\begin{align}\label{enerconven}
	\begin{split}
		e ({\bm k},s)&= \sqrt{{\bm k}^2 + {m^*}^2}+g_v V_0 -\frac{1}{2}g_\rho\,b_0 \\
		&=  E(\bm k,s)+g_v V_0 - \frac{1}{2}g_\rho\,b_0 .
	\end{split}
\end{align}
Note that where ``$e$'' should refer to the base of the natural logarithm it would be clear from the context within which we use it. Also note that we will always refer to the charge of a particle as ``$q$'' with the addition of subscript from table \ref{tab:sub} to indicate the particular particle.
\subsection{Nomenclature}\label{sec:nom}
In this work we will frequently refer to some very specific concepts, which we will define here. 
%
\subsubsection{Nuclear matter}
Nuclear matter is pure hadronic matter. Within the context of this work this may either refer to matter consisting of neutrons and various mesons, which can also be called neutron matter. Or it might refer to some mix of protons, neutrons or mesons, but not including leptons. 
\subsubsection{Neutron star matter}
Neutron star matter refers to the type of matter we assume to be present in the interior of a neutron star. This matter is charge neutral and beta-equilibrated. It therefore consists of a mix of protons, neutrons, mesons and leptons (electrons and/or muons).
\subsubsection{Baryon species}
The only baryons we will consider are protons and neutrons. However, baryon {\em species} will refer to the distinction made for the (two) possible magnetic dipole projections on the $z$-axis.  
These projections will be denoted by different values of $\lambda$ with
\begin{align}\label{lambda}
	\lambda = \pm 1.
\end{align}
We introduce $\lambda$ since, as we will show, spin is not a good quantum number of our magnetised matter Hamiltonian. If this was not the case, we would have referred to {\em spin} or {\em spins} instead of {\em species}. Thus {\em baryon species} will refer to four distinct types of particles, two protons and two neutrons where each type of neutron (proton) has a different orientation of its dipole moment. We will also use species in the context of individual baryons, e.g. proton species, which will refer to only protons with different values of $\lambda$. 
\subsubsection{Filling configuration}
The {\em filling configuration} refers to the way in which the baryon species contribute to the total baryon and/or proton and neutron densities, so that the 
energy density of the system is at a minimum. 
\subsection{Dirac matrices}
We will use the Dirac representation of the Dirac matrices as given by Itzykson and Zuber \cite{zuber} where
\begin{subequations}
	\begin{eqnarray}
	\alpha^i &=& 
				\left[
						\begin{array}{cc}
							0      &   \sigma^i\\
							\sigma^i &   0				 
						\end{array}	
					\right]\mbox{, and}	\label{pd1}\\		
		\beta &=& 
				\left[
					\begin{array}{cc}
						\dblone_2 &   0\\
						0       &   -\dblone_2				 
					\end{array}	
				\right].\label{pd2}
	\end{eqnarray}
\end{subequations}
In the above $\bm\sigma$ are the Pauli-matrices,
\begin{subequations}\label{pauli-spin}
	\begin{eqnarray}
			\sigma^{1} &=& 
					\left[
						\begin{array}{cc}
								0 & 1\\
								1 & 0				 
						\end{array}	
					\right],			\label{pauli1}\\
			\sigma^{2} &=&
					\left[\begin{array}{cc}
								0   & -i\\
								i  &  0				 
							\end{array}	
						\right]\mbox{, and}\label{pauli2}\\
			\sigma^{3} &=&
					\left[\begin{array}{cc}
								1  &   0\\
								0  &  -1				 
							\end{array}	
						\right],\label{pauli3}			
	\end{eqnarray}
\end{subequations}
and $\dblone_2$ the $2\times2$ identity matrix.\\
\\
The $\gamma$ matrices are
	\begin{eqnarray}
		\gamma^\mu&=&(\gamma^0,\bm\gamma)=(\beta,\beta\bm\alpha).
	\end{eqnarray}
The tensor $\sigma^{\mu\nu}$ is defined in terms of the $\gamma$ matrices,
\begin{eqnarray}
	\sigma^{\mu\nu} = \frac{i}{2}\left[\gamma^\mu,\gamma^\nu\right],
\end{eqnarray}
while its components have the property that
\begin{eqnarray}
	\sigma^{ij} = \epsilon_{ijk}\Sigma^k,\label{sigmaxpnd}
\end{eqnarray}
where 
${\Sigma^i}$ is part of the nucleon spin operator, defined in terms of the Pauli-spin matrices, $\bm {\sigma}$, as 
\begin{eqnarray}
	 {\bm \Sigma}= 
		\left[
			\begin{array}{cc}
				 {\bm \sigma} &0\\
				0 &  {\bm \sigma}
			\end{array}	
		\right].
\end{eqnarray}
As such $\Sigma^3 = \Sigma_z$ is
\begin{eqnarray}
	\Sigma^3= 
		\left[
			\begin{array}{cc}
				\sigma^3 &0\\
				0 & \sigma^3
			\end{array}	
		\right]
		=\left[
			\begin{array}{cccc}
				1&0&0 &0\\
				0&-1&0 &0\\
				0&0&1 &0\\
				0&0&0 &-1\\
			\end{array}	
		\right].\label{eqnSigmaz}
\end{eqnarray}%
\subsection{Minskowski space metric}
The metric for flat space-time (Minkowski space), $\eta^{\mu\nu}$ is taken as
\begin{eqnarray}
	\eta = 
	{\left[\begin{array}{cccc}
		1&0&0&0\\
		0&-1&0&0\\0&0&-1&0\\0&0&0&-1
	\end{array}\right].}\label{metric}
\end{eqnarray}

\chapter{Unmagnetised nuclear and neutron star matter}\label{chap:unmag}
The average density of a neutron star can be estimated from its mass-radius relationship. The ``canonical" neutron star has a mass on the order of 1.5 M$_{\odot}$ (solar masses) and a radius of about 12 km. Such a star's average density would be around $10^{14}$ g/cm$^3$, which is about the density of saturated nuclear matter. Thus it can be assumed that, at least in part, a neutron star consists of dense nuclear matter \cite{csg}. \\
\\
In this chapter, which for the most part is based on \cite{diener}, our description of $\mbox{unmagnetised}$ neutron star matter will be given. We assume that the star is composed of a charge neutral and $\beta$-equilibrated mix of protons, neutrons, electrons and muons. These particles and their interactions will be described within the context of Quantum Hadrodynamics and the relativistic mean-field approximation. 
\section{Quantum Hadrodynamics}
Quantum Hadrodynamics (QHD), also known as the Walecka-model, is a relativistic description of nuclei and nuclear matter with hadronic degrees of freedom, i.e. mesons mediate the interaction between baryons \cite{walecka}. In this description of unmagnetised neutron star matter, protons and neutrons (baryons) interact via the exchange of scalar (sigma), vector (omega) and isovector (rho) mesons. The meson exchanges are described by coupling the meson fields to the baryon densities, or currents, in the Lagrangian. The coupling strengths are fixed at the values that reproduce various properties of saturated nuclear matter (as discussed in section \ref{sec:satnuc}). QHD parameter sets are distinguished by different values of the coupling strengths as well as the presence of various self-couplings of the mesons fields. Various parameter sets are described in the literature, but for this study we will use the QHD1 \cite{walecka}, NL3 \cite{NL3} and FSUGold \cite{FSU1} parameter sets. \\
\\
QHD has been extensively used to study the properties of nuclei and nuclear matter (for a review see \cite{recentprogress}), as well as neutron star matter and the neutron star equation of state (see for instance \cite{csg} and \cite{webertxt}). The equation of state is the relation between the matter's pressure and energy density as a function of density. In general, different descriptions of neutron star matter (or combinations thereof) are referred to as different {\em equations of state}.  QHD is of course only one approach in describing the neutron star equations of state. Other equations of state can also include more exotic particles such as hyperons, koan condensates and/or quark matter (for a recent review of various equations of state see \cite{Lattimer06}). As mentioned previously, in this work the baryon contribution to neutron star equation of state is restricted to protons, neutrons and mesons. 
\subsection{Properties of saturated nuclear matter}\label{sec:satnuc}
The challenge in describing matter at high densities is to develop a model that not only describes matter at high densities, but also the properties of matter 
observed at normal densities. The philosophy of QHD is to constrain the various coupling constants in such a way that the calculated values of various symmetric nuclear matter properties match the observed ones.\\
\\
Symmetric nuclear matter, or just nuclear matter, is an idealised system that stems from one of the original models of the nucleus, the liquid-drop model \cite{csg}. The properties of nuclear matter are inferred from the experimentally observed properties of finite nuclei. However, since these properties cannot be directly observed, there is some disagreement as to what the exact values are. 
\subsubsection{Saturation density}
The short-ranged, strong nuclear interaction is the dominant interaction between nucleons. It is essentially attractive, which is necessary to form stable nuclei, but repulsive at short distance ($\leq 0.4$ fm) \cite{csg}.  However, this interaction does not have infinite range and above a certain density the nucleus/ nuclear matter will become unstable. The saturation density marks the point at which the pressure in the nuclear system is zero and the binding energy is at a minimum. \\
\\
For nuclear matter the saturation density is given as $0.153$ fm$^{-3}$ in \cite{csg} and $0.16$ fm$^{-3}$ in \cite{webertxt}.
\subsubsection{Binding energy}
In a general sense the binding energy of a system is the energy expended, or required, to form a bound system. For stable systems the binding energy is negative and thus the system is at an energy state lower than that of the energy sum of the components. At the saturation density the binding energy of the system will be at a minimum, since the system will be in its most stable (lowest energy) state.\\
\\
The binding energy of nuclear matter is given as $-16.3$ MeV/nucleon in \cite{csg} and $-16.0$ MeV/nucleon in \cite{webertxt}.
\subsubsection{Compression modulus}\label{comp}
The compression modulus defines the curvature of the equation of state at saturation and is related to the high density behaviour of the equation of state \cite{csg}. A {\em stiff} equation of state refers to the situation when the system's pressure rapidly increases with an increase in (energy) density. In the case of a {\em soft} equation of state, the pressure increases more gradually as a function of the (energy) density \cite{csg}. \\
\\
The compression modulus $K$ is defined as 
\begin{eqnarray}
	K \equiv 9\left[\rho^{2}\frac{d^{2}}{d\rho^{2}}\left(\frac{\epsilon}{\rho}\right)\right]_{\rho=\rho_{0}}\label{K}
\end{eqnarray}
and gives an indication of the stiffness of the equation of state, since it is essentially the derivative of the pressure. The value of $K$ has been estimated to be 234 MeV (with some uncertainty) \cite{csg}. However, \cite{webertxt} states that the value of $K$ is around 265 MeV.
\subsubsection{Symmetry energy}\label{asym}
Stable nuclei with low proton number ($Z$) prefer a nearly equivalent neutron number ($N$). As $Z$ increases the (repulsive) Coulomb interaction between the protons also increases. As can be seen on any table of nuclides, stable nuclei diverge from $N=Z$ (isospin symmetric) nuclei to ones with a $N>Z$ as $Z$ increases. This preference for neutrons is described by the symmetry energy.\\
\\
As a measure of the symmetry energy, the symmetry energy coefficient $a_{4}$ was defined. This coefficient stems from the liquid-drop model of the nucleus and refers to the contribution made by the isospin asymmetry to the energy of the nucleus \cite{waleckatext}. In the semi-empirical mass formula (also known as the droplet formula for nuclear masses), $a_4$ is the coefficient of the
\begin{eqnarray}
	\frac{(N-Z)^2}{A}
\end{eqnarray}
contribution to the mass of the nucleus \cite{csg}, where
	$A=N+Z$.
This coefficient is given by
\begin{eqnarray}
	a_{4} = \frac{1}{2}\left(\frac{\partial^{2}}{\partial t^{2}}\frac{\epsilon}{\rho_b}\right)_{t=0}\mbox{ with }\left(t\equiv\frac{\rho_{n} -
		 \rho_{p}}{\rho_b}\right)\label{a4}
\end{eqnarray}
and
\begin{itemize}
	\item $\epsilon$ is the energy density of the system, while 
	\item $\rho_b$ refers to the baryon density:
	\begin{eqnarray}\label{totbar}
		\rho_b = \rho_p + \rho_n
	\end{eqnarray}
	where $\rho_p$ and $\rho_n$ are the proton and neutron densities respectively.
\end{itemize}
Thus the smaller the value of $a_4$, the more asymmetric the system tends to be. The value of $a_4$ is estimated to be between $31$ and $33$ MeV according to \cite{Lattimer07}, while \cite{csg} and \cite{webertxt} state the value of $a_4$ to be $32.5$ MeV (without specifying any uncertainty).
\subsection{QHD Formalism}\label{sec:QHD}
The most general nuclear matter Lagrangian density, that can encompass the QHD1, NL3 and FSUGold parameter sets, is \cite{diener}
\begin{align}
	\begin{split}\label{fsuglagrangian}
			{\cal L} \ =&\ \bar{\psi}{ (x)}\Big[\gamma^{\mu}\big(i\partial_{\mu}
			- g_{v}V_{\mu}{ (x)} 
			- \frac{g_\rho}{2}{\bm\tau}\cdot{\bm b}_\mu{ (x)}\big)
			- \big(m-g_{s}\phi{ (x)}\big)\Big]\psi{ (x)} \\
			&+\, \frac{1}{2}\partial_{\mu}\phi{ (x)}\partial^{\mu}\phi{ (x)} - \frac{1}{2}m_s^{2}\phi^{2}{ (x)} 
			- \frac{\kappa}{3!}\big(g_s\phi{ (x)}\big)^3 - \frac{\lambda}{4!}\big(g_s\phi{ (x)}\big)^4\\
			&-\, \frac{1}{4}V^{\mu\nu}V_{\mu\nu} + \frac{1}{2}m_\omega^{2}V^{\mu}{ (x)}V_{\mu}{ (x)} + 
			\frac{\zeta}{4!}\big(g_v^2V^\mu{ (x)} V_\mu{ (x)}\big)^2\\
			&-\, \frac{1}{4}{\bm b}^{\mu\nu}\cdot{\bm b}_{\mu\nu} +
			 \frac{1}{2}m_{\rho}^{2}{\bm b}^{\mu}{ (x)}\cdot{\bm b}_{\mu}{ (x)}
			+\, \Lambda_v\big(g_v^2V^\mu{ (x)} V_\mu{ (x)}\big)\big(g_\rho^2{\bm b}^{\mu}{ (x)}\cdot{\bm b}_{\mu}{ (x)}\big),
	\end{split}
\end{align}
where the field tensors have been defined as
\begin{subequations}
	\begin{eqnarray}
		V_{\mu\nu} &=& \partial_\mu V_\nu{ (x)} - \partial_\nu V_\mu{ (x)}\mbox{, and} \\
		{\bm b}_{\mu\nu} &=& \partial_\mu{\bm b}_{\nu}{ (x)} - \partial_\nu{\bm b}_{\mu}{ (x)},
	\end{eqnarray}
\end{subequations}
and
\begin{itemize}
	\item $m$ the nucleon mass (proton and neutron mass are taken to be equal),
	\item $\psi{ (x)}$ the isodoublet baryon field 
	\begin{eqnarray}\label{isodoub}
			\psi{ (x)} = \left[
				\begin{array}{c}
					\psi_p{ (x)}\\
					\psi_n{ (x)}				 
				\end{array}	
				\right],
		\end{eqnarray}
		where $\psi_p{ (x)}$ is the proton field and $\psi_n{ (x)}$ is the neutron field,
	\item $\phi{ (x)}$ the sigma (scalar) meson field with coupling constant $g_s$,
	\item $V^\mu{ (x)}$ the omega (vector) meson field with coupling constant $g_v$, and
	\item ${\bf b}^\mu{ (x)}$ is the Lorentz vector field denoting the three isospin components of the rho meson fields,
	\begin{eqnarray}
		{\bf b}^\mu{ (x)} = \big(b_1^\mu{ (x)}, b_2^\mu{ (x)}, b_3^\mu{ (x)}\big)\,,
	\end{eqnarray}
	with coupling constant $g_\rho$.\\
	The charged rho meson fields ($\rho^\pm$) can be constructed in terms of the first two components of 
	${\bm b}^\mu{ (x)}$ as \cite{csg}
		\begin{eqnarray}
			b^\mu_\pm{ (x)} = \frac{1}{\sqrt{2}}\big(b^\mu_1{ (x)} \pm b^\mu_2{ (x)}\big) \label{oprho},
		\end{eqnarray}
		while
	\item ${\bm \tau} = (\tau_1, \tau_2, \tau_3)$ is the isospin operator. This operator is described in terms of the 
	Pauli 2$\times$2 spin-matrices as
		\begin{eqnarray}
			{\bm \tau}  = \left[
						\begin{array}{cc}
							{\bm \sigma} & 0\\
							0 & {\bm \sigma}				 
						\end{array}	
					\right].		
		\end{eqnarray}
		Since $\psi{ (x)}$ (\ref{isodoub}) is an isodoublet spinor consisting of two 4$\times$1 Dirac spinors is has the 
		total dimension of $\psi{ (x)}$ is 8$\times$1. Therefore, $\bm \tau$ is in actual fact given by
			\begin{eqnarray}
				{\bm \tau} = {\bm \sigma} \otimes \dblone_4,
			\end{eqnarray}
	and hence the explicit expression for $\tau_3$ is
	\begin{eqnarray}
		\tau_3 
		= 
		\sigma_3\otimes 
		\dblone_4
		= \left[\begin{array}[h]{cc}\dblone_4 & 0\\0 & -\dblone_4\end{array}\right]\label{tau3}.
	\end{eqnarray}
	The eigenvalues of $\tau_3$ are $\tau_0$ with
	\begin{eqnarray}
		\tau_0 = 
		\left\{\begin{array}{cc}
			1&\mbox{for protons}\\
			-1&\mbox{for neutrons}
		\end{array}\right..
	\end{eqnarray}
\end{itemize}
The Lagrangian is constructed by including the free-field Lagrangians for all fields (representing different particles) present in the description. As will become clear from the equations of motion of the different meson fields, the meson (boson) fields are coupled to the different baryon (fermion) densities and currents in the simplest way (one boson exchange) such that the baryons are the source of the meson fields. The self-coupling terms in the meson fields were introduced to achieve a better match between the calculated and observed properties of nuclear matter at the nuclear saturation density \cite{recentprogress}.
\subsection{Photon field}
In general the Coulomb interaction is not included in the description of neutron star matter and for this reason $A^0$ is always chosen to be zero (see section \ref{sec:eqm} for more details) \cite{diener}. As we are dealing with unmagnetised matter in this chapter the effect of the photon field will not be considered here.
\subsection{Equations of motion}
Using the Euler-Lagrange equation \cite{goldstein},
\begin{eqnarray}
	\partial_\nu\left(\frac{\partial {\cal L}}{\partial (\partial_\nu \phi_\alpha)}\right) - 
	\frac{\partial {\cal L}}{\partial \phi_\alpha} = 0\label{EL},
\end{eqnarray}
where $\phi_\alpha$ refers to a general field, the equations of motion of the different fields are
\begin{subequations}\label{EQM}
	\begin{align}
			\partial_{\mu}\partial^{\mu}\phi{ (x)} + m_s^{2}\phi{ (x)} + \frac{\kappa}{2!}g_s^3\phi{ (x)}^2 + \frac{\lambda}{3!}g_s^4\phi{ (x)}^3  
			&= g_{s}\bar{\psi}{ (x)}\psi{ (x)}\label{EQM1},\\
			\partial_{\mu}V^{\mu\nu} + m_\omega^{2}V^{\nu}{ (x)} + \frac{\zeta}{3!}g_v^4V_\nu{ (x)}^2 V^\nu{ (x)} + 
			2\Lambda_vg_v^2V^\nu(x) g_\rho^2{\bm b}^{\mu}{ (x)}\cdot{\bm b}_{\mu}{ (x)} &=
			g_{v}\bar{\psi}{ (x)}\gamma^{\nu}\psi{ (x)}\label{EQM2},\\
			\begin{split}
			\partial_{\mu}{\bm b}^{\mu\nu} + m_{\rho}^{2}{\bm b}^{\nu}{ (x)} + 
			2\Lambda_v g_v^2V^\nu{ (x)} V_\nu{ (x)} g_\rho^2{\bm b}^\nu{ (x)}
			=\ &\frac{g_\rho}{2}\bar{\psi}{ (x)}\gamma^{\nu}{\bm \tau}\psi{ (x)} 
			\end{split}\label{EQM3},\\
\mbox{and \ \ }
			 \Big[\gamma^{\mu}\big(i\partial_{\mu}-g_{v}V_{\mu}{ (x)} - \frac{g_\rho}{2}{\bm \tau}\cdot{\bf b}_\mu{ (x)}\big)
			-(m-g_{s}\phi{ (x)})\Big]\psi{ (x)} &= 0.\label{EQM4}
	\end{align}
\end{subequations}
If the self-coupling terms in equations (\ref{EQM1}) to (\ref{EQM3}) are ignored, equation (\ref{EQM1}) is the Klein-Gordon equation with scalar source term, while equations (\ref{EQM2}) and (\ref{EQM3}) are the Proca equation for massive vector boson coupled to a conserved baryon current. Equation (\ref{EQM4}) is the Dirac equation with scalar and vector field introduced in a minimal fashion \cite{recentprogress}. \\
\\
Obtaining solutions to these equations can be very difficult, since they are non-linear and coupled. Thus the solutions will have to be approximated. We will use the relativistic mean-field approximation to do just that. 
\section{Relativistic mean-field approximation}\label{sec:RMF}
Since the coupling constants in QHD are large, a perturbative expansion, as employed in theories like Quantum Electrodynamics and high energy Quantum Chromodynamics, is not feasible. Instead the meson (boson) fields are replaced by their ground state expectation values, which are classical fields. This is called the relativistic mean-field (RMF) approximation, also known as the Relativistic Hartree approximation.\\
\\
Considering field operators, the RMF approximation is the same as Fourier expanding the boson operators and only keeping the zeroth modes (as only these modes survives when the expectation value is taken with regards to a translational invariant ground state). 
Since only the zero modes are considered, these solutions must be the ones corresponding to a minimum in the energy.\\
\\
The RMF approximated equations of motion of the meson fields will be solved self-consistently. 
Self-consistency is a central theme of the RMF approximation and our calculation: we will initially assume the ground state to have certain properties and based on these assumptions the very same properties of the ground state will be evaluated. Self-consistency is achieved when the calculated properties match the original assumptions.\\
\\
In this chapter we will assume that the RMF ground state is translational as well as rotational invariant. We proceed by making the RMF approximation based on these assumptions and then evaluate whether these properties are indeed present in the RMF ground state. 
\subsection{Boson operators}
The RMF approximation implies that \cite{FSU1}
\begin{subequations}\label{rmfreplace}
\begin{eqnarray}
		\phi{ (x)} &\longrightarrow& \left\langle \phi{ (x)}\right\rangle = \phi_{0},\\
		V^{\mu}{ (x)} &\longrightarrow& \left\langle V^\mu{ (x)}\right\rangle = g^{\mu 0}V_{0},\mbox{ and}\\
		{\bm b}^{\mu}{ (x)} &\longrightarrow& \left\langle b^{\mu}_a{ (x)}\right\rangle = g^{\mu 0}\delta_{a3}b_0.
\end{eqnarray}
\end{subequations}
The spatial components of the vector boson fields ($V^{\mu}$ and ${\bm b}^{\mu}$) vanish due to the rotational symmetry of the ground state since in such a ground state there can be no preferred direction. As mentioned already, this symmetry is assumed to be present, since at this stage we cannot show it explicitly (the ground state will be discussed in section \ref{ssec:gs}).\\
\\
Furthermore only the third component of ${\bm b}^{\mu}$, that describes the neutral rho meson $\rho^0$, survives. This is because the first two components of ${\bm b}^\mu$ can be written in terms of raising and lowering operators of the charged rho meson fields (\ref{oprho}), hence only the third component has a non-vanishing expectation value in the RMF approximation \cite{csg}. 
\subsection{Fermion operators and sources}
In the RMF approximation only the boson operators get replaced by their expectation values and 
$\psi{ (x)}$ remains an operator. Since the baryons densities are the sources of meson fields in equations (\ref{EQM1}) to (\ref{EQM3}) these sources have to be replaced by their normal-ordered ground state expectation values in the RMF approximation in order to be consistent with (\ref{rmfreplace}). Therefore the following substitutions, where $\left|\Phi\right\rangle$ is the ground state, also need to be made:
\begin{subequations}\label{nucRMF}
	\begin{eqnarray}
		\bar\psi{ (x)}\psi{ (x)} &\longrightarrow &
		\left\langle\Phi\right| {\bm :}\bar\psi{ (x)}\psi{ (x)}{\bm :}\left|\Phi\right\rangle = 
		\left\langle \bar{\psi}\psi\right\rangle,\\
	\bar\psi{ (x)}\gamma^\mu\psi{ (x)} &\longrightarrow& 
	\left\langle\Phi\right| {\bm :}\bar\psi{ (x)}\gamma^\mu\psi{ (x)}{\bm :}\left|\Phi\right\rangle = 
	\left\langle \bar{\psi}\gamma^0\psi\right\rangle\label{rmfomega0},\mbox{ and}\\
	\bar\psi{ (x)}\gamma^\mu\tau_a\psi{ (x)} &\longrightarrow &
	\left\langle\Phi\right| {\bm :}\bar\psi{ (x)}\gamma^\mu\tau_a\psi{ (x)}{\bm :}\left|\Phi\right\rangle =
	 \left\langle \bar{\psi}\gamma^0\tau_3\psi\right\rangle.
\end{eqnarray}
\end{subequations}
The normal-ordered ground state expectation value is taken since we will ignore the contribution of the filled negative energy baryon states, as the vacuum has a (infinite!) constant energy. This is known as the no-sea approximation \cite{NL3}.\\
\\
To be consistent with (\ref{rmfreplace}), the expectation values of the spatial components of the vector currents must also be zero. For a rotational invariant ground state this property is obvious: rotating any vector current by $\pi$ radians will give the negative of the original current, but since the ground state (source of the current) is rotationally invariant this must be equal to the original value of the current. Thus vector currents must be zero.
\subsection{Equations of motion and baryon spectrum}\label{ssec:rmfeqm}
In the RMF approximation, the equations of motion (\ref{EQM}) reduce to
\begin{subequations}\label{MFTFSUEQM}
	\begin{eqnarray}
			g_s\phi_0 &=& \frac{g_{s}^2}{m_s^2}\left[\left\langle\bar\psi\psi\right\rangle - 
			\frac{\kappa}{2}(g_s\phi_0)^2 - \frac{\lambda}{6}(g_s\phi_0)^3\right],\label{FSUsigmaEQM}\\
			g_{v}V_0 &=& \frac{g_{v}^2}{m_\omega^2}\left[\left\langle \psi^{\dagger}\psi\right\rangle - 
			\frac{\zeta}{6}(g_vV_0)^3 -  2\Lambda_v(g_vV_0)(g_\rho b_0)^2\right],\label{FSUomegaEQM}\\
			g_{\rho}b_0 &=&
			 \frac{g_\rho^2}{m_\rho^2}\left[\frac{1}{2}\left\langle\psi^\dagger\tau_3\psi\right\rangle - 
			 2\Lambda_v(g_vV_0)^2(g_\rho b_0)\right],\mbox{ and}\label{FSUrhoEQM}\\
			 0 &=& 
			\left[i\gamma^{\mu}\partial_{\mu}-g_{v}\gamma^0V_{0} - \frac{g_\rho}{2}\tau_3\gamma^0b_0
			- (m-g_{s}\phi)\right]\psi{ (x)}.\label{FSUbaryon}
	\end{eqnarray}	
\end{subequations}
Of particular interest is equation (\ref{FSUbaryon}) which, in essence, is the free Dirac equation with modified mass and energy. Thus we assume the solution for $\psi{ (x)}$ is of the form
\begin{eqnarray}
	\psi{ (x)} = \psi({\bm k},s)\,e^{i{\bm k}\cdot {\bm x} - ie({\bm k},s)t}\,.\label{Kexpand}
\end{eqnarray}
Here $\psi({\bm k},s)$ is the four component Dirac spinor ($s$ denotes the spin index) and $e({\bm k},s)$ the energy associated with specific momentum state\footnote{In natural units the momentum and wave vectors are equivalent.}, denote by ${\bm k}$, with spin $s$ \cite{walecka}. Substituting equation (\ref{Kexpand}) into equation (\ref{FSUbaryon}) yields
\begin{eqnarray}
	\big(\,-\gamma_ik^i + \gamma_0e({\bm k},s) - g_v\gamma_0V_0 - \frac{g_\rho}{2}\tau_3\gamma_0b_0 - (m-g_s\phi_0)\big)\psi ({\bm k},s) = 0\,.\label{HD}
\end{eqnarray}
Reverting to the notation of the Dirac matrices (${\bm \alpha}$ and $\beta$), as well as considering only one of the baryon species in the isospin doublet (\ref{isodoub}), equation (\ref{Kexpand}) can be re-written as
\begin{align}\label{nucmat}
	\begin{split}	
	\Big(e({\bm k},s) - g_vV_0-\frac{g_\rho}{2}\tau_3b_0\Big) \psi({\bm k},s) 
	\ &=\ \ \Big({\bm \alpha}\cdot{\bm k} + \beta m^*\Big)\, \psi({\bm k},s)\\
	&=\ \left(
			\begin{array}{cc}
				m^*\dblone_2&{\bm \sigma}\cdot{\bm k}\\
				{\bm \sigma}\cdot{\bm k}&-m^*\dblone_2			 
			\end{array}	
		\right)\psi({\bm k},s)\\
	&=\  \ E({\bm k},s)\ \psi({\bm k},s),
	\end{split}
\end{align}
using the convention established by (\ref{enerconven}) and where 
$m^*$ is the reduced nucleon mass:
\begin{eqnarray}\label{mstar}
	m^* = (m-g_s\phi_0).
\end{eqnarray}
From (\ref{nucmat}) we can deduce that $\psi({\bm k},s)$ will be of the form
\begin{eqnarray}
	 \psi({\bm k},s)\propto
	 \ \left(
		\begin{array}{c}
			\phi({\bm k},s)\\
			\chi({\bm k},s)
		\end{array}	
	\right)
\end{eqnarray}
and it can be easily shown that it is indeed
\begin{eqnarray}
	 \chi({\bm k},s)=\frac{{\bm \sigma}\cdot{\bm k}}{E({\bm k},s)+m^*}\phi(s)
\end{eqnarray}
with
\begin{eqnarray}
		\phi(s) = 
		\left\{\begin{array}{cc}
			\left[\begin{array}{c}
			1\\
			0
		\end{array}\right]&\mbox{for }s=1\\
		\\
			\left[\begin{array}{c}
			0\\
			1
		\end{array}\right]&\mbox{for }s=-1
		\end{array}\right.
\end{eqnarray}
representing the two spin species.
As discussed in \cite{diener}, 
the eigenvalues of $\psi({\bm k},s)$ are 
\begin{eqnarray}
	e ({\bm k},s)= g_v V_0 + \frac{1}{2}g_\rho\,\tau_0\,b_0 +\sqrt{{\bm k}^2 + {m^*}^2}.\label{ekfsu}
\end{eqnarray}
%
%
\subsection{General densities}
Since we are considering the system in the mean-field approximation $\psi{ (x)}$ is not the quantity of interest but rather the various (average) densities, the sources of the different meson fields, in (\ref{MFTFSUEQM}). As the ground state is assumed to be translational invariant these densities will not depend 
on $x$ and hence from this point onwards the $x$-dependency of $\psi{ (x)}$ will be suppressed.\\
\\
The average density of a general operator $\Gamma$ is calculated by considering the individual contributions from all the occupied momentum states in the form of \cite{csg}
\begin{eqnarray}\label{gendensity}
	\left\langle\psi^\dagger\Gamma
	\psi\right\rangle  &=& 
	\sum_s\int\frac{d{\bm k}}{(2\pi)^3}\left(\psi^\dagger\Gamma
	\psi\right)_{{\bm k},s}\,
	\Theta\big[\,\mu-e({\bm k},s)\big],\label{GENdens}
\end{eqnarray}
where
\begin{itemize}
	\item $\Gamma$ can be any operator related to a specific density in (\ref{MFTFSUEQM}), 
	\item $e({\bm k},s)$ are the positive single-particle energies, since the negative energy (anti-particle) states are not considered, 
	\item $\mu$ is the chemical potential/Fermi energy\footnote{In the zero temperature case the Fermi energy and the chemical potential are equivalent.}, 	
	\item $\Theta[\,\mu-e({\bm k},s)]$ is a step function with 
	\begin{eqnarray}
		\Theta[\,\mu-e({\bm k},s)] = 
		\left\{\begin{array}{cc}
			1&\mbox{if}\  e({\bm k},s)\leq\mu\\
			0&\mbox{if}\ e({\bm k},s)>\mu
		\end{array}\right.\mbox{, and}	
	\end{eqnarray}
	\item $\left(\psi^\dagger\Gamma\psi\right)_{{\bm k},s}$ is the single particle expectation value with regards to $\Gamma$ of the single particle spinors which are normalised to one, so that $\left(\psi^\dagger\psi\right)_{{\bm k},s}=1\ \forall \ {\bm k},s$.
\end{itemize}
\subsection{RMF ground state and vector densities}\label{ssec:gs}
We can now return to the question of whether the assumptions and the substitutions made in (\ref{rmfreplace}) are indeed consistent with the assumed properties of the ground state. These assumptions are that the ground state is
\begin{itemize}
	\item translational invariant,
	\item rotational invariant,
	\item static, and
	\item has definite spin and parity.
\end{itemize}
These points, as well as the properties of the ground state that support them, have been well documented in the literature (see \cite{recentprogress} and references therein). However, we have to belabour this point in the light of the coming chapters, where the rotational invariance of the ground state will be broken due to the presence of the magnetic field. In order to construct the baryon ground state the baryon operator $\psi{ (x)}$ must be known. However, it is quite tedious to construct and since we actually only need to know the characteristics of the ground state, it would be preferable if its properties can be deduced in some other way. Since the baryons spectrum reflects the properties of the ground state, once we know the spectrum we can deduce all the necessary characteristics.\\
\\
The main question is therefore: what does the assumption that the meson fields are classical, time-independent fields imply about the ground state? This question is of importance since the meson fields are coupled to baryon sources (densities). Since we have a plane-wave solution for $\psi{ (x)}$ the temporal dependency vanishes when densities of the form of (\ref{gendensity}) is considered.\\
\\
Regarding the symmetry of the ground state: from the energies (\ref{ekfsu}) there is no preferences for a given direction and the energy is only dependent on the magnitude of $\bm k$, which is indicative of rotational invariance. To investigate the vector densities which we set to zero in (\ref{nucRMF}), namely ones with the form of 
	$\left\langle\bar\psi\gamma^i\psi\right\rangle$,
from (\ref{gendensity}) we need to consider $\left(\bar\psi\gamma^i\psi\right)_{{\bm k},s}$, which can be shown to be
\begin{align}
	\begin{split}\label{vecdens}
		\left(\bar\psi\gamma^i\psi\right)_{{\bm k},s}\ &=
		\left(\psi^\dagger{\bm \alpha}\psi\right)_{{\bm k},s}\\
		&=
		\frac{\bm k}{\sqrt{{\bm k}^2 + {m^*}^2}}\ .
	\end{split}
\end{align}
Since the integral runs over all occupied states that have an energy lower than the Fermi energy, the boundaries of the integral can be expressed in terms of the Fermi momentum ${\bm k}^F$. 
Consequently 
the integration is performed for ${\bm k}\leq{\bm k}^F$. 
Thus the integral relating to (\ref{vecdens}) will be zero, since an uneven integrand is integrated over a symmetric interval.\\
\\
As all spatial vector currents are zero, we can deduce that our assumption of RMF approximation and its implications are valid and we indeed have a translational invariant ground state with rotational symmetry.
\subsection{Calculating particle densities}
From (\ref{GENdens}) the particle density, $\rho$, can be constructed using the orthogonality of $\psi{ (x)}$:
\begin{eqnarray}
	\rho&=&\left\langle \psi^\dagger\psi\right\rangle\nonumber\\ 
	 &=& \sum_s\int\frac{d{\bm k}}{(2\pi)^3}\left(\psi^\dagger\psi\right)_{{\bm k},s}\,
	\Theta\big[\,\mu-e({\bm k},s)\big]\nonumber\\
	 &=&\sum_s\frac{1}{(2\pi)^3}\int_0^{k_F} d^3k\nonumber\\
	&=& \sum_s\frac{1}{(2\pi)^3}\int_0^{k_F} dk\,4\pi k^2\nonumber\\
	&=& \frac{2}{3\pi^2}{k_F}^3\,.\label{rhoB}
\end{eqnarray}
%
However, calculating other densities from the explicit construction of $\psi{ (x)}$ is cumbersome.\\
\\
A less labour-intensive method is described in \cite{csg}. This method relies on the fact the RMF approximation seeks out the lowest energy state of the system, which is of course the ground state. Thus, instead of calculating densities from the construction of the matrix elements pertaining to the particular density, 
the energy density is simply minimised with regards to a choice of variable, i.e.
\begin{eqnarray}\label{EQMeps}
	\frac{\partial {\epsilon}}{\partial \phi_\alpha} = 0.
\end{eqnarray}
This point is further illustrated in appendix \ref{ap:RMF}. Consequently the only thing that needs to be constructed explicitly is the energy density, which is part of the equation of state.
\section{Equation of state}
For the purpose of investigating neutron stars in this study the equation of state is the main quantity of interest. Knowing the relationship between the energy density and the pressure of the matter in the interior of the star as a function of density will enable one to calculate the mass-radius relationship of the star.\\
\\
The internal properties of any energy-mass distribution (i.e. matter) are described by the energy-momentum tensor ($T^{\mu\nu}$) of the distribution. 
In general the energy-momentum tensor of a static, spherically symmetric perfect fluid (no viscosity or heat conduction) moving with a velocity ${\bf v}$ is \cite{csg, walecka}
\begin{eqnarray}
	T^{\mu\nu} = -P\eta^{\mu\nu} + (P + \epsilon)u^\mu u^\nu\label{genEMT},
\end{eqnarray}
where
\begin{itemize}
	\item $\epsilon$ is the energy density, 
	\item $P$ is the pressure,
	\item $\eta^{\mu\nu}$ the metric tensor of Minkowski space, and
	\item $u^\mu$ is the four-velocity:
\begin{eqnarray}
	u^\mu &=& \frac{dx^\mu}{d\tau} \nonumber\\
		&=& \sqrt{1-{\bm v}^2}\,\big(1,v^1,v^3,v^3\big) \label{u4velo}
\end{eqnarray}
and 
$u^\mu u_\mu = 1$ \cite{csg}.
\end{itemize}
The Minkowski (flat space) metric $\eta^{\mu\nu}$ is used, since it can be deduced that the change in the curvature of space-time in the interior of the star is such that on the length scale of nucleon interactions the metric is locally flat \cite{csg}.\\ 
\\
As discussed in \cite{diener}, Noether's theorem relates $T^{\mu\nu}$ to the Lagrangian density, $\cal{L}$, as
\begin{eqnarray}
	T^{\mu\nu} &=& \frac{\partial\cal{L}}{\partial (\partial_\mu \phi_\alpha)}\,
	\partial^\nu \phi_\alpha - \cal{L}\eta^{\mu\nu}. \label{EMT}
\end{eqnarray}
Since the fields $\phi_\alpha$ are operators, in the mean-field approximation they also have to be replaced by their ground state expectation values. Thus, considering the general expression of $T^{\mu\nu}$ (\ref{genEMT}), (\ref{EMT}) becomes
\begin{align}
	\begin{split}
		\left\langle T^{\mu\nu}\right\rangle &= \left\langle \frac{\partial\cal{L}}{\partial (\partial_\mu \phi_\alpha)}\,
		\partial^\nu \phi_\alpha - \cal{L}\eta^{\mu\nu}\ \right\rangle \label{EMTi}\\
		&= -P\eta^{\mu\nu} + (P + \epsilon)u^\mu u^\nu
	\end{split}
\end{align}
if a static, spherical symmetric fluid moving with velocity $\textbf{v}$ is consider \cite{walecka}. If $\textbf{v} = 0$, then 
\begin{subequations}\label{FSUeos1}
	\begin{eqnarray}
		\epsilon &=& \left\langle T^{00}\right\rangle \nonumber\\ &=& 
		\left\langle i\bar{\psi}\gamma_{0}\partial_{0}\psi\right\rangle - 
		\left\langle {\cal L}\right\rangle\label{FSUeps1},\mbox{ and}\\
		P&=& \frac{1}{3}\left\langle T^{ii} \right\rangle\nonumber\\&=&
		\frac{1}{3}\left\langle i\bar{\psi}\gamma^{i}\partial_{i}\psi\right\rangle + \left\langle {\cal L}\right\rangle 
		\label{FSUpres1},
	\end{eqnarray}
\end{subequations}
where
$\left\langle {\cal L}\right\rangle$ is the ground state expectation value of $\cal L$.\\
\\ 
The pressure can also be thermodynamically linked to $\epsilon$ and $\rho$ through the first law of thermodynamics as \cite{walecka}
\begin{eqnarray}
	P=\rho^2 \frac{\partial}{\partial \rho}\left(\frac{\epsilon}{\rho}\right),
\end{eqnarray}
which can be shown to be equivalent to
\begin{eqnarray}
	P = \sum_i \mu_i \rho_i -\epsilon\label{pres}
\end{eqnarray}
where $i$ labels the chemical potentials and densities of all particles present in the energy density.
\subsection{Energy density}
Using the expansion of $\psi$ (\ref{Kexpand}) the first term in $\epsilon$ from (\ref{FSUeps1}) reduces to
\begin{align}
	\begin{split}	
	\left\langle i\bar{\psi}\gamma_{0}\partial_{0}\psi\right\rangle
	&= \sum_\alpha \sum_s\int\frac{d{\bm k}}{(2\pi)^3}\,e_\alpha({\bm k},s)\,\Theta[\,\mu_\alpha-e_\alpha({\bm k},s)]\\ 
	&=  \sum_\alpha \sum_s\int\frac{d{\bm k}}{(2\pi)^3}\,\big(g_v V_0 + \frac{g_\rho}{2}\,\tau_0\,b_0 +\sqrt{{\bm k}^2 
	+ {m^*}^2}\big)\,\Theta[\,\mu_\alpha-e_\alpha({\bm k},s)]\\
	&=  \sum_\alpha\sum_s\left(\int_0^{k^F_\alpha}\frac{d{\bm k}}{4\pi^3}\sqrt{{\bm k}^2 + {m^*}^2}\right)
	+g_v V_0(\rho_n+\rho_p) + \frac{g_\rho}{2}b_0\big(\rho_p-\rho_n\big)\label{epsFSUexp}	,
	\end{split}
\end{align}
where
\begin{itemize}
	\item the sum over $\alpha$ refers to protons and neutrons, and
	\item $k^F_\alpha$ is the magnitude of the Fermi momentum.
\end{itemize}
From (\ref{FSUeps1}) the energy density is
\begin{eqnarray}\label{epsi}
	\epsilon\!
	&=& \sum_\alpha\sum_s\left(\int_0^{k^F_\alpha}\frac{d{\bm k}}{4\pi^3}\,\sqrt{{\bm k}^2 + {m^*}^2}\right)
	+g_v V_0(\rho_n+\rho_p) + \frac{g_\rho}{2}b_0\big(\rho_p-\rho_n\big)\\
	&\,&+ \frac{1}{2}m_s^{2}\phi^{2}_0 + \frac{\kappa}{3!}\big(g_s\phi_0\big)^3 + \frac{\lambda}{4!}\big(g_s\phi_0\big)^4- \frac{1}{2}m_\omega^{2}V_0^2 
	- \frac{\zeta}{4!}\big(g_vV_0\big)^4- \frac{1}{2}m_{\rho}^{2}b_0^2 - \Lambda_v\big(g_vV_0\big)^2\big(g_\rho b_0\big)^2\nonumber
\end{eqnarray}
since $\left\langle {\cal L}\right\rangle$ is given by 
\begin{align}
	\begin{split}
		\left\langle {\cal L}\right\rangle \ 
		=&\ - \frac{1}{2}m_s^{2}\phi^{2}_0 
		- \frac{\kappa}{3!}\big(g_s\phi_0\big)^3 - \frac{\lambda}{4!}\big(g_s\phi_0\big)^4
		+ \frac{1}{2}m_\omega^{2}V_0^2 + \frac{\zeta}{4!}\big(g_vV_0\big)^4 \\
		&\ + \frac{1}{2}m_{\rho}^{2}b_0^2 + \Lambda_v\big(g_vV_0\big)^2\big(g_\rho b_0\big)^2\label{varL}
	\end{split}
\end{align}
when the expectation value of ${\cal L}$ (\ref{fsuglagrangian}) is calculated using the RMF ground state and (\ref{FSUbaryon}) is also considered.\\
\\
Once the energy density is known, the pressure and the scalar density, $\ave{\bar\psi\psi}$ of (\ref{FSUsigmaEQM}), can be constructed. Note that from (\ref{GENdens}) the other densities in (\ref{MFTFSUEQM}) are simply the total baryon densities and the isospin density (difference between the proton and neutron densities), namely
\begin{subequations}
	\begin{eqnarray}
			\left\langle \psi^{\dagger}\psi\right\rangle &=& \rho_p+\rho_n\ =\ \rho_b\mbox{,\  and}\\
			\left\langle\psi^\dagger\tau_3\psi\right\rangle&=&\rho_p-\rho_n.\label{isorho}
	\end{eqnarray}	
\end{subequations}
\subsection{Pressure}
From (\ref{pres}), using (\ref{epsi}) for $\epsilon$, the pressure in nuclear matter is given by
\begin{eqnarray}\label{presi}
	P\!
	&=&  \sum_\alpha\sum_s\left(
	\mu_\alpha\rho_\alpha-\int_0^{k^F_\alpha}\frac{d{\bm k}}{4\pi^3}\,\sqrt{{\bm k}^2 + {m^*}^2}\right)\\
	&\,&- \frac{1}{2}m_s^{2}\phi^{2}_0 - \frac{\kappa}{3!}\big(g_s\phi_0\big)^3 - \frac{\lambda}{4!}\big(g_s\phi_0\big)^4 + \frac{1}{2}m_\omega^{2}V_0^2 
	+ \frac{\zeta}{4!}\big(g_vV_0\big)^4 + \frac{1}{2}m_{\rho}^{2}b_0^2 + \Lambda_v\big(g_vV_0\big)^2\big(g_\rho b_0\big)^2.\nonumber
\end{eqnarray}
\subsection{Scalar density}
Deriving (\ref{FSUsigmaEQM}) using (\ref{EQMeps}):
\begin{eqnarray}
	\frac{\partial {\epsilon}}{\partial \phi_0} = 0=\sum_\alpha\sum_s\left(\int_0^{k^F_\alpha}\frac{d{\bm k}}{4\pi^3}\frac{\partial}{\partial \phi_0} \sqrt{{\bm k}^2 + {m^*}^2}\right)+ m_s^{2}\phi_0 + \frac{\kappa}{2}g_s^3\phi_0^2 + \frac{\lambda}{6}g_s^4\phi_0^3\label{scal}
\end{eqnarray}
it is deduced that
\begin{eqnarray}
	\left\langle\bar\psi\psi\right\rangle=\sum_\alpha\sum_s\int_0^{k^F_\alpha}\frac{d{\bm k}}{4\pi^3}\frac{m^*}{\sqrt{{\bm k}^2 + {m^*}^2}}
\end{eqnarray}
by comparing the above to (\ref{FSUsigmaEQM}) and keeping in mind that $m^*=m-g_s\phi_0$.\\
\\
Up to this point we have only dealt with the hadron contributions to the star's equation of state. However, as will become apparent in the next section, other particles also need to be considered when equilibrated systems are investigated. 
\section{Equilibrium conditions}\label{sec:eqm}
A neutron star is 
stabilised against gravitational collapse by the degeneracy pressure of the nuclear matter in the star's interior \cite{csg}. Therefore, since a neutron star is bound by gravity and not the nuclear strong force, a star consisting out of only positively charged (protons) and neutral particles (neutrons and mesons\footnote{The charged rho mesons are not considered in the RMF approximation, see section \ref{sec:RMF}}) would not be stable: the range of Coulomb potential is much greater than that of the nuclear potential and 
a charged star would be ripped apart by the Coulomb repulsion. Hence the star must be charge neutral and thus leptons must also be considered in our description of neutron star matter. Electrons as well as muons (heavy electrons) will be included. 
\subsection{Leptons and neutrinos}
Muons will be assumed to be present if the Fermi energy of the electron reaches the muon rest mass energy of 105.658 MeV \cite{PDGmuon}. Energetic muons decays to electrons via 
\begin{eqnarray}
	\mu^{-} \rightarrow e^{-} + \bar{\nu}_{e} + \nu_\mu\,.
\end{eqnarray}
Chemical equilibrium with regards to the above reaction implies that
\begin{eqnarray}
	\mu_e = \mu_\mu\label{muoneqm},
\end{eqnarray}
where
$\mu_e$ and $\mu_\mu$ are the electron and muon chemical potentials respectively; $\bar{\nu}_{e}$ is the electron anti-neutrino and $\nu_\mu$ are muon neutrino. The effect of the neutrinos are not considered in this study since the neutrinos are very weakly interacting and assumed to simply diffuse out of the system \cite{csg}.
\subsection{Charge neutrality}
Equating the proton and lepton densities ensures charge neutrality:
\begin{eqnarray}
	\rho_p = \rho_{l}=\rho_e+\rho_\mu,
\end{eqnarray}
where $\rho_p$ refers to the proton density and $\rho_{l}$ to the lepton density. 
\subsection{$\beta$-equilibrium}
Since neutron stars are stable, long-lived objects we are interested in general equilibrium configurations of the star. The outer crust of the star is assumed to be composed of iron atoms in a lattice \cite{BPS} (for a modern calculation of the crustal equation of state, see \cite{Rus}). As the density increase the matter will become more neutron-rich, since it becomes energetically favourable for protons and electrons to undergo inverse $\beta$-decay:
\begin{eqnarray}\label{invbetadec}
		p + e^{-} \rightarrow n + \nu_{e}\,.
\end{eqnarray}
Further into the crust the neutron-drip line is reached and neutrons will start to be unbound and leach from the nuclei, marking the start of the inner crust of the star. Nuclei in the inner crust are still confined to a lattice permeated by a free neutron fluid. Due to the competition between Coulomb and the nuclear (strong) interaction at the densities in the inner crust the nuclei may assume various shapes. This matter phase is referred to in the literature as the {\em pasta phase}, but is beyond the scope of this work; for one of the most recent articles see \cite{pasta}. The transition from the inner crust to core is where all structure breaks down and hence the core of the star essentially consists of a liquid of neutrons, protons and leptons (although the protons and leptons will constitute a minority).\\
\\
Since free neutrons have a short lifetime compared to that of the star (about 10 minutes under normal conditions \cite{firestone}) before undergoing $\beta$-decay,
\begin{eqnarray}\label{betadecay}
	n \rightarrow p + e^{-} + \bar{\nu}_{e}\,,
\end{eqnarray}
the equilibrium state of a closed, dense and time-evolved system, such as a neutron star interior, would be $\beta$-equilibrated and thus in equilibrium with regards to (ignoring the neutrinos)
\begin{eqnarray}
	n \rightleftharpoons p + e^{-} \,.
\end{eqnarray}
This will entail that the relations between the chemical potentials of the particles must be
\begin{eqnarray}
	\mu_n = \mu_p + \mu_e\label{beta},
\end{eqnarray}
where $\mu_n$ denotes the neutron chemical potential.\\
\\
Considering these equilibrium conditions we are now in a position to calculate the equation of state of a stable neutron star consisting out of protons, neutrons, electrons and muons. 
\section{Nuclear matter observables}
Most of the nuclear matter properties we can calculate directly from the expressions in section \ref{sec:satnuc}. However, for the symmetry energy we will derive a simplified expression. 
\subsection{Symmetry energy}
As shown in \cite{diener} for unmagnetised nuclear matter the symmetry energy coefficient (\ref{a4}) can be written in terms of the magnitude of the Fermi momentum ($k^F$) of the baryons as
\begin{eqnarray}
	a_{4} = \frac{(k^F)^2}{6\sqrt{(k^F)^2 + m^{*2}}}+\frac{(k^F)^3}{12\pi^2}\left(\frac{g_\rho^2}{m_\rho^2 + 2\Lambda_vg_\rho^2(g_vV_0)^2}\right)\label{a4unmag}
\end{eqnarray}
since in unmagnetised nuclear matter $k^F$ is the same for protons and neutrons. However, a more general expression, that can also be applied to magnetised matter, is obtained when $a_4$ is expressed in term of the Fermi energies of the baryons. In appendix \ref{ap:a4} we show that in this case, $a_4$ is
\begin{eqnarray}
	a_{4} = \frac{1}{4}\left(\left.\frac{d \mu'_n}{dt}\right|_{t=0}-\left.\frac{d \mu'_p}{dt}\right|_{t=0}\right)
	+\frac{1}{8}\left(\frac{g_\rho^2\, \rho_b}{m_\rho^2 + 2\Lambda_vg_\rho^2(g_vV_0)^2}\right)\label{a4mag},
\end{eqnarray}
where
\begin{itemize}
	\item $\mu'$ is the Fermi energy equivalent of $E(\bm k,s)$ of (\ref{enerconven}), i.\,e. for neutrons
	\begin{eqnarray}
		\mu'_n = \mu_n - g_v V_0 + \frac{1}{2}g_\rho\,b_0,\mbox{ and}
	\end{eqnarray}
	\item $t$ is once again 
	\begin{eqnarray}\label{a4t}
		t\equiv\frac{\rho_{n} -\rho_{p}}{\rho_b}.\nonumber
	\end{eqnarray}
\end{itemize}
\section{Neutron star matter}\label{sec:NSmat}
Including the leptons, the most general RMF Lagrangian density describing neutron star matter is
\begin{eqnarray}
		{\cal L} \!&=&\bar{\psi}\Big[i\gamma^{\mu}\partial_{\mu} - g_{v}\gamma^{0}V_{0}
			- \frac{g_\rho}{2}\gamma^{0}\tau_3b_0
			- \big(m-g_{s}\phi_0\big)\Big]\psi+
			\sum_l \bar{\psi}_l\Big(i\gamma^{\mu}\partial_{\mu} - m_l\Big)\psi_l\\
			&\,&- \frac{1}{2}m_s^{2}\phi^{2}_0 
			- \frac{\kappa}{3!}\big(g_s\phi_0\big)^3 - \frac{\lambda}{4!}\big(g_s\phi_0\big)^4
			+ \frac{1}{2}m_\omega^{2}V_0^2
			+ \frac{\zeta}{4!}\big(g_vV_0\big)^4 
			+ \frac{1}{2}m_{\rho}^{2}b_0^2 + \Lambda_v\big(g_vV_0\big)^2\big(g_\rho b_0\big)^2\nonumber	\label{NSL}
\end{eqnarray}
with sum over $l$ implying the lepton species. \\
\\
Since the values of the different coupling constants are fixed (but dependent on which parameter set use) the only unknown quantities are the meson and fermion fields. When the equations of motion of the different fields as well as the imposed equilibrium conditions are considered it is clear that there is only one free parameter which can be arbitrarily specified: the total baryon density. Thus all observables will be calculated as a function of the total baryon density, $\rho_b$ (\ref{totbar}). 
%
%
%
\subsection{Energy density}
Including the leptons will not modify the energy density substantially.  From the single particle baryon energies (\ref{ekfsu}) the lepton single particle energies can be deduced to be
\begin{eqnarray}
	e_l ({\bm k},s)=  \sqrt{{\bm k}^2 + {m_l}^2}.
\end{eqnarray}
For the energy density of neutron star matter we include the contribution of the leptons to the $\epsilon$ for baryons (\ref{epsi}) and it becomes
\begin{align}
	\begin{split}\label{epsii}
	\epsilon
	=&
	\sum_\alpha\sum_s\left(\int_0^{k^F_\alpha}\frac{d{\bm k}}{4\pi^3}\,\sqrt{{\bm k}^2 + {m^*}^2}\right)+ 
	\sum_l\sum_s\left(\int_0^{k^F_l}\frac{d{\bm k}}{4\pi^3}\,\sqrt{{\bm k}^2 + {m_l}^2}\right)\\
	&+ \frac{1}{2}m_s^{2}\phi^{2}_0 +
	\frac{\kappa}{3!}\big(g_s\phi_0\big)^3 + \frac{\lambda}{4!}\big(g_s\phi_0\big)^4+g_v V_0(\rho_n+\rho_p) - \frac{1}{2}m_\omega^{2}V_0^2\\
	&- \frac{\zeta}{4!}\big(g_vV_0\big)^4 + \frac{g_\rho}{2}b_0\big(\rho_p-\rho_n\big)- \frac{1}{2}m_{\rho}^{2}b_0^2 - 
	\Lambda_v\big(g_vV_0\big)^2\big(g_\rho b_0\big)^2.
	\end{split}
\end{align}
\subsection{Pressure}
Using the equilibrium conditions introduced in section \ref{sec:eqm} the expression for the pressure (\ref{pres}) simplifies to
\begin{align}
	P\ =&\ \mu_n(\rho_n+\rho_p)-\epsilon\label{presiii}\\
	=&\ \sum_\alpha\left(\mu_n\rho_\alpha-\sum_s\int_0^{k^F_\alpha}\frac{d{\bm k}}{4\pi^3}\,\sqrt{{\bm k}^2 + {m^*}^2}\right)
	-\sum_l\sum_s\left(\int_0^{k^F_l}\frac{d{\bm k}}{4\pi^3}\,\sqrt{{\bm k}^2 + {m_l}^2}\right)\label{presii}\\
	&- \frac{1}{2}m_s^{2}\phi^{2}_0 - \frac{\kappa}{3!}\big(g_s\phi_0\big)^3 - \frac{\lambda}{4!}\big(g_s\phi_0\big)^4 + \frac{1}{2}m_\omega^{2}V_0^2 
	+ \frac{\zeta}{4!}\big(g_vV_0\big)^4 + \frac{1}{2}m_{\rho}^{2}b_0^2 + \Lambda_v\big(g_vV_0\big)^2\big(g_\rho b_0\big)^2.\nonumber
\end{align}
\section{Summary}
The equation of state of unmagnetised neutron star matter in general equilibrium was derived. These calculations will be the base for deriving the equation of state of magnetised neutron star matter in general equilibrium. As a first step in that direction the equation of state of magnetised neutron matter will be derived in the following chapter.

\chapter{Ferromagnetism in neutron matter}\label{chap:ferroneu}
The aim of this chapter is to establish a description of magnetised neutron matter.  In particular ferromagnetised neutron matter is considered, which will serve as a first approximation to ferromagnetised neutron star matter. Our description essentially considers the interaction of a single (neutron) magnetic dipole moment with the collective dipole moment of the 
system (the magnetisation). This interaction is described in a Lorentz invariant manner through an appropriate 
Lagrangian density of the system.\\
\\
The possibility of neutron matter undergoing a ferromagnetic phase transition will be investigated by adjusting the strength of the coupling between a magnetic field and the neutron's magnetic dipole moment. We will show that magnetising neutron matter induces a magnetisation in the matter, which is coupled to the magnetic field. We will assume that the system is in a ferromagnetic state when the magnetic response of the system is equal to the magnetic field that is needed to induce this response. Furthermore, the presence of this magnetic field must correspond to a minimum in the total energy density of the system.
%
%
%
%
\section{Magnetic interaction with neutrons}\label{sec:neumag}
Neutrons are neutral particles with a non-zero magnetic dipole moment and spin (of $\frac{1}{2}$). Since these particles are neutral, the origin of the magnetic dipole moment must lie with moving charges (quarks) within the particle \cite{gross}. To investigate the magnetic interaction of the neutron the appropriate (fundamental) coupling would be to couple the magnetic field directly to the charged quarks and observe the result on the scale of the neutron. However, quark degrees of freedom are not tractable on the scale of nucleons (protons and neutrons) \cite{walecka}, which are the degrees of our system. Hence we include an effective interaction, where the magnetic field couples to the spin of the neutron, to our system's Lagrangian density. \\
\\
For this purpose we use
\begin{eqnarray}
	-\frac{g_n}{2}\bar{\psi}{ (x)}\sigma^{\mu\nu}F_{\mu\nu}\psi{ (x)}\label{sigmacoup},
\end{eqnarray}
where
\begin{itemize}
\item $F_{\mu\nu} = \partial_\mu A_\nu{ (x)}- \partial_\nu A_\mu{ (x)}$ is the electromagnetic field tensor,
\item $\sigma^{\mu\nu} = \frac{i}{2}\left[\gamma^\mu,\gamma^\nu\right]$ are the generators of the Lorentz group in the Dirac space \cite{gross}, and
\item $g_n$ is the coupling constant (with units of the magnetic dipole moment). 
\end{itemize}
This is the simplest way to couple the magnetic field to the spin of a particle \cite{gross}. This coupling was already used by Broderick et al. \cite{Brod00} to investigate the magnetic properties of neutron star matter. However, in this study $g_n$ will not only be taken at the value that reproduces the observed neutron magnetic dipole moment at normal densities. Rather $g_n$ will also be taken at larger values 
in order to investigate the behaviour of the system at increased values of the neutron dipole moment.\\
\\
Note that this coupling is the same as the one that gives rise to the anomalous contribution to the magnetic dipole moment of an electron. Since electrons are fundamental particles, these anomalous corrections come from higher order effects in the coupling to the photon field and are thus quantum corrections of the order $\hbar$ \cite{P+S}. However, these quantum corrections are not the type of contributions we are considering when studying the neutron coupling to the magnetic field. Rather we are concerned with those due to the fact that the neutron is composed of more fundamental charged particles. Broderick et al. \cite{Brod00} referred to including $-\frac{g_n}{2}\bar{\psi}{ (x)}\sigma^{\mu\nu}F_{\mu\nu}\psi{ (x)}$ in their Lagrangian density as including the {\em anomalous} contribution to the magnetic dipole moment. This designation has been used frequently by various authors whose work is based on \cite{Brod00}. Since the modification of the baryon dipole moment has a different origin than for electrons we will rather not use the term {\em anomalous} when referring to these contributions to the dipole moments. Instead we will either refer to (\ref{sigmacoup}) by name or as the {\em dipole coupling}. \\
\\
The coupling in (\ref{sigmacoup}) is a contraction between two tensors (thus a Lorentz scalar) and, ignoring $\bar{\psi}(x)$ and $\psi(x)$, has the unit of energy/ (length)$^{-1}$. Therefore it can be included in the standard QHD Lagrangian for nuclear matter (\ref{fsuglagrangian}), with the addition of the free-field electromagnetic Lagrangian density of $-\frac{1}{4}F^{\mu\nu}F_{\mu\nu}$. 
\section{Magnetised neutron matter}\label{sec:neu}
To simplify the description of magnetised neutron matter, apart from the interaction between the neutrons and the magnetic field only the interactions between neutrons and scalar (sigma) as well as vector (omega) mesons are considered.
This will be done in the context of the simplest QHD parameter set, QHD1.
\subsection{Gauge field, $A^\mu$}\label{sec:Amu}
We will choose a free-falling frame of reference (a choice that can always be made) so that the time-component of the magnetic vector potential, $A^\mu$, is zero and that the magnetic field lies in the $z$-direction. We will also assume that the magnetic field is constant, i.e. ${\bm B}=B\hat{z}$. All these assumptions are encompassed by our choice of $A^{\mu}$, where
\begin{eqnarray}
	A^\mu=(0,0,Bx,0)\label{Amu}.
\end{eqnarray}
We will use this choice of $A^\mu$ throughout this work.
\subsection{Lagrangian density and equations of motion}
The Lagrangian we use to describe such a system is based on that of unmagnetised nuclear matter (\ref{fsuglagrangian}), but including the terms mentioned in section \ref{sec:neumag}:
\begin{eqnarray}\label{Lbfield}
		{\cal L} &=& \bar{\psi}{ (x)}\Big[\gamma^{\mu}(i\partial_{\mu}-g_{v}V_{\mu}{ (x)}) -\frac{g_n}{2}
		\sigma^{\mu\nu}F_{\mu\nu} - 
		\Big(m_n-g_{s}\phi{ (x)}\Big)\Big]\psi{ (x)}\nonumber\\
		&\,& +\, \frac{1}{2}\partial^{\mu}\phi{ (x)}\partial_{\mu}\phi{ (x)} -\frac{1}{2}
		 m_{s}^{2}\phi^{2}{ (x)}-\frac{1}{4}V^{\mu\nu}V_{\mu\nu} + \frac{1}{2}m_\omega^{2}V^{\mu}{ (x)}V_{\mu}{ (x)}
		 -\frac{1}{4}F^{\mu\nu}F_{\mu\nu}.
\end{eqnarray}
Using the Euler-Lagrange equation, the equation of motion for the different fields can be shown to be
\begin{subequations}\label{EQMn}
	\begin{eqnarray}
		\partial_{\mu}\partial^{\mu}\phi{ (x)} + m_{s}^{2}\phi{ (x)} &=& g_{s}\bar{\psi}{ (x)}\psi{ (x)},\\
		\partial_{\mu}V^{\mu\nu} + m_\omega^{2}V^{\nu}{ (x)} &=& g_{v}\bar{\psi}{ (x)}\gamma^{\nu}\psi{ (x)},\\
		\partial _\mu \left(F^{\mu\nu} + 
		\frac{g_n}{2}\bar{\psi}{ (x)}\sigma^{\mu\nu}\psi{ (x)}\right)&=&0,\mbox{ and}\label{Maxwell}\\
		\Big[\gamma^{\mu}(i\partial_{\mu}-g_{v}V_{\mu}{ (x)}) -
		\frac{g_n}{2}\sigma^{\mu\nu}F_{\mu\nu} - 
		\big(m_n-g_{s}\phi{ (x)}\big)\Big]\psi{ (x)} &=& 0\label{DiracB}.
	\end{eqnarray}
\end{subequations}
The equation of motion of the meson fields do not differ from those of unmagnetised matter, while equation (\ref{DiracB}) is the one for a fermion in a magnetic field. Expanding equation (\ref{Maxwell}) in its non-covariant form we have that
\begin{subequations}\label{neumax}
	\begin{eqnarray}
		\bm \nabla\cdot\left({\bm E}-\frac{g_n}{2}\bar{\psi}{ (x)}i\bm \alpha\psi{ (x)}\right)&=&0\mbox{ and,}\label{neumaxa}\\
		\bm \nabla\times\left(\bm B+g_n\bar{\psi}{ (x)}\bm\Sigma\psi{ (x)}\right)
		&=&\frac{\partial}{\partial t}\left({\bm E}-\frac{g_n}{2}\bar{\psi}{ (x)}i\bm \alpha\psi{ (x)}\right).\label{neumaxb}
	\end{eqnarray}
\end{subequations} 
In this chapter we only consider (charge neutral) neutron matter. Consequently there are no free charges or currents in the system. Hence (\ref{neumaxa}) and (\ref{neumaxb}) are the applicable versions of Maxwell's equations for magnetised matter (\ref{maxwell}). 
Thus we relate
\begin{subequations}\label{neumaxnc}
	\begin{eqnarray}
		-\frac{g_n}{2}\bar{\psi}{ (x)}i\bm \alpha\psi{ (x)}&=&\bm P,\mbox{ and}\\
		-g_n\bar{\psi}{ (x)}\bm\Sigma\psi{ (x)}&=&\bm M\label{nm}
	\end{eqnarray}
\end{subequations}
to the polarisation and the magnetisation of magnetised neutron matter. Since we are dealing with neutral matter, $\bm E$ must be zero and we will ignore (\ref{neumaxa})\footnote{${\bm E}=0$ does not necessarily imply that $\bm P=0$, but we can show that in the RMF approximation this is indeed the case.}. \\
\\
From $A^\mu$ (\ref{Amu}) we establish that ${\bm B}=B\hat{z}$. Then, considering the properties of $\sigma^{\mu\nu}$ (\ref{sigmaxpnd}), we have that
\begin{subequations}\label{sred}
	\begin{align}
		\begin{split}
			\sigma^{ij}F_{ij} &= -2{\bm\Sigma}\cdot {\bm B}\\ &=-2{\Sigma_z}{B},\mbox{ and}
		\end{split}\\
		-\frac{1}{4}F^{\mu\nu}F_{\mu\nu}&=-\frac{1}{2}B^2.
	\end{align}	
\end{subequations}
where $\Sigma_z$ is the $z$ component of the nucleon spin operator (\ref{eqnSigmaz}).
Using the above 
modifies the equations of motion (\ref{EQMn}), and expansions thereof, to
\begin{subequations}\label{QHD1EQM}
	\begin{eqnarray}
		\partial_{\mu}\partial^{\mu}\phi{ (x)} + m_{s}^{2}\phi{ (x)} &=& g_{s}\bar{\psi}{ (x)}\psi{ (x)}, \\
		\partial_{\mu}V^{\mu\nu} + m_\omega^{2}V^{\nu}{ (x)} &=& g_{v}\bar{\psi}{ (x)}\gamma^{\nu}\psi{ (x)},\\
		\bm \nabla\times\left(\bm B+g_n\bar{\psi}{ (x)}\bm\Sigma\psi{ (x)}\right)&=&0\,\footnotemark,\label{redB}\mbox{ and}\\
		\Big[\gamma_{\mu}(i\partial^{\mu}-g_{v}V^{\mu}{ (x)}) +g_n\,B\,\Sigma_z - 
		\big(m_n-g_{s}\phi{ (x)}\big)\Big]\psi{ (x)} &=& 0\label{simDENB}.
	\end{eqnarray}
\end{subequations}
\footnotetext{In order to establish a comparison later, we choose to keep (\ref{redB}) general. }
These equations govern the behaviour of the different degrees of freedom in the system and in particular equation (\ref{redB}) establishes the magnetic response of the system. As such, it must also be satisfied by a self-generating (ferromagnetic) magnetic field. 
In order to investigate ferromagnetism in the system 
we need to establish the properties of $\psi{ (x)}$ and the ground state. This will now be done next by considering magnetised neutrons in the RMF approximation.
\section{Relativistic mean-field approximation}\label{sec:RMFn}
The RMF approximation (of section \ref{sec:RMF}) will also be employed here. Again we assume the ground state to be translational invariant. However, the magnetic field breaks the overall rotational invariance of the ground state.
Therefore we will assume that ground state is rotationally invariant in the plane perpendicular to the magnetic field and has reflection symmetry in the direction of the magnetic field.\\
\\
This is a weaker symmetry than in the unmagnetised case, where the ground state has full rotational symmetry. We will show that these assumptions indeed hold in the RMF approximation for magnetised matter and are sufficient for our purposes.  
\subsection{Particle operators and sources}
As before the meson field operators are replaced by the ground state expectation values, thereby becoming classical fields
\begin{subequations}\label{MFTn}
\begin{eqnarray}
	\phi{ (x)} \longrightarrow \left\langle \Phi\left|\phi{ (x)}\right|\Phi\right\rangle = 
	\left\langle \phi\right\rangle = \phi_{0},\mbox{ and}\\
	V^\mu{ (x)} \longrightarrow \left\langle \Phi\left|V^\mu{ (x)}\right|\Phi\right\rangle = 
	\left\langle V^\mu\right\rangle = V^0,
\end{eqnarray}
\end{subequations}
while $\psi{ (x)}$ remains an operator.\\
	\\
Regarding the sources of the meson fields, the RMF approximation again necessitates the following substitutions
\begin{subequations}\label{RMF}
	\begin{eqnarray}
		\bar\psi{ (x)}\psi{ (x)} &\longrightarrow &
		\left\langle \Phi\left|{\bm :}\bar\psi{ (x)}\psi{ (x)}{\bm :}\right|\Phi\right\rangle
		= 	\left\langle \bar\psi\psi\right\rangle,\mbox{ as well as }\\
		\bar\psi{ (x)}\gamma^{\nu}\psi{ (x)} &\longrightarrow &
		\left\langle \Phi\left|{\bm :}\bar\psi{ (x)}\gamma^{\nu}\psi{ (x)}{\bm :}\right|\Phi\right\rangle
		= 	\left\langle \bar\psi\gamma^{0}\psi\right\rangle\label{rmfomega}.
	\end{eqnarray} 
As was the case in chapter \ref{chap:unmag}, the spatial components of the vector field and density will be zero. Although the symmetry of the ground state is weaker, in the $xy$-	plane the vector currents will vanish due to the rotational symmetry in this plane. In the $z$-direction the average of all vector currents will be zero due to the reflection symmetry, since the currents in different directions will cancel each other.\\
	\\ 
We proceed by evaluating the equation of motion of the neutrons (\ref{simDENB}) to establish the properties of the ground state within the context of the RMF approximation together with our choice of $A^\mu$. If these hold (which we will show) then the source of the magnetic field will also be influenced, since the density to which it couples must be calculated in the same ground state as the rest of the system. Therefore the following replacement (since at this stage it cannot be strictly motivated) should also be made, if the system is to be treated in a consistent manner,
	\begin{eqnarray}
		\bar\psi{ (x)}{\bm \Sigma}\psi{ (x)} &\longrightarrow &
		\left\langle \Phi\left|{\bm :}\bar\psi{ (x)}{\bm \Sigma}\psi{ (x)}{\bm :}\right|\Phi\right\rangle
		= 	\left\langle \bar\psi\Sigma_z\psi\right\rangle.\label{uGSuN}
	\end{eqnarray} 
\end{subequations}
Due to the conditions imposed by the choice of $A^\mu$ only the $z$-component survives in (\ref{uGSuN}). However, it is more convenient to refer to the full vector quantity $\left\langle \bar\psi{\bm \Sigma}\psi\right\rangle$ with the understanding that
\begin{eqnarray}
	\left\langle \bar\psi{\bm \Sigma}\psi\right\rangle
	= 	\left\langle \bar\psi\Sigma_z\psi\right\rangle.
\end{eqnarray}
\subsection{Equations of motion and neutron spectrum}
Under the RMF assumptions 
and using the same definition of $m^*$ (\ref{mstar}) the baryon and meson equations of motion (\ref{QHD1EQM}) become
\begin{subequations}\label{MFT1EQM}
	\begin{align}
		g_{s}\phi_0 &= \frac{g_{s}^2}{m_{s}^{2}}\left\langle \bar\psi\psi\right\rangle\label{MFT1sigma}, \\
		g_{v}V_{0} &= \frac{g_{v}^2}{m_\omega^{2}}\left\langle\psi^\dagger\psi\right\rangle,\label{MFTomN}\\
		\bm \nabla\times\left(\bm B+g_n\left\langle \bar{\psi}\bm\Sigma\psi\right\rangle\right)&= 0\label{redBn},\mbox{ and}\\
		\Big[i\gamma^{\mu}\partial_{\mu}-g_{v}\gamma^0V_{0} + g_nB\Sigma_z - 
		m^*\Big]\psi{ (x)}  &= 0\label{MFT1nucleon}.
	\end{align}
\end{subequations}
The equation of motion of the magnetic field (\ref{redBn}) will be discussed in section \ref{rmfbn}. It should be noted that it 
contains information regarding the origin of the specific (ferro)magnetic field which we would like to employ in this study. Note that 
the general solution that would satisfy (\ref{redBn}), which are curl-free fields, is restricted by the RMF assumption of a translational invariant ground state to be any constant magnetic field as 
explicitly assumed 
through our choice of $A^\mu$ (\ref{Amu}). \\
\\
As in the previous chapter the ansatz that
\begin{eqnarray}
	\psi{ (x)} = \psi({\bm k},\lambda)\,e^{ik\cdot x} =  \psi({\bm k},\lambda)\,e^{i{\bm k}\cdot {\bm x} - ie({\bm k},\lambda)t}\label{bansatz},
\end{eqnarray}
where $\psi({\bm k},\lambda)$ are the eigenstates of the Hamiltonian and $e({\bm k},\lambda)$ the single particle energies, is made. Ignoring the label $\lambda$ for the moment, the neutron Hamiltonian can be deduced from equation (\ref{MFT1nucleon}) to be
\begin{eqnarray}
	H_D&=&\gamma_0{\bm \gamma}\cdot \bm k-g_n\gamma_0 B\,\Sigma_z + 
		\gamma_0\,m^*+g_{v}V^{0}.\label{hamil1}
\end{eqnarray}
One should note that the spin operator, 
\begin{eqnarray}
	\bm S=\frac{\bm\Sigma}{2}\label{spinop},
\end{eqnarray}
and the Hamiltonian (\ref{hamil1}) does not commute. It implies that spin would not be a good quantum number of the eigenstates of $H_D$. Instead of spin we choose $\lambda$, the orientation of the particle's magnetic dipole moment with respect to the magnetic field, as a good quantum number and label for the energy states. This choice will be obvious once the spectrum of (\ref{hamil1}) is derived.\\
\\
Re-writing the Hamiltonian (\ref{hamil1}) as a matrix 
\begin{align}
	\begin{split}
		H_D\,\psi({\bm k},\lambda)  
		 =&\left[
				\begin{array}{cc}
					g_n B\sigma_z +m^*&{\bm \sigma}\cdot{\bm k}\\
				{\bm \sigma}\cdot{\bm k}&-g_n B\sigma_z -m^*\\
				\end{array}	
			\right]\psi({\bm k},\lambda)+g_{v}V_{0}\,\psi({\bm k},\lambda)\\
			=&\ e({\bm k},\lambda)\,\psi({\bm k},\lambda)
	\end{split}
\end{align}
it can be solved (details of which is given in appendix \ref{ap:landau}) to yield the spectrum
\begin{eqnarray}
	e({\bm k},\lambda)-g_vV^0&=& \pm\sqrt{g_n^2 B^2+k_{z}^2+k_{\bot}^2+{m^*}^2 \pm 2g_n B\sqrt{k_{\bot}^2+{m^*}^2}}\nonumber\\
	&=& \pm\sqrt{\left(\sqrt{k_{\bot}^2+{m^*}^2}+\lambda g_n B\right)^2+k_{z}^2} \label{singlepatE},
\end{eqnarray}
where
\begin{itemize}
	\item $\lambda = \pm 1$ distinguishes the different single particle energies, and
	\item 
	$k_{\bot}^2 = k_{x}^2+k_{y}^2$, i.e. the sum of the squares of the components of $\bm k$ perpendicular to $\bm B$.
\end{itemize}
As is the case for solutions of the free Dirac field \cite{diener}, both positive and negative energies (referring to particles and anti-particles) are acceptable solutions. The energy gap between the particle and anti-particle states are influenced by the scalar mesons as well as the magnetic field, while the vector meson contributes a global shift in the energy.
%
%
\subsection{RMF ground state}
From the single particle energies (\ref{singlepatE}) it is clear that rotational invariance of the ground state is broken due to the presence of the magnetic field. \\
\\
The ground state is defined by the filling of all the positive energy states with energy below that of the Fermi energy. In $\bm k$-space the Fermi energy defines a surface that encompass all the states in the system. This surface reflects the symmetry of the ground state and we 
can deduce from (\ref{singlepatE}), the Fermi surface becomes an ellipsoid with axial (cylindrical) symmetry. \\
\\
In terms of our calculations the symmetry of the ground state features in the bounds of the integrals performed to calculate any densities.  For a rotational invariance ground state the integrals were characterised by a single Fermi momentum (${\bm k}_F$). 
However, in this case the direction perpendicular to $\bm B$ (denoted by $k_\bot$) and that parallel to $\bm B$ ($k_z$) becomes distinct. For the perpendicular direction 
there is no preferred orientation and rotational symmetry is preserved. In the $z$-direction 
there is no directional preference, since the energy depends only on the magnitude $k_z$. Thus, as was assumed earlier, the rotational symmetry of the ground state in the unmagnetised case was replaced by rotational symmetry in the $xy$-plane with reflection symmetry in the direction of $\hat{z}$.\\
\\
Note that each choice of $\lambda$ constitutes an independent, axially symmetric Fermi surface in $\bm k$-space. 
\subsection{RMF magnetic field}\label{rmfbn}
From (\ref{nm}) and (\ref{redBn}) we can relate ${\bm M}$ to $-g_n\left\langle \bar{\psi}\bm\Sigma\psi\right\rangle$. Note that equation (\ref{redBn}) only states that
\begin{align}\label{magbn}
	\begin{split}
			\bm \nabla\times\bm H\ =&\ 0\\
			=&\ \bm \nabla\times\left(\bm B-\bm M\right),
	\end{split}
\end{align}
which does not imply that ${\bm H}=0$, and consequently that $\bm B=\bm M$. However, for boundary conditions appropriate to an uniformly magnetised cylinder with no free charges or currents, 
${\bf B}={\bf M}$ \cite{griff}. Our magnetised ground state reflects this cylindrical symmetry and since we are dealing with neutral matter, 
${\bf B}={\bf M}$ should indeed hold. We can show that for ferromagnetised neutron matter the magnetic field that minimises the energy density satisfies (\ref{redBn}) in the form
\begin{align}
	 B &= -g_n\left\langle \bar\psi\Sigma_z\psi\right\rangle\label{MFT1Bz}.
\end{align}
Thus, as one would expect, the ferromagnetic field in neutron matter can only be the result of the magnetisation of neutrons.
\section{Densities}
Similar to unmagnetised matter in section \ref{sec:RMF}, all the nucleon densities on the right hand side of equations (\ref{MFT1sigma}), (\ref{MFTomN}) and (\ref{MFT1Bz}) still involve an integral over all occupied momentum states. The general expression for densities (\ref{GENdens}) still holds. However, as we have already established, the rotational invariance of the ground state is broken and the densities have to be calculated more carefully. The symmetry of the ground state is reflected in the bounds of the density integrals and thus imposed 
by the Heaviside step function, $\Theta$. Thus the bounds imposed by the step function will differ, compared to the unmagnetised case.\\
\\
Therefore, the general expression for densities (\ref{GENdens}) is modified by considering sums over $\lambda$ instead of spin,
\begin{eqnarray}
	\left\langle\psi^\dagger\Gamma
	\psi\right\rangle  &=& 
	\sum_\lambda\int\frac{d{\bm k}}{(2\pi)^3}\left(\psi^\dagger\Gamma
	\psi\right)_{{\bm k},\lambda}\,
	\Theta\big[\,\mu-e({\bm k},\lambda)\big]\label{GENdensB}
\end{eqnarray}
and should be evaluated using cylindrical coordinates. This implies that all integrals over $\bm k$ becomes double integrals
\begin{eqnarray}
	\int d\bm k = \int\int 2\pi k_\bot dk_\bot dk_z. 
\end{eqnarray}
Due to the double integral the contributions of the two directions need to be written as functions of each other to be able to solve the integrals over $\bm k$. For simplicity the bounds on the $k_\bot$-integral will be written in terms of $k_z$,
\begin{subequations}
	\begin{eqnarray}
		k_{\bot,\lambda}^F(k_z, \lambda) &=& 
		\sqrt{\Big(\sqrt{(\mu-g_vV^0)^2-k_z^2}-\lambda g_n B\Big)^2 - {m^*}^2},\mbox{ and}\\
		k_{z}^F(\lambda) &=& \sqrt{(\mu-g_vV^0)^2-\big(m^*+\lambda g_n B\big)^2},
	\end{eqnarray}
\end{subequations}
so that the general expression for densities in magnetised matter (\ref{GENdensB}) becomes
\begin{eqnarray}
	\left\langle\psi^\dagger\Gamma\psi\right\rangle  
	&=& 
	\sum_\lambda\int_{-k_{z}^F(\lambda)}^{k_{z}^F(\lambda)}\int_0^{k_{\bot,\lambda}^F(k_z, \lambda) } 
	\frac{2\pi k_\bot}{(2\pi)^3}
	\left(\psi^\dagger\Gamma\psi\right)_{{\bm k},\lambda}dk_\bot dk_z.
\end{eqnarray}
Note that results obtained in the case of unmagnetised matter (utilising spherical symmetry) are equivalent to that of magnetised matter (cylindrical symmetry) when $B$ is taken to be zero.
\subsection{Vector densities}
Despite the change in type of symmetry the arguments presented in section \ref{ssec:gs}, as to why the spatial components of the RMF vector boson fields are zero, still holds: perpendicular to the magnetic field the ground state is rotational invariant, whereas in the $z$-direction there is no preference for directions parallel or anti-parallel to the magnetic field. Thus the ground state expectation value of any vector quantity will be zero. This can also be shown by explicit construction of $\psi{ (x)}$. 
\section{Equation of state}
The (RMF) ground state expectation value of the Lagrangian density, $\left\langle {\cal L}\right\rangle$, can be constructed from (\ref{Lbfield}) and simplified by considering (\ref{MFT1nucleon}).  
Using $\left\langle {\cal L}\right\rangle$ the energy density and the pressure of magnetised neutron matter can be calculated along the same lines as 
in chapter \ref{chap:unmag}.
\subsection{Energy density}
The energy density can be calculated from the energy-momentum tensor (\ref{FSUeps1}). In this case $\left\langle i\bar{\psi}\gamma_{0}\partial_{0}\psi\right\rangle$ is given by
\begin{align}
	\begin{split}
		\left\langle i\bar{\psi}\gamma_{0}\partial_{0}\psi\right\rangle
		=& \sum_\lambda\int\frac{d{\bm k}}{(2\pi)^3}\,e({\bm k},s)\,\Theta[\,\mu-e({\bm k},\lambda)]\\ 
		=&  \sum_\lambda\int\frac{d{\bm k}}{(2\pi)^3}\,\left(
		g_v V_0 
		+\sqrt{\left(\sqrt{k_{\bot}^2+{m^*}^2}+\lambda g_n B\right)^2+k_{z}^2}\,\right)
		\,\Theta[\,\mu-e({\bm k},\lambda)].
	\end{split}
\end{align}
Thus the energy density of magnetised neutron matter in QHD1 is 
\begin{align}
	\begin{split}
		\epsilon
		=&  \sum_\lambda\int_{-k_{z}^F(\lambda)}^{k_{z}^F(\lambda)}\int_0^{k_{\bot,\lambda}^F(k_z, \lambda) } 
		\frac{k_\bot}{(2\pi)^2}
		\sqrt{k_{z}^2+\left(\sqrt{k_{\bot}^2+{m^*}^2}+\lambda g_n B\right)^2}dk_\bot dk_z\\
		&+ \frac{1}{2}m_s^{2}\phi^{2}_0 +g_v V_0\rho_n - \frac{1}{2}m_\omega^{2}V_0^2 +\frac{1}{2}B^2.\label{epsBN}
	\end{split}
\end{align}
\subsection{Pressure}
Using the general expression of the pressure (\ref{pres}) together with equation (\ref{epsBN}), the pressure of magnetised pure neutron matter is 
\begin{align}
	\begin{split}
		P
		=&\,  \mu_n\rho_n-\sum_\lambda\int_{-k_{z}^F(\lambda)}^{k_{z}^F(\lambda)}\int_0^{k_{\bot,\lambda}^F(k_z, \lambda) } 
		\frac{k_\bot}{(2\pi)^2}
		\sqrt{\left(\sqrt{k_{\bot}^2+{m^*}^2}+\lambda g_n B\right)^2+k_{z}^2}\,dk_\bot dk_z\\
		&- \frac{1}{2}m_s^{2}\phi^{2}_0 -g_v V_0\rho_n + \frac{1}{2}m_\omega^{2}V_0^2 -\frac{1}{2}B^2.\label{presBN}
	\end{split}
\end{align}
\section{Scalar density}
From equation (\ref{MFT1sigma}) the meson field couples to the scalar density of the nucleon field $\left\langle\bar{\psi}\psi\right\rangle$, which needs to be calculated. Once $\epsilon$ is known, 
(\ref{scal}) can again be used to calculate the scalar density. This leads to 
\begin{eqnarray}\label{scalneuB}
	\left\langle\bar{\psi}\psi\right\rangle = 
	\frac{1}{(2\pi)^3}\sum_\lambda
	\int_{-k_{z}^F(\lambda)}^{k_{z}^F(\lambda)}\int_0^{k_{\bot,\lambda}^F(k_z, \lambda) } 
	\frac
	{ m^* \left( \sqrt{{k_\bot}^2+{m^*}^2}+\lambda g_n B\right)\left(2\pi k_\bot\right)dk_\bot dk_z}
	{\sqrt{{k_\bot}^2+{m^*}^2}\sqrt{{k_z}^2 + \left(\sqrt{{k_\bot}^2+{m^*}^2}+\lambda g_n B\right)^2}}\ \label{uGu}
\end{eqnarray}
from which the $\phi_0$ can be calculated.
\section{Magnetisation}
From equation (\ref{MFT1Bz}) it is clear that the source of an internally generated magnetic field is the magnetisation, which we related to $\langle\bar{\psi}\Sigma_z\psi\rangle$. The calculation of this density follows the same lines as that of the scalar density. The quantity of interest are the values of $\bm B$ that will minimise the energy density at a constant neutron density. Thus calculating
\begin{eqnarray}\label{eqmBN}
	\frac{\partial {\epsilon}}{\partial B} = 0\label{B=0}
\end{eqnarray}
will yield the equation of motion of $B$ and thus an expression for $\left\langle\bar{\psi}\Sigma_z\psi\right\rangle$. Comparing the expression obtained from (\ref{B=0}) to the equation of motion of $\bm B$ (\ref{MFT1Bz}) yields
\begin{eqnarray}
	\left\langle\bar{\psi}\Sigma_z\psi\right\rangle = 
	\frac{1}{(2\pi)^3}\sum_\lambda
	\int_{-k_{z}^F(\lambda)}^{k_{z}^F(\lambda)}\int_0^{k_{\bot,\lambda}^F(k_z, \lambda) } 
	\frac
	{\lambda\left(\sqrt{{k_\bot}^2+{m^*}^2}+\lambda g_n B\right)\left(2\pi k_\bot\right)}
	{\sqrt{{k_z}^2 + \left(\sqrt{{k_\bot}^2+{m^*}^2}+\lambda g_n B\right)^2}}\,dk_\bot dk_z\label{uGDu}
\end{eqnarray}
and hence an expression for the self-consistent calculation of $\bm B$. The validity of (\ref{uGDu}) can be, and was also, confirmed through the explicit construction of $\psi$ and evaluation of the expectation value on the left of (\ref{uGDu}).
\section{Equilibrium conditions}
Pure neutron matter is a first approximation of neutron star matter. Thus the condition of beta-equilibrium will be ignored in this instance. Since neutrons are neutral particles, pure neutron matter satisfies all the other conditions for neutron star matter mentioned in section \ref{sec:eqm}. All calculations of neutron matter and neutron star matter properties will done while the neutron density is kept constant.\\
\\
Since the neutron stars are long-lived objects, they are considered to be in their lowest energy state \cite{csg}. Therefore an internally magnetised neutron star should be in a lower energy state than the unmagnetised star. This condition is automatically satisfied by the ferromagnetic phase, since the phase transition will only take place if the energy will be lowered.
\section{Ferromagnetism in neutron matter}\label{sec:ferroneu}
Since no distinction is made between neutron matter and neutron star matter in this chapter, this discussion is also directly applicable to the description of neutron star matter.\\
\\
Ferromagnetism is a property of solids that requires no external field to maintain magnetisation \cite{griff}. The system undergoes a ferromagnetic phase transition when it is energetically favourable to align the otherwise randomly orientated magnetic dipole moments, thus magnetising the system. As will be shown in the following chapters neutrons are the dominant particles within our model for neutron star matter, even when protons and leptons are also considered. Thus any discussion on magnetisation of the star will deal predominantly with the magnetisation of neutron matter. \\
\\
The effect of including the coupling (\ref{sigmacoup}) to the Lagrangian of neutron matter is to couple the magnetic field to the dipole moment of the neutrons. If ${\bm B}\neq 0$ the spin degeneracy in the single particle energies (\ref{singlepatE}) are lifted and depending on the sign of $\bm B$ one choice of $\lambda$ will have a lower energy than the other. Filling these lower energy states will not only lower the total energy of the system, but also result in individual dipole moments not paired with a counterpart of opposite dipole moment orientation with regards to $\bm B$ (choice of $\lambda$). These unpaired dipole moments will in turn induce an effective magnetic dipole moment in the system (magnetisation), resulting in a magnetic field being generated. When the induced magnetisation matches the original field $\bm B$, then the system will be in a stable, lower energy (compared to unmagnetised matter) ferromagnetic state.  \\
\begin{figure}
	\centering
		\includegraphics[width=1.\textwidth]{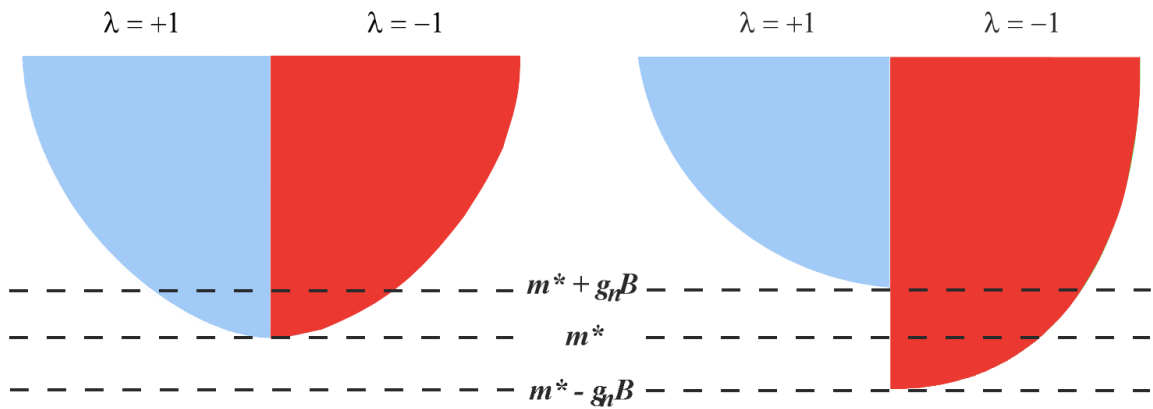}
	\caption[Illustration of one quadrant of the Fermi surface of each species ($\lambda\pm1$) along the $k_\bot=0$ plane.]{Illustration of one quadrant of the Fermi surface of each species ($\lambda\pm1$) along the $k_\bot=0$ plane. On the left $\bm B=0$ and on the right $\bm B>0$.  
	The unpaired dipole moments, due to the gap of $2g_n B$ between the lowest energy states of the two species, are the source of the magnetic field.
	 }
	\label{fig:disks}
\end{figure}
\\
The magnetic field $\bm B$ of the ferromagnetic phase in neutron matter is calculated by self-consistently solving the equation of motion of $\bm B$ (\ref{MFT1Bz}). The calculation aims to match a seed magnetic field $\bm B'$ to the magnetic field $\bm B$ generated by the unmatched dipole moments (induced by $\bm B'$). 
When $\bm B' = \bm B\neq 0$ the system has undergone a phase transition to the ferromagnetic state. In general a phase transition is accompanied by the breaking of some symmetry. In this case it is the rotational invariance (spherical symmetry) of the ground state that is broken, since the magnetic field induces a specific direction/orientation. Note that $\bm B=0$ will always satisfy equation (\ref{MFT1Bz}), but equation (\ref{MFT1Bz}) will only give non-zero solution of $\bm B$ if it lowers the total energy of the system.
\subsection{Medium effects on the coupling $g_n$}\label{runcoup}
In the units used in this study (see section \ref{units}) the magnetic dipole moment, $\mu^{(dip)}_n$, of the neutron is
\begin{eqnarray}
	\mu^{(dip)}_n=\kappa_n\mu^{(dip)}_N=\kappa_n\frac{q_p}{2 m}\label{mun},
\end{eqnarray}
where $\mu_N$ is the nuclear magneton and $\kappa_n=-1.913$  \cite{PDGmuon}. As shown in appendix \ref{ap:dipole}, to reproduce the magnetic dipole moment at normal densities $g_n$ must be equal to
\begin{eqnarray}
	g_n=-\frac{\kappa_n\mu^{(dip)}_N}{2}=g_n^{\scriptscriptstyle(0)}= 0.0305,\label{gnnormal}
\end{eqnarray}
in the units used in this dissertation.\\
\\
The value quoted above for the neutron dipole moment, and thus also the value of $g_n=g_n^{\scriptscriptstyle(0)}$, is for a free neutron. In general this value would change in a medium mainly due to the fact that the neutron is a compound object whose properties will be affected by the medium and as such it will be density dependent. Unfortunately this density dependence is not known experimentally and is also difficult to compute theoretically as it will entail a full blown non-perturbative QCD calculation. Due to these difficulties, the approach we take here is to simply treat $g_n$ as a parameter and to enquire whether a ferromagnetic phase is at all possible for a reasonable value of $g_n$. With this we mean that the value of the neutron magnetic dipole moment must lie between the value in (\ref{mun}) and the other extreme would be the value obtained by simply adding the magnetic dipole moments of the constituent quarks. The latter is what one would expect at the very high densities when the (internal) quark degrees of freedom, rather than baryon degrees of freedom itself, becomes the relevant degrees of freedom (asymptotic freedom). \\
\\
To estimate this upper limit we will define the quark magneton (in analogy to the nuclear magneton) in terms of the charge of the quark $q_Q$ and its mass $m_Q$ as
\begin{eqnarray}\label{qmu}
	\mu^{(dip)}_Q=\frac{q_Q}{2 m_Q}.
\end{eqnarray}
Using the mass and charge of the {\em up} and {\em down} quarks, $\mu^{(dip)}_u$ and $\mu^{(dip)}_d$ are calculated.  Combining the effect of these quark magnetons as $\mu^{(dip)}_{udd}$, where
\begin{eqnarray}
	\mu^{(dip)}_{udd}=(2\mu^{(dip)}_d + \mu^{(dip)}_u),
\end{eqnarray}
an estimate of the magnetic dipole moment of the constituent particles of the neutron can be gained. Comparing $\mu^{(dip)}_{udd}$ to $\mu^{(dip)}_N$ will yield an idea of the possible increase in the neutron's magnetic dipole moment at high density where the quarks are expected to be the relevant degrees of freedom. Since
\begin{eqnarray}
	\frac{\mu^{(dip)}_{udd}}{\mu^{(dip)}_N}\cong500
\end{eqnarray}
it shows that an increase in $g_n$ of the order of tens of times $g_n^{\scriptscriptstyle(0)}$ may not be unfeasible at high density. For an increase of a factor of $x$ in the neutron magnetic dipole moment $g_n$ has to be adjusted to (\ref{gnadj}):
\begin{eqnarray}\label{gnAdj}
	g_n=-x\frac{\kappa_n\mu^{(dip)}_N}{2}.
\end{eqnarray}
The behaviour of the system at higher values of $g_n$ will be investigated in chapters \ref{chap:resferro} and \ref{chap:resnuc}. 
\section{Summary}
The formalism for coupling a magnetic field to the dipole moment of a neutron was derived. Using this formalism the mechanisms to explore the possible ferromagnetic phase of neutron matter by adjusting the coupling strength between the neutrons and the magnetic field was developed. This formalism will serve as a basis to further develop the model that will enable us to study the ferromagnetic phase of neutron star matter. This will be done in the next chapter.

\chapter{Ferromagnetism in neutron star matter}\label{chap:beta}
The aim of this chapter is to add charged particles to our description of ferromagnetism in neutron matter. We do this to obtain a description of ferromagnetism in beta-equilibrated, charge neutral nuclear, as well as neutron star, matter. 
\section{Magnetic interaction with charged particles}\label{sec:mmpat}
The electromagnetic potential $A^\mu$ couples to both the charge and spin of charged particles. To include both these effects in our description we add
\begin{eqnarray}
	\bar{\psi}_p{ (x)}\left(-q_p\gamma^\mu A_\mu-\frac{g_p}{2}\sigma^{\mu\nu}F_{\mu\nu}\right)\psi_p{ (x)}\label{chargedcoup}
\end{eqnarray}
to the Lagrangian describing protons. Here we couple the magnetic potential to the proton wave function with the charge as coupling strength, while the electromagnetic field tensor is coupled to the spin operator with coupling strength that, together with the standard contribution to the magnetic dipole moment from the charge coupling, determines the proton's total magnetic dipole moment.\\
\\
In contrast, for the leptons only the term
\begin{eqnarray}
	-\psi_l{ (x)}q_l\gamma^\mu A_\mu\psi_l{ (x)}
\end{eqnarray}
is added to the Lagrangian. Including the second term in (\ref{chargedcoup}) to the lepton Lagrangian density will describe the \textsl{anomalous} contribution to the lepton magnetic dipole moment. The anomalous dipole moment has its origin in higher order interaction of photons with leptons \cite{P+S}.  However, these corrections are of the order $\hbar$ and thus small compared the possible change in the baryon magnetic dipole moments which are due to the internal structure of the baryons. Furthermore, Mao et al. \cite{mao2} found that including this coupling in the description of magnetised matter only has a small influence. Thus we will not include the \textsl{anomalous} term to the magnetic field's interaction with the leptons. 
\section{Magnetised neutron star matter}
As mentioned in chapter \ref{chap:unmag} our description of neutron star matter includes only protons, neutrons, electrons and muons. The Lagrangian density we employ to describe such a magnetised system is\footnote{Note that since $\psi{ (x)}$ (\ref{isodoub}) is an isodoublet spinor, $F^{\mu\nu}\sigma_{\mu\nu}$ should technically be $\dblone_2\otimes F^{\mu\nu}\sigma_{\mu\nu}$ and 
$g_b= 
		\left[\begin{array}[h]{cc}
			g_p\dblone_4& 0\\
			0 & g_n\dblone_4
		\end{array}\right]$ so that 
$-\bar{\psi}{ (x)}\frac{g_b}{2}F^{\mu\nu}\sigma_{\mu\nu}\psi{ (x)}=
-\bar{\psi}_p{ (x)}\frac{g_p}{2}F^{\mu\nu}\sigma_{\mu\nu}\psi_p{ (x)}
-\bar{\psi}_n{ (x)}\frac{g_n}{2}F^{\mu\nu}\sigma_{\mu\nu}\psi_n{ (x)}.$
}
\begin{eqnarray}
\label{LBB}
				{\cal L}&=&
				\bar{\psi}{ (x)}
				\left[\gamma^{\mu}\left(i\partial_{\mu}-q_b\frac{1+\tau_3}{2} A_\mu{ (x)}- g_{v}V_{\mu}{ (x)} 
				- \frac{g_\rho}{2}{\bm \tau}\cdot{\bf b}_\mu{ (x)}\right)
				-\frac{g_b}{2}F^{\mu\nu}\sigma_{\mu\nu}-(m-g_{s}\phi{ (x)})\right]\psi{ (x)}\nonumber \\
				&&+\, \frac{1}{2}\partial_{\mu}\phi\partial^{\mu}\phi{ (x)} - \frac{1}{2}m_s^{2}\phi^{2}{ (x)} 
				- \frac{\kappa}{3!}\big(g_s\phi{ (x)}\big)^3 - \frac{\lambda}{4!}\big(g_s\phi{ (x)}\big)^4 \nonumber \\
				&&-\, \frac{1}{4}V^{\mu\nu}V_{\mu\nu} + \frac{1}{2}m_\omega^{2}V^{\mu}{ (x)}V_{\mu}{ (x)} + 
				\frac{\zeta}{4!}\big(g_v^2V^\mu{ (x)} V_\mu{ (x)}\big)^2 \\
				&&-\, \frac{1}{4}{\bm b}^{\mu\nu}\cdot{\bm b}_{\mu\nu} +
				 \frac{1}{2}m_{\rho}^{2}{\bm b}^{\mu}{ (x)}\cdot{\bm b}_{\mu}{ (x)}
				+\, \Lambda_v\big(g_v^2V^\mu{ (x)} V_\mu{ (x)}\big)
				\big(g_\rho^2{\bm b}^{\mu}{ (x)}\cdot{\bm b}_{\mu}{ (x)}\big)\nonumber \\
				&&-\frac{1}{4}F^{\mu\nu}F_{\mu\nu}+\sum_l \bar{\psi}_l{ (x)}
				\Big(\gamma^{\mu}\left(i\partial_{\mu}-q_l A_\mu{ (x)}\right) - m_l\Big)\psi_l{ (x)}\nonumber,
\end{eqnarray}
where 
\begin{itemize}
	\item $q_b$ is the baryon charge,
	\item $g_b$ the general baryon magnetic field coupling,
	\item $\psi_l(x)$ and $q_l$ are the lepton wave functions and charges respectively.
\end{itemize}
Using the Euler-Lagrange equation the equations of motion for the different fields can be shown to be
\begin{subequations}\label{EQMB}
	\begin{align}
		\partial_{\mu}\partial^{\mu}\phi{ (x)} + m_s^{2}\phi{ (x)} + \frac{\kappa}{2!}g_s^3\phi{ (x)}^2 
		+ \frac{\lambda}{3!}g_s^4\phi{ (x)}^3  
		=\ &g_{s}\bar{\psi}{ (x)}\psi{ (x)},\label{EQMB1}\\
		\partial_{\mu}V^{\mu\nu} + m_\omega^{2}V^{\nu}{ (x)} + \frac{\zeta}{3!}g_v^4V_\nu{ (x)}^2 V^\nu{ (x)} + 
		2\Lambda_vg_v^2V^\nu g_\rho^2{\bm b}^{\mu}{ (x)}\cdot{\bm b}_{\mu}{ (x)}
		=\ &g_{v}\bar{\psi}{ (x)}\gamma^{\nu}\psi{ (x)},\label{EQMB2}\\
		\partial_{\mu}{\bm b}^{\mu\nu} + m_{\rho}^{2}{\bm b}^{\nu}{ (x)}\, +\, 
		2\Lambda_v g_v^2V^\nu{ (x)} V_\nu{ (x)} g_\rho^2{\bm b}^\nu{ (x)} 
		=\ &\frac{g_\rho}{2}\bar{\psi}{ (x)}\gamma^{\nu}{\bm \tau}\psi{ (x)},\label{EQMB3}\\ 
		\begin{split}
			\partial _\mu \Big(F^{\mu\nu} + 
			g_b\bar{\psi}{ (x)}\sigma^{\mu\nu}\psi{ (x)}\Big)
			=\ \bar{\psi}&{ (x)}\gamma^{\nu}q_b \frac{1+\tau_3}{2}\psi{ (x)}\\
		 	+&\sum_l\bar{\psi}_l{ (x)}q_l\gamma^{\nu}\psi_l{ (x)},
		\end{split}\label{EQMB4}\\
		\Big[\gamma^{\mu}(i\partial_{\mu}-q_lA_\mu{ (x)})-m\Big]\psi_l{ (x)}
		=&\ 0,\mbox{ and}\label{EQMB5}\\
		\Biggr[\gamma^{\mu}\Big(i\partial_{\mu}-q_b\frac{1+\tau_3}{2} A_\mu{ (x)}- g_{v}V_{\mu}{ (x)} 
		- \frac{g_\rho}{2}{\bm \tau}\cdot{\bf b}_\mu{ (x)}\Big)
		-\frac{g_b}{2}F^{\mu\nu}\sigma_{\mu\nu} -(&m-g_{s}\phi{ (x)})\Biggr]\psi{ (x)}=0.\label{DEBB}
	\end{align}
\end{subequations}
Equations (\ref{EQMB1} - \ref{EQMB3}) are the familiar equations that describe the most general (in terms of parameter sets) meson fields, 
as was also found in \cite{diener}. Equation (\ref{EQMB5}) is the Dirac equation for a fermion in magnetic field and equation (\ref{DEBB}) just a slightly modified version thereof. Equation (\ref{EQMB4}) differs from the corresponding version in neutron matter (\ref{Maxwell}) since in this case the baryon charges also couples to the magnetic field. Expanding (\ref{EQMB4}) we have
\begin{subequations}\label{promax}
	\begin{align}
		\bm \nabla\cdot\left({\bm E}-\frac{g_b}{2}\bar{\psi}{ (x)}i\bm \alpha\psi{ (x)}\right)=&\  q_b\psi^\dagger{ (x)}\frac{1+\tau_3}{2}\psi{ (x)}
		+\sum_l q_l\psi_l^\dagger{ (x)}\psi_l{ (x)},\mbox{ and}\label{promaxncb}\\
		\begin{split}
			\bm \nabla\times\left(\bm B+g_b\bar{\psi}{ (x)}\bm\Sigma\psi{ (x)}\right)
			=&\ \frac{\partial}{\partial t}\left({\bm E}-\frac{g_b}{2}\bar{\psi}{ (x)}i\bm \alpha\psi{ (x)}\right)\\
			&+q_b\bar{\psi}{ (x)}{\bm\gamma}\frac{1+\tau_3}{2} \psi{ (x)}
		 	+\sum_lq_l\bar{\psi}_l{ (x)}{\bm\gamma} \psi_l{ (x)}.\label{magNS}
		\end{split}
	\end{align}
\end{subequations}
Since we are also dealing with charged particles, we would expect non-zero free charge densities and currents to be present in the system. Expanding $\psi{ (x)}$ 
we can relate the following quantities, by comparing (\ref{promax}) to (\ref{maxwell}) as well as taking our cue from (\ref{neumaxnc}), to the magnetisation, polarisation, and the free charge densities and current of magnetised neutron star matter:
\begin{subequations}\label{promaxnc}
	\begin{eqnarray}
		-\frac{g_n}{2}\bar{\psi}_n{ (x)}i\bm \alpha\psi_n{ (x)}-\frac{g_p}{2}\bar{\psi}_p{ (x)}i\bm \alpha\psi_p{ (x)}&=&\bm P,\\
		-g_n\bar{\psi}_n{ (x)}\bm\Sigma\psi_n{ (x)}-g_p\bar{\psi}_p{ (x)}\bm\Sigma\psi_p{ (x)}&=&\bm M,\label{magNSmag}\\
		q_p\psi_p^\dagger{ (x)}\psi_p{ (x)}+\sum_l q_l\psi_l^\dagger{ (x)}\psi_l{ (x)}&=&\rho_f,\mbox{ and}\\
		q_p\bar{\psi}_p{ (x)}{\bm\gamma} \psi_p{ (x)}+\sum_lq_l\bar{\psi}_l{ (x)}{\bm\gamma} \psi_l{ (x)}
		&=&\bm J_f.
	\end{eqnarray}
\end{subequations}
Since we omitted the $\sigma^{\mu\nu}F_{\mu\nu}$ coupling for the leptons, they have no contribution to the magnetisation and polarisation of neutron star matter and so these quantities only depend on the baryons.
\subsection{Gauge field, $A^\mu$}\label{sec:amu}
In all our calculations the same choice of the gauge field (\ref{Amu}) will be employed. This means that the electric field, $\bm E$ (gradient of the zeroth component of $A^\mu$), will always be zero, even in the case when the matter is charged (nuclear matter). \\
\\
In nuclear matter only the baryons are considered and thus this type of matter will be positively charged due to the presence of the protons. However, we are only interested in the nuclear interactions and furthermore a non-zero electric field simply produces a shift in the single particle energies as well as the energy density of the system. Thus by taking $\bm E=0$ we ignore these contributions as background effects. However, we still include the coupling of the magnetic field to the charge of the protons since this coupling has a fundamental influence on the system. This influence will be clear once we have cast the system in a more mathematically manageable form. 
\section{Relativistic mean-field approximation}
To achieve this, 
the RMF approximation of sections \ref{sec:RMF} and \ref{sec:RMFn} will once again be made here. 
\subsection{Particle operators and sources}
Since we assume the interaction of the mesons with magnetic field to be negligible when compared to that of the magnetic and nucleon fields, the three equations for the different meson field operators are exactly those of unmagnetised matter (\ref{EQM1} - \ref{EQM3}). Once again these fields are replaced by their ground state expectation values:
\begin{subequations}\label{MFTB}
\begin{align}
	\phi{ (x)} &\longrightarrow \left\langle \Phi\left|\phi{ (x)}\right|\Phi\right\rangle = 
	\left\langle \phi\right\rangle = \phi_{0},\\
	V^\mu{ (x)} &\longrightarrow \left\langle \Phi\left|V^\mu{ (x)}\right|\Phi\right\rangle = 
	\left\langle V^\mu\right\rangle = V^0,\mbox{as well as}\\
	{\bm b}^{\mu}{ (x)} &\longrightarrow \left\langle b^{\mu}_a{ (x)}\right\rangle = g^{\mu 0}\delta_{a3}b_0.
\end{align}
\end{subequations}
%
Also, as was done for unmagnetised matter (\ref{nucRMF}), this entails that the meson field's nucleon source terms should be replaced by their normal-ordered ground state expectation values:
\begin{subequations}\label{rmfdens}
	\begin{eqnarray}
		\bar\psi(x)\psi{ (x)} &\longrightarrow &
		\left\langle \Phi\left|{\bm :}\bar\psi{ (x)}\psi{ (x)}{\bm :}\right|\Phi\right\rangle
		= 	\left\langle \bar\psi\psi\right\rangle,\\
		\bar\psi{ (x)}\gamma^{\nu}\psi{ (x)} &\longrightarrow &
		\left\langle \Phi\left|{\bm :}\bar\psi{ (x)}\gamma^{\nu}\psi{ (x)}{\bm :}\right|\Phi\right\rangle
		= 	\left\langle \bar\psi\gamma^{0}\psi\right\rangle,\label{vecBB}\\
		\bar\psi{ (x)}\gamma^\mu\tau_a\psi{ (x)} &\longrightarrow &
		\left\langle\Phi\right| {\bm :}\bar\psi{ (x)}\gamma^\mu\tau_a\psi{ (x)}{\bm :}\left|\Phi\right\rangle =
	 	\left\langle \bar{\psi}\gamma^0\tau_3\psi\right\rangle,\mbox{ and}\\
		\bar\psi{ (x)}{\bm \Sigma}\psi{ (x)} &\longrightarrow &
		\left\langle \Phi\left|{\bm :}\bar\psi{ (x)}{\bm \Sigma}\psi{ (x)}{\bm :}\right|\Phi\right\rangle
		= 	\left\langle \bar\psi{\bm \Sigma}\psi\right\rangle
		= 	\left\langle \bar\psi\Sigma_z\psi\right\rangle.
	\end{eqnarray} 
\end{subequations}
As was the case in the previous chapters, 
the spatial components of the baryon vector current $\left\langle \bar\psi\gamma^{\nu}\psi\right\rangle$, as well as isospin current $\left\langle\bar\psi\gamma^\mu\tau_a\psi\right\rangle$, are zero due to the symmetry we assume to be present in the ground state. For neutrons these assumptions are the same as the ones made in the previous chapter. However, for protons we are now dealing with the Landau problem in the $xy$-plane and an essentially free particle in the $z$-direction and have to consider our assumptions more carefully.\\
\\
Albeit a bit different, we assume the symmetries 
of magnetised neutron ground state to also be present for protons: rotational symmetry in the $xy$-plane and reflection symmetry in the $z$-direction. A hand-waving justify our assumption: there is still no coupling between the magnetic field and the baryon momenta that will establish a further directional preference, other than that established by the choice of $A^\mu$. In other words, the magnetic field still does not force the system to distinguish between, for example, the positive and negative $z$-directions.\\ 
\\
In confirming these assumptions we will follow the same modus operandi as before.
\subsection{Equations of motion}
Considering the properties of the $\sigma_{\mu\nu}$ tensor (\ref{sred}) with our choice for $A^\mu$ (\ref{Amu})
in the RMF approximation, the equations of motion of the different fields (\ref{EQMB}) become
\begin{subequations}\label{MFT1EQMB}
	\begin{align}
		g_s\phi_0 &= \frac{g_{s}^2}{m_s^2}\left[\left\langle\bar\psi\psi\right\rangle - 
			\frac{\kappa}{2}(g_s\phi_0)^2 - \frac{\lambda}{6}(g_s\phi_0)^3\right],\label{FSUsigmaEQMB}\\
		g_{v}V_0 &= \frac{g_{v}^2}{m_\omega^2}\left[\left\langle \psi^{\dagger}\psi\right\rangle - 
		\frac{\zeta}{6}(g_vV_0)^3 -  2\Lambda_v(g_vV_0)(g_\rho b_0)^2\right],\label{FSUomegaEQMB}\\
		g_{\rho}b_0 &=
		\frac{g_\rho^2}{m_\rho^2}\frac{\left\langle\psi^\dagger\tau_3\psi\right\rangle}{2}\Biggr/\left[1 - 
		2\frac{g_\rho^2}{m_\rho^2}\Lambda_v(g_vV_0)^2\right]\!,\label{FSUrhoEQMB}\\
			0 &=\bm \nabla\times\left(\bm B+g_b\left\langle \bar\psi{\bm \Sigma}\psi\right\rangle\right),\label{FSUEQMB}\\
		%
		0&= \Big[\gamma^{\mu}(i\partial_{\mu}-q_lA_\mu)-m\Big]\psi_l{ (x)},\mbox{ and}\label{lepEQM}\\
		\Biggr[\gamma^{\mu}\left(i\partial_{\mu}-q_b\frac{1+\tau_3}{2}A_\mu\right)&- \frac{g_\rho}{2}\gamma^0\tau_3b_0-g_{v}\gamma^0V_{0} + g_b\bm\Sigma \!\cdot\!\bm B - 
		m^*\Biggr]\psi(x) = 0\label{MFT1nucleonB}.
	\end{align}
\end{subequations}
Since there are no free currents equation (\ref{FSUEQMB}) is obtained from equation (\ref{magNS}) with the additional constraints of charge neutrality, which entails that (\ref{promaxncb}) vanishes in our system. In section \ref{sec:nsmag} the implications of (\ref{FSUEQMB}) will be discussed.\\
\\
In the case of the electron (\ref{lepEQM}) and baryon (\ref{MFT1nucleonB}) equations of motion, $A^\mu$ was not simplified in order to make a general derivation of magnetised fermion spectrum. 
\subsection{Magnetised fermion operator and spectrum}
The spectrum of a system of magnetised neutrons interacting only via the exchange of vector and scalar mesons has already been derived (\ref{singlepatE}). Including the presence of the isovector rho mesons in equation (\ref{MFT1nucleonB}), the magnetised neutron spectrum for positive energy neutrons becomes
\begin{eqnarray}\label{Fn}
	e_n({\bm k},\lambda)
	= \sqrt{\left(\sqrt{k_{\bot}^2+{m^*}^2}+\lambda g_n B\right)^2+k_{z}^2} + g_vV^0-\frac{g_\rho b_0}{2}.
\end{eqnarray}
For the charged particles we consider the protons as the generic case from which the spectrum of the electrons and muons can also be deduced. For protons the modified Dirac equation (\ref{MFT1nucleonB}) becomes
\begin{align}
	\Big[\gamma^{\mu}\left(i\partial_{\mu}-q_pA_\mu\right)+ \frac{g_\rho}{2}\gamma^0b_0-g_{v}\gamma^0V_{0} + g_b\bm\Sigma \!\cdot\!\bm B - 
	m^*\Big]\psi{ (x)} = 0\label{eneuBB},
\end{align}
which we re-write as
\begin{eqnarray}
	\left(i\partial_0-g_vV_0 -\frac{g_\rho}{2}\tau_0b_0\right)\psi{ (x)}=
	\Big({\bm \alpha}\!\cdot\!\left({\bm p}-q_p{\bm A}\right)-g_b\beta \bm\Sigma \!\cdot\!\bm B+\beta m^*\Big)\psi{ (x)}.
\end{eqnarray}
In appendix \ref{ap:landau} we show for our choice of the gauge field (\ref{Amu}) the above equation describes a one dimensional quantum mechanical harmonic oscillator in the $xy$-plane (the spectrum is degenerate in the $x$-direction) and what is essentially a free particle in the $z$-direction. The spectrum of such a system is given by
\begin{eqnarray}\label{Fp}
	e(k_z,\lambda,n)=\sqrt{k_z^2+\left(\sqrt{{m^*}^2+2\,\alpha\,q_p B n}+\lambda  g_p B\right)^2}+ g_vV^0+\frac{g_\rho b_0}{2},
\end{eqnarray}
where
\begin{itemize}
\item $k_z$ is the momentum in the $z$-direction,
\item $n$ is an integer labelling the Landau levels, where 
	\begin{eqnarray}\label{llcon}
		n=\left(n'+\frac{1}{2}-\alpha\,\frac{\lambda }{2}\right)\mbox{ with } n'=0,1,2,3...
	\end{eqnarray}
	using the convention introduced in appendix \ref{ap:landau} (\ref{LLenercon}) to discern the level counting scheme,
\item $\alpha =\frac{q_p B}{|q_p B|}$, and once again
\item  $\lambda = \pm 1$ is the eigenvalue of $\sigma_z$.
\end{itemize}
This is the three dimensional version of the Landau spectrum where the energy is independent of the momenta in the $x$- and $y$-directions\footnote{Technically it is actually the position in the $x$-direction which is of importance due to our choice of $A^\mu$ and the momentum in the $x$-direction has no meaning in this context, since the quantum numbers labelling a state are $n,k_z,k_y$ and $\lambda$.}. 
Instead levels are labelled by $n$ which, like in the case of the harmonic oscillator, labels the quantised energy levels.
\begin{figure}
	\centering
		\includegraphics[width=1.0\textwidth]{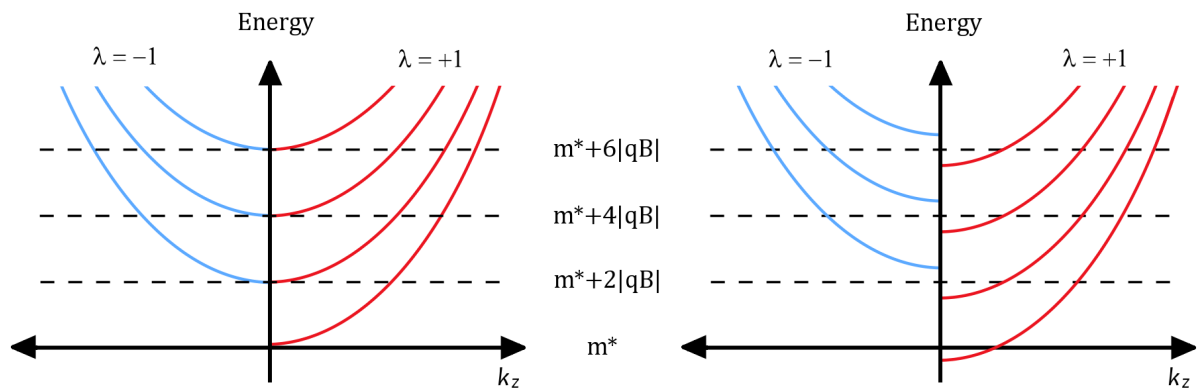}
	\caption[Illustration of the Landau levels occupied by protons in a magnetic field for $g_p=0$ and $g_p\neq 0$.]{Illustration of the Landau levels (in one half of the Fermi surface for each choice of $\lambda$) occupied by protons in a magnetic field pointing in the positive $z$-direction. On the left the spectrum is without the inclusion of the $-\frac{g_p}{2}\bar{\psi}\sigma^{\mu\nu}F_{\mu\nu}$ coupling to the magnetic field, while on the right it is included. For further clarification of the 
labelling scheme, see section \ref{ap:land:spec} as well as figure \ref{fig:LLap}. }
	\label{fig:LLap2}
\end{figure}
The fourth quantum number labelling the energy states in such a system is $k_y$, which would label the energy states within a specific Landau level. However, in this case the energy is degenerate for $k_y$ and throughout this work we will assume that all occupied Landau levels are completely filled.  Figure \ref{fig:LLap2} is an illustration of the dispersion relation (\ref{Fp}). To illustrate the degeneracy in $k_y$ one can imagine each filled level as a sheet coming out of the page.
\subsection{Magnetised fermion ground state}
Once again we should confirm that the assumptions made about the RMF ground state indeed holds. As before, the main assumptions are that the ground state is time-independent and translationally invariant.\\
\\
The ground state is clearly time-independent, but at first glance the state does not seem to be translationally invariant. As is shown in appendix \ref{ap:landau}, the choice of gauge induces an spatial dependency to the Dirac spinor (\ref{chiansatz}). However, this does not break translational invariance if any translational of the system is accompanied by a gauge transformation. This is clear from the particle spectrum (\ref{Fp}), which is independent of position.\\
\\
Equation (\ref{Fp}) also only depends on the magnitude $k_z$ and thus the reflection symmetry in the $z$ direction is present in the Fermi surface, as assumed. As the single particle energies are independent of the $x$- and $y$-directions, the Fermi surface for each choice of $n$ and $\lambda$ also do not depend on them. Thus all vector currents will vanish, as was assumed in (\ref{rmfdens}).
\subsection{RMF magnetic field}\label{sec:nsmag}
At the start of this discussion we should reiterate that the ferromagnetic field, ${\bm B}=B\hat{z}$, is the magnetic field that minimises the total energy density of the system. We consider our system to be a constantly magnetised system, a condition which we imposed through our choice of gauge field $A^\mu$. As such the system is unconcerned with the origin of the magnetic field and (\ref{FSUEQMB}) is a statement regarding the magnetic field in the bulk of the system. Employing the RMF version of (\ref{magNS}) we rewrite (\ref{FSUEQMB}), so that
\begin{align}\label{rmfbeqm}
	\begin{split}
		\bm \nabla\times{\bm H}&=\bm \nabla\times\left({\bm B}-{\bm M}\right)=0.
	\end{split}
\end{align}
Note that, as for neutron matter, (\ref{rmfbeqm}) does not imply that ${\bm H}=0$, and therefore that ${\bm B}={\bm M}$, 
but simply that, together with the RMF assumptions, both ${\bm B}$ and ${\bm M}$ are uniform in the bulk. \\
\\
To find the constant value of the magnetic field requires (\ref{rmfbeqm}) to be supplemented by boundary conditions. For neutron matter we took the boundary conditions to be those appropriate for a uniform magnetised cylinder with no free currents leading to 
\begin{align}
	\begin{split}
		{\bm H}&={\bm B}-{\bm M}=0.
	\end{split}
\end{align}
However, with the inclusion of charged particles, it is less clear which boundary conditions should be imposed. Indeed, it is well known that in quantum Hall systems the introduction of boundaries will lead to edge currents flowing at the boundaries. Such currents are completely consistent with the conditions of charge neutrality and vanishing free currents in the bulk, yet the edge currents will contribute to the constant magnetic field in the bulk, regardless of how far away the boundary is. Therefore, unless specific detail about the physics at the boundary is specified, these boundary conditions and the computation of the magnetic field remain ambiguous. \\
\\
To avoid this difficulty, we abandon equation (\ref{rmfbeqm}) as the condition from which the magnetic field should be computed self-consistently and rather require that the magnetic field minimises the energy density in the bulk. Generally we can of course not expect that this will coincide with (\ref{rmfbeqm}) unless the latter is supplemented with appropriate boundary conditions. This is indeed what is found upon detailed computation of these two conditions (note that these conditions do, however, coincide for neutron matter). \\
\\
However, the minimisation condition seems to be the more realistic one in real physical system as one does not realistically expect that boundary effects very far away will have such a major impact on bulk properties, mainly due to screening effects in realistic systems. We shall therefore take the minimisation condition as the appropriate condition to compute the bulk ferromagnetic properties of neutron star matter.\\
\\
The issue raised above will of course have ramifications for the mean-field description of truly spatially confined physical systems such as neutron stars. We will return to this issue in chapter \ref{chap:resferro} when the most na\"ive implementation of the mean-field approximation for neutron stars will be discussed together with the salient points that need to be addressed more carefully in such systems.\\
\\
Therefore, in order to proceed, with our calculation of the ferromagnetic field we must construct the energy density.
\section{Densities of charged particles in a magnetic field}
In order to do so we will construct the magnetised charged particle densities. All results derived for charged particles in this section will refer to protons. However, these results will be derived in such a way that the expressions for leptons can be obtained by simply replacing the proton charge and mass by the lepton ones.
\subsection{Density}\label{sec:densBB}
Since the magnetised single proton energies (\ref{Fp}) are independent of $k_y$ the system is degenerate in that momentum direction.  Hence the Pauli exclusion principle does not apply and for each value of $n$ and $k_z$ the system can accommodate an infinite number of particles.
However, the occupation of each level is influenced by the strength of the magnetic field. The fundamental magnetic flux, $\Phi_0$, defines the magnetic flux quanta that contains only one energy state \cite{yobi}.  Thus the number of particles per unit area in each level is $B$ divided by $\Phi_0$. As shown in appendix \ref{apsec:Lp}, this contributes a factor of 
$|q_p B|/2\pi$ to the density for each value of $n$ and $k_z$. \\
\\
Each level has a cut-off in terms of $k_z$ denoted by $k_z^F(n,\lambda)$. For protons with a Fermi energy $\mu_p$, 
 $k_z^F(n,\lambda)$ is calculated from (\ref{Fp}) as the real and positive solutions of
\begin{align}\label{kLL}
	\begin{split}
		{k_z^F}^2(n,\lambda)
		&=\left(\mu_p- g_vV^0-\frac{g_\rho b_0}{2}\right)^2-\left(\sqrt{{m^*}^2+2\,|q_p B|n}+\lambda  g_p B\right)^2\\
		&={E_p^F}^2-\left(\sqrt{{m^*}^2+2\,|q_p B|n}+\lambda  g_p B\right)^2\!,
	\end{split}
\end{align}
where
\begin{itemize}
	\item 
		$E_p^F=\mu_p- g_vV^0-g_\rho b_0/2$, and
	\item $\alpha\,q_p B$ has been dropped, without loss of generality, for $|q_p B|$.
\end{itemize}
Thus the particle density is the total contribution of all allowed values of $k_z$, summed for all levels. As illustrated in appendix \ref{apsec:Lp}, it is given by (\ref{Lrho}), 
\begin{subequations}\label{rhoLL}
	\begin{eqnarray}
		\rho_p
		&=&\sum_{\lambda,n}\frac{|q_p B|}{4\pi^2}\int({\psi}^\dagger\psi)_{n,\lambda}\Theta\big[\,\mu-e(k_z,\lambda,n)\big]dk_z\label{fdens}\\
		&=&\sum_{\lambda,n}\frac{|q_p B|}{4\pi^2}\int^{k_z^F(n,\lambda)}_{-k_z^F(n,\lambda)}dk_z\label{kdens}\\
		&=&\frac{|q_p B|}{2\pi^2}\sum_{\lambda,n} k_z^F(n,\lambda),
	\end{eqnarray}
\end{subequations}
where $({\psi}^\dagger\psi)_{n,\lambda}$ are the single particle expectation values which are normalised to 1. 
%
%
\subsection{Energy density}
Following from the structure of the proton density and using the ansatz (\ref{assump}) made for the proton wave function, the proton contribution $\epsilon_p$ to the energy density can be deduced as
\begin{align}\label{epsLLp}
	\begin{split}
		\epsilon_p
		&=\sum_{\lambda,n}\frac{|q_p B|}{4\pi^2}\int(i\bar{\psi}\gamma_0\partial_0\psi)_{n,\lambda}\Theta\big[\,\mu-e(k_z,\lambda,n)\big]dk_z\\
		&=\sum_{\lambda,n}\frac{|q_p B|}{4\pi^2}\int^{k_z^F(n,\lambda)}_{-k_z^F(n,\lambda)}
		(i\bar{\psi}\gamma_0\partial_0\psi)_{n,\lambda}dk_z\\
		&=\frac{|q_p B|}{2\pi^2}\sum_{\lambda,n} \int^{k_z^F(n,\lambda)}_{0}
		e_p(k_z,\lambda,n)\,dk_z\\
		&=\frac{|q_p B|}{2\pi^2}\sum_{\lambda,n} \int^{k_z^F(n,\lambda)}_{0}
		E_p\,dk_z+g_vV^0\rho_p+\frac{g_\rho b_0}{2}\rho_p,
	\end{split}
\end{align}
where $E_p$'s dependence on $k_z,\lambda$ and $n$ has been suppressed for brevity's sake.\\
\\
Thus from the energy-momentum tensor the energy density (\ref{FSUeps1}) of magnetised proton matter interacting via the exchange of mesons is
\begin{align}
	\begin{split}
		\epsilon 
		\,=\,&\sum_{\lambda,n}\frac{|q_p B|}{4\pi^2}\int E_p\,\Theta\big[\,\mu_p-e_p\big]dk_z
		+ \frac{1}{2}m_s^{2}\phi^{2}_0 +\frac{\kappa}{3!}\big(g_s\phi_0\big)^3 + \frac{\lambda}{4!}\big(g_s\phi_0\big)^4\\
		& +g_vV^0\rho_p - \frac{1}{2}m_\omega^{2}V_0^2 - \frac{\zeta}{4!}\big(g_vV_0\big)^4 +\frac{g_\rho b_0}{2}\rho_p
		- \frac{1}{2}m_{\rho}^{2}b_0^2\\ &- 
		\Lambda_v\big(g_vV_0\big)^2\big(g_\rho b_0\big)^2+\frac{1}{2}B^2.
	\end{split}
\end{align}
Having derived the particle density for charged fermions in a magnetic field we can turn our attention to the other source terms in the equations of motion of the meson fields (\ref{MFT1EQMB}). The omega meson couples to the baryon density and the rho meson couples to the isospin density (the difference between the proton density and the neutron density), thus we only need to still derive an expression for the proton scalar density in the presence of a magnetic field.
\subsection{Scalar density}
Since the magnetised neutron contribution of $\left\langle\bar\psi\psi\right\rangle$ has already been derived (\ref{scalneuB}), we will focus on the proton contribution here. Using the same technique as before, the proton contribution can be deduced taking the derivative of the proton energy density to $m^*$. Taking this derivative of (\ref{epsLLp}) results in 
\begin{eqnarray}\label{scalrhoBB}
		\left\langle\bar\psi_p\psi_p\right\rangle
		=\sum_{\lambda,n}\frac{|q_p B|}{4\pi^2}\int^{k_z^F(n,\lambda)}_{-k_z^F(n,\lambda)}
			\frac
			{{m^*}\left(\sqrt{{m^*}^2+2|q_p B| n}+\lambda  g_p B\right)}
			{\sqrt{{m^*}^2+2|q_p B| n}\,\sqrt{k_z^2+\left(\sqrt{{m^*}^2+2|q_p B| n}+\lambda  g_p B\right)^2}}\,
			dk_z.
\end{eqnarray}
Thus the scalar density of magnetised nuclear star matter is 
\begin{align}
	\left\langle\bar\psi\psi\right\rangle\,=&\,\left\langle\bar\psi_n\psi_n\right\rangle+\left\langle\bar\psi_p\psi_p\right\rangle\\
	\begin{split}
		=&\,\frac{1}{(2\pi)^3}\sum_\lambda
			\int_{-k_{z}^F(\lambda)}^{k_{z}^F(\lambda)}dk_z
			\int_0^{k_{\bot}^F(k_z,\lambda)}
			\frac
			{ m^* \left( \sqrt{{k_\bot}^2+{m^*}^2}+\lambda g_n B\right)\left(2\pi k_\bot\right)}
			{\sqrt{{k_\bot}^2+{m^*}^2}\sqrt{{k_z}^2 + \left(\sqrt{{k_\bot}^2+{m^*}^2}+\lambda g_n B\right)^2}}\ dk_\bot	\\
		&+\,\sum_{\lambda,n}\frac{|q_p B|}{4\pi^2}\int^{k_z^F(n,\lambda)}_{-k_z^F(n,\lambda)}
			\frac
			{{m^*}\left(\sqrt{{m^*}^2+2|q_p B| n}+\lambda  g_p B\right)}
			{\sqrt{{m^*}^2+2|q_p B| n}\,\sqrt{k_z^2+\left(\sqrt{{m^*}^2+2|q_p B| n}+\lambda  g_p B\right)^2}}\,
			dk_z.\nonumber
	\end{split}
\end{align} 
\section{Equilibrium conditions}\label{sec:eqmB}
For the magnetised interior of a neutron star consisting of protons, neutrons, electrons and muons we will assume that the equilibrium conditions described in section \ref{sec:eqm} hold.\\
\\
These conditions are that, at a fixed baryon density, the star is assumed to be charge neutral and in beta-equilibrium while the muon states will only be populated once the electron's Fermi energy reaches the rest mass of the muons. Additionally 
we will assume that the calculated magnetic field must be the one that minimises the energy of the system (at a fixed baryon density). All equations derived in section \ref{sec:eqm} are still valid for magnetised matter; however, since the charged baryon density is dependent on the magnetic field the condition of constant baryon density has to be considered carefully. \\
\\
As illustrated in the expression of the particle densities (\ref{rhoLL}), either the Fermi energy (\ref{fdens}) or the Fermi momentum (\ref{kdens}) can be used to constrain the densities. As was done in the previous chapters, the constraint that the baryon density is constant will also be enforced here. This usually also meant that the Fermi energy for a given total baryon density was independent of the other parameters and/or variables. However, due to the pre-factor of $\frac{|q_p B|}{4\pi^2}$,
the charged fermion (proton, electron and/or muon) particle densities acquire an explicit dependence on the magnetic field.  In turn these particles' Fermi energies (Fermi momenta) will also acquire a dependence on the magnetic field, since it is the total baryon density that is kept constant and not the individual proton or neutron densities. \\
\\ 
Due to this dependency on the magnetic field it would be simpler to constrain densities through the use of the Heaviside step function (which limits the single particle energies to be equal or less than the Fermi energy), than by using the Fermi momentum. From (\ref{rhoLL}) any charged particle density in its simplest form, in terms of the Heaviside step function, is
\begin{eqnarray}
		\rho
		&=&\sum_{\lambda,n}\frac{|q_p B|}{4\pi^2}\int\Theta\big[\,\mu-e(k_z,\lambda,n)\big]dk_z
\end{eqnarray}
and thus the total derivative of the density to $B$ is
\begin{eqnarray}
		\frac{d{\rho}}{d B}
		&=&\frac{d}{d B}\sum_{\lambda,n}\frac{|q_p B|}{4\pi^2}\int\Theta\big[\,\mu-e(k_z,\lambda,n)\big]dk_z\nonumber\\
		&=&\frac{|q_p|}{4\pi^2}\sum_{\lambda,n}\int\Theta\big[\,\mu-e(k_z,\lambda,n)\big]dk_z
		+\frac{|q_p B|}{4\pi^2}\sum_{\lambda,n}\int\delta\!\left(\,\mu-e\left(k_z,\lambda,n\right)\right)d k_z\,\frac{d{\mu}}{d B}\nonumber\\
		&=&\frac{\rho}{B}+\frac{\partial {\rho}}{\partial \mu}\frac{d{\mu}}{d B}.
\end{eqnarray}
Consequently for a non-zero magnetic field the constraint that the baryon density must be a constant translates to
\begin{align}
	\begin{split}
		\frac{d{\rho_b}}{d B}\, =\, 0 \, 
		&=\,\frac{d{\rho_n}}{d B}+\frac{d{\rho_p}}{d B} \\
		&=\,\frac{d{\rho_n}}{d B}+\frac{{\rho_p}}{B}+ \frac{\partial \rho_p}{\partial \mu_p}\frac{d{\mu_p}}{d B}.
	\end{split}
\end{align} 
These additional constraints imposed on magnetised baryonic matter will be used in the calculation of the magnetic field. However, in order to proceed, the equation of state must first be derived. 
%
%
%
\section{Equation of state}\label{sec:EoSB}
Having already derived the proton's contribution to the energy density of magnetised neutron star matter we can now derive general expressions for its energy density and pressure.
\subsection{Energy density}
From equation (\ref{epsLL}) the contribution of the leptons to the energy density is simply
\begin{align}
	\begin{split}
		\epsilon_l
		&=\sum_{l,\lambda,n}\frac{|q_l B|}{4\pi^2}\int^{k_z^F(n,\lambda)}_{-k_z^F(n,\lambda)}
		(i\bar{\psi_l}\gamma_0\partial_0\psi_l)_{n,\lambda}dk_z\\
		&=\frac{|q_l B|}{2\pi^2}\sum_{l,\lambda,n}\int^{k_z^F(n,\lambda)}_{0}
		e_l(k_z,\lambda,n)\,dk_z,
	\end{split}
\end{align}
where $l$ refers to either electrons or muons. The lepton spectrum can be deduced from the proton spectrum (\ref{Fp}) by ignoring the contribution from the meson fields and replacing the proton's mass and charge by the lepton's\footnote{Ignoring the meson contribution as well as the mass difference, the proton and lepton spectrum are identical, except that the $n=0$ Landau level will be occupied for the opposite choice of $\lambda$.}
\begin{align}
	\begin{split}
	e_l(k_z,\lambda,n)
	&=\, \sqrt{k_z^2+{m_l}^2+2|q_l B|n}.	
	\end{split}
\end{align}
Adding together the contributions from all the particles in the system the energy density of such a magnetised system is 
\begin{align}
	\begin{split}\label{epsLL}
		\epsilon 
		\,=\, &\sum_{\lambda,n}\frac{|q_p B|}{4\pi^2}\int e_p\,\Theta\big[\,\mu_p-e_p\big]dk_z
		+\sum_{l,\lambda,n}\frac{|q_l B|}{4\pi^2}\int e_l\,\Theta\big[\,\mu_l-e_l\big]dk_z\\
		&+ \sum_\lambda\int\frac{d{\bm k}}{(2\pi)^3}\,e_n\,\Theta[\,\mu_n-e_n] 
		+ \frac{1}{2}m_s^{2}\phi^{2}_0 +\frac{\kappa}{3!}\big(g_s\phi_0\big)^3 + \frac{\lambda}{4!}\big(g_s\phi_0\big)^4\\
		& - \frac{1}{2}m_\omega^{2}V_0^2 - \frac{\zeta}{4!}\big(g_vV_0\big)^4 - \frac{1}{2}m_{\rho}^{2}b_0^2 - 
		\Lambda_v\big(g_vV_0\big)^2\big(g_\rho b_0\big)^2+\frac{1}{2}B^2
	\end{split}\\
	\begin{split}\label{epsLL2}
		\,=\,&\sum_{\lambda,n}\frac{|q_p B|}{4\pi^2}\int E_p\,\Theta\big[\,\mu_p-e_p\big]dk_z
		+\sum_{l,\lambda,n}\frac{|q_l B|}{4\pi^2}\int e_l\,\Theta\big[\,\mu_l-e_l\big]dk_z\\
		&+ \sum_\lambda\int\frac{d{\bm k}}{(2\pi)^3}\,E_n\,\Theta[\,\mu_n-e_n] 
		+ \frac{1}{2}m_s^{2}\phi^{2}_0 +\frac{\kappa}{3!}\big(g_s\phi_0\big)^3 + \frac{\lambda}{4!}\big(g_s\phi_0\big)^4\\
		& +g_vV^0(\rho_p+\rho_n) - \frac{1}{2}m_\omega^{2}V_0^2 - \frac{\zeta}{4!}\big(g_vV_0\big)^4 +\frac{g_\rho b_0}{2}(\rho_p-\rho_n) 
		- \frac{1}{2}m_{\rho}^{2}b_0^2\\ &- 
		\Lambda_v\big(g_vV_0\big)^2\big(g_\rho b_0\big)^2+\frac{1}{2}B^2,
	\end{split}
\end{align}
where the meson contributions to the Fermi energy of the baryons have been separated out of the integrals in the second line. 
\subsection{Pressure}
The pressure is once again calculated 
using equation (\ref{pres}) which, as in section \ref{sec:eqm}, can be simplified 
by considering the imposed equilibrium conditions and therefore becomes
\begin{eqnarray}
	P&=&\, \mu_n(\rho_n+\rho_p)-\epsilon\nonumber\\
		&=&\,\mu_n(\rho_n+\rho_p)-\sum_{\lambda,n}\frac{|q_p B|}{4\pi^2}\int E_p\,\Theta\big[\,\mu_p-e_p\big]dk_z
		-\sum_{l,\lambda,n}\frac{|q_l B|}{4\pi^2}\int e_l\,\Theta\big[\,\mu_l-e_l\big]dk_z\nonumber\\
		&\,&- \sum_\lambda\int\frac{d{\bm k}}{(2\pi)^3}\,E_n\,\Theta[\,\mu_n-e_n] 
		- \frac{1}{2}m_s^{2}\phi^{2}_0 - \frac{\kappa}{3!}\big(g_s\phi_0\big)^3 - \frac{\lambda}{4!}\big(g_s\phi_0\big)^4 
		+ \frac{1}{2}m_\omega^{2}V_0^2 \nonumber\\
		&\,&+ \frac{\zeta}{4!}\big(g_vV_0\big)^4 
		+ \frac{1}{2}m_{\rho}^{2}b_0^2 + \Lambda_v\big(g_vV_0\big)^2\big(g_\rho b_0\big)^2-\frac{1}{2}B^2.\label{epsLLpres}
\end{eqnarray}

\section{Calculating the ferromagnetic field}\label{sec:BBB}
Having derived an expression for the energy density the constraint that the ferromagnetic field should be the one that minimises the energy density can be imposed. Thus taking the total derivative with regard to the magnetic field and setting it to zero an expression for the ferromagnetic field is obtained, i.e.
\begin{eqnarray}
	\frac{d {\epsilon}}{d B} = 0.
\end{eqnarray}  
If such a non-zero magnetic field that minimises the energy density exists, then the system will be in a ferromagnetic state. As shown in section \ref{sec:RMF}, this is equivalent to the method used in the RMF approximation to derive the equations of motion of the different fields (\ref{EQMeps}) as well as the ferromagnetic field in the case of pure neutron matter (\ref{eqmBN}).\\
\\
However, as already shown, the particle and energy densities of charged fermions are dependent on the magnetic field due to the factor arising from the filled Landau levels of $\frac{|q B|}{4\pi^2}$. Thus the total derivative of $\epsilon$ with regards to $B$ has to be taken to derive an equation of motion for the magnetic field $B$, which translates to
\begin{align}
	\begin{split}\label{totalB}
		\frac{d{\epsilon}}{d B}\, &=\, 0\\
		&=\,\frac{\partial {\epsilon}}{\partial B}+\sum_i\frac{\partial {\epsilon}}{\partial \rho_i}\frac{d{\rho_i}}{d B}
		+\sum_i\frac{\partial {\epsilon}}{\partial \mu_i}\frac{d{\mu_i}}{d B},
	\end{split}
\end{align}  
where $i$ refers to the different fermions in the system. Using the expression for $\epsilon$ of magnetised neutron star matter (\ref{epsLL2}) the equation above becomes 
\begin{eqnarray}
		0	&=&
		\sum_{\lambda,n}\frac{|q_p B|}{4\pi^2}\int \frac{\partial E_p}{\partial B}\,\Theta\big[\,\mu_p-e_p\big]dk_z
		+\sum_{l,\lambda,n}\frac{|q_l B|}{4\pi^2}\int \frac{\partial e_l}{\partial B}\,\Theta\big[\,\mu_l-e_l\big]dk_z\nonumber\\
		&\,&+\sum_\lambda\int\frac{d{\bm k}}{(2\pi)^3}\,\frac{\partial E_n}{\partial B}\,\Theta[\,\mu_n-e_n]+\frac{\epsilon_p}{B}+\sum_l\frac{\epsilon_l}{B}
		+ B + g_vV^0\frac{d \rho_b}{d B}+\frac{g_\rho b_0}{2}\left(\frac{d \rho_p}{d B}-\frac{d \rho_n}{d B}\right) \nonumber\\
		&\,&+\,\sum_{\lambda,n}\frac{|q_p B|}{4\pi^2}\int E_p\,\delta\!\left(\,\mu_p-e_p\right)dk_z\,\frac{d{\mu_p}}{d B}
		+\sum_{l,\lambda,n}\frac{|q_l B|}{4\pi^2}\int e_l\,\delta\!\left(\,\mu_l-e_l\right)dk_z\,\frac{d{\mu_l}}{d B}\nonumber\\
		&\,&+\,\sum_\lambda\int\frac{d{\bm k}}{(2\pi)^3}\,E_n\,\delta\!\left(\mu_n-e_n\right)\,\frac{d{\mu_n}}{d B}.
\end{eqnarray}
Using the equilibrium conditions and derived expressions of sections \ref{sec:eqm} and \ref{sec:eqmB} this can be simplified as
\begin{eqnarray}
		-B
		&=&
		\sum_{\lambda,n}\frac{|q_p B|}{4\pi^2}\int \frac{\partial E_p}{\partial B}\,\Theta\big[\,\mu_p-e_p\big]dk_z
		+\sum_{l,\lambda,n}\frac{|q_l B|}{4\pi^2}\int \frac{\partial e_l}{\partial B}\,\Theta\big[\,\mu_l-e_l\big]dk_z\nonumber\\
		&\,&+\,\sum_\lambda\int\frac{d{\bm k}}{(2\pi)^3}\,\frac{\partial E_n}{\partial B}\,\Theta[\,\mu_n-e_n]+\frac{\epsilon_p}{B}+\sum_l\frac{\epsilon_l}{B}
		 + g_vV^0\frac{d \rho_b}{d B}+\frac{g_\rho b_0}{2}\left(\frac{d \rho_p}{d B}-\frac{d \rho_n}{d B}\right) \nonumber\\
		&\,&+\,E_p^F\frac{\partial \rho_p}{\partial \mu_p}\frac{d{\mu_p}}{d B}
		+\sum_l e_l^F\frac{\partial \rho_l}{\partial \mu_l}\frac{d{\mu_l}}{d B}+\,E_p^F\frac{\partial \rho_n}{\partial \mu_n}\frac{d{\mu_n}}{d B}\\
		&=&
			\sum_{\lambda,n}\frac{|q_p B|}{4\pi^2}\int \frac{\partial E_p}{\partial B}\,\Theta\big[\,\mu_p-e_p\big]dk_z
			+\sum_{l,\lambda,n}\frac{|q_l B|}{4\pi^2}\int \frac{\partial e_l}{\partial B}\,\Theta\big[\,\mu_l-e_l\big]dk_z\nonumber\\
		&\,&+\sum_\lambda\int\frac{d{\bm k}}{(2\pi)^3}\,\frac{\partial E_n}{\partial B}\,\Theta[\,\mu_n-e_n]
			+\frac{\epsilon_p}{B}+\sum_l\frac{\epsilon_l}{B}
			-\left(E_n^F-g_\rho b_0\right)\frac{\rho_p}{B}
\end{eqnarray}
where $E_n^F$ and $E_p^F$ refer to the contribution of only the baryons to the Fermi energy. From the neutron dipole density, (\ref{uGDu}), and using the expression for $E_p$ from (\ref{Fp}) (keeping in mind the convention of (\ref{enerconven})), the ferromagnetic field in neutron star matter will be solutions of 
\begin{eqnarray}\label{bpfull}
		B
		&=&
			-\sum_{\lambda,n}\frac{|q_p B|}{4\pi^2}\int^{k_z^F(n,\lambda)}_{-k_z^F(n,\lambda)}
			\frac
			{\left(\sqrt{{m^*}^2+2|q_p B| n}+\lambda  g_p B\right)}
			{\sqrt{k_z^2+\left(\sqrt{{m^*}^2+2|q_p B| n}+\lambda  g_p B\right)^2}}
			\left(\frac{\alpha q_p n}{\sqrt{{m^*}^2+2|q_p B| n}}+\lambda  g_p\right)
			dk_z\nonumber\\
		&\,&
			-\frac{1}{(2\pi)^3}\sum_\lambda\int_{-k_{z}^F(\lambda)}^{k_{z}^F(\lambda)}dk_z
			\int_0^{k_{\bot}^F(k_z,\lambda)}
			\frac
			{\lambda g_n\left(\sqrt{{k_\bot}^2+{m^*}^2}+\lambda g_n B\right)\left(2\pi k_\bot\right)}
			{\sqrt{{k_z}^2 + \left(\sqrt{{k_\bot}^2+{m^*}^2}+\lambda g_n B\right)^2}}\,dk_\bot\nonumber\\
		&\,&
			-\sum_{l,\lambda,n}\frac{|q_l B|}{4\pi^2}\int^{k_z^F(n,\lambda)}_{-k_z^F(n,\lambda)}
			\frac{\alpha q_l n}
			{\sqrt{k_z^2+{m^*}^2+2|q_l B| n}}\,dk_z
			-\frac{\epsilon_p}{B}-\sum_l\frac{\epsilon_l}{B}+\left(E_n^F-g_\rho b_0\right)\frac{\rho_p}{B}.
\end{eqnarray}
\subsection{Medium effect on the coupling $g_p$}\label{runcoupp}
Our investigation of ferromagnetism in neutron star matter will proceed along the same lines as our investigation into the ferromagnetic phase of neutron matter. Similar to what was done in the case of neutron matter in chapter \ref{chap:ferroneu}, we will investigate the possibility of the ferromagnetic phase by adjusting the general baryon dipole coupling constant $g_b$. 
In this section we will derive an upper limit for the proton dipole coupling constant $g_p$.\\ 
\\
As shown in appendix \ref{ap:dipole} the magnetic dipole moment stems from the fact that the proton is charged (which contributes the factor of $2$), as well as from the fact that protons, like neutrons, are composite particles. In the units used in this study the magnetic dipole moment, $\mu^{(dip)}_p$, of the proton is (\ref{magdip}):
\begin{eqnarray}
	\mu^{(dip)}_p=(2+\kappa_p)\frac{q_p}{2 m}=(2+\kappa_p)\mu^{(dip)}_N\label{mup},
\end{eqnarray}
where $\mu^{(dip)}_N$ is the nuclear magneton and $\kappa_p=0.793$ (\ref{normaldip}).  To reproduce the correct value of the magnetic dipole moment at normal densities $g_p$ must be equal to
\begin{eqnarray}
	g_p=-\frac{\kappa_p\mu^{(dip)}_N}{2}=g_p^{\scriptscriptstyle(0)}= -0.0126,\label{gpnormal}
\end{eqnarray}
in the units used in this dissertation.\\
\\
As motivated for neutrons in section \ref{runcoup}, we also assume that the proton dipole moment will change as the density increases, due to composite nature of the particle. We will again establish an upper limit for the value of the proton magnetic dipole moment by considering the effective dipole moment of the (asymptotic) free quarks that make up the proton. The proton has two {\em up} quarks and one {\em down} quarks. Recalling the definition of the quark magneton (\ref{qmu}) the effective dipole moment of these three free quarks are
\begin{eqnarray}
	\mu^{(dip)}_{uud}=(2\mu^{(dip)}_u + \mu^{(dip)}_d).
\end{eqnarray}
Comparing this to the normal proton dipole moment (ignoring the prefactors),
\begin{eqnarray}
	\frac{\mu^{(dip)}_{uud}}{\mu^{(dip)}_N}\cong1800
\end{eqnarray}
shows that an increase in $g_p$ of the order of tens of times $g_p^{\scriptscriptstyle(0)}$ may not be unreasonable at high densities. For an increase of a factor of $x$ in the proton magnetic dipole moment $g_p$ has to be adjusted to
\begin{eqnarray}\label{gpAdj}
	g_p=x\times g_p^{\scriptscriptstyle(0)}=-\frac{x(2+\kappa_p)-2}{2}\mu^{(dip)}_N
\end{eqnarray}
according to (\ref{gpadj}).\\
\\
Having derived these equations we are now in a position to investigate the ferromagnetic phase of neutron star matter.

\section{Summary}
Based on the previous two chapters the formalism for investigating ferromagnetism in neutron star matter consisting of protons, neutrons, electrons and muons was derived. Since the particle density of charged fermions in a magnetic field is dependent on the magnetic field, care had to be taken in deriving expressions for different observables. Note that the formalism described here can also be used to study the response of neutron star matter in an external magnetic field.\\
\\
In the following chapters we will present the results obtained using this formalism to study the ferromagnetic phase in different types of neutron star matter as well as the response of these types of matter when an external magnetic field is applied.

\chapter{Results: Magnetised nuclear matter}\label{chap:resnuc}
In the first of two chapters concerning the results of this study, we will focus on the properties of magnetised symmetric nuclear matter. We calculated the properties mentioned in section \ref{sec:satnuc} in a non-zero magnetic field for different values of $g_b$ (the dipole coupling constant). The values obtained using different QHD parameters sets are compared to establish the differences in behaviour of these parameter sets' description of magnetised nuclear matter. \\
\\
We will not consider ferromagnetic fields in this chapter and all magnetic fields are assumed to be due to external sources. We assume a magnetic field, ${\bm B}=B\hat{z}$, which is constant throughout the system. Hence we will exclusively refer to the magnetic field in terms of its magnitude, $B$.

\section{Quantum Hadrodynamics}
As mentioned in chapter \ref{chap:unmag} the different QHD parameter sets employed here are QHD1, NL3 and FSUGold. Although these parameter sets give a fairly similar description for matter at normal densities, their behaviour at high density differ considerably (see \cite{diener} and references therein for more details). Since the exact properties of cold dense matter are not known, we are unable to distinguish which are the more applicable parameter sets and thus for comparison all three sets were used.\\
\\
Table \ref{tab:westcoupling} gives the values of the different coupling constants within these parameter sets while in table \ref{tab:westmasses} the various masses are listed.
\begin{table*}[bbh]
	\centering
		\begin{tabular}{cccccccc}
			\hline\hline
			Model & $g_s^2$ & $g_v^2$ & $g_\rho^2$& $\kappa$ [MeV] & $\lambda$ & $\zeta$ & $\Lambda_v$ \\
			\hline
			QHD1 \cite{recentprogress}  & 109.6 &  190.4 & 0.0 & 0.0 & 0.0 & 0.0 & 0.0 \\
			NL3 \cite{FSU}  & 104.3871 & 165.5854 & 79.6000 & 3.8599 & -0.01591 & 0.00 & 0.00 \\
			FSUGold \cite{FSU} & 112.1996 & 204.5469 & 138.4701& 1.4203 & +0.0238  & 0.0600 & 0.0300 \\
			\hline
		\end{tabular}
		\caption[Coupling constants of different QHD parameter sets. ]
		{\bf Coupling constants of different QHD parameter sets. All coupling constants are dimensionless, except for $\kappa$ which is given in MeV\footnotemark.}
	\label{tab:westcoupling}
\end{table*}
\footnotetext{In tables or graphs where the unit of an observable is listed, it will always be listed in square brackets.}
\begin{table*}[tbh]
	\centering
		\begin{tabular}{cccccc}
			\hline\hline
			Model & $M_{\mbox{\footnotesize{proton}}}$& $M_{\mbox{\footnotesize{neutron}}}$& $m_s$& $m_\omega$& $m_\rho$ \\
			\hline
			QHD1 \cite{recentprogress}  &  939 & 939 & 520 & 783 \\
			NL3 \cite{FSU} & 939 & 939 & 508.1940 & 782.5  & 763\\
			FSUGold \cite{FSU}& 939 & 939 & 491.5000 & 782.5  & 763\\
			\hline
		\end{tabular}
	\caption{Particle masses [MeV] of the different parameter sets.}
	\label{tab:westmasses}
\end{table*}
%
%
%
\begin{table}[tbh]
	\centering
		\begin{tabular}{|lcccc|}
		\hline
		Parameter set&\multicolumn{2}{c}{$\rho_{\mbox{\footnotesize 0}}$ [fm$^{-3}$]}&\multicolumn{2}{c|}{E$_b$ [MeV]}\\
		&Published&Calculated&Published&Calculated\\
		\hline\hline
		Observed \cite{csg}&0.153&&-16.3&\\
		QHD1 \cite{recentprogress}	&0.148	&0.148&	-15.75&	-15.73\\		
		NL3	\cite{FSUL}&0.148	&	0.148&-16.24	&-16.24\\
		FSUGold	\cite{FSUL}&0.148	&0.148&	-16.30&	-16.28\\
		\hline
		\end{tabular}
	\caption[Comparison between the published and calculated values of the saturation density and the binding energy per nucleon of nuclear matter.]{Comparison between the published and calculated values of the saturation density of nuclear matter as well as the binding energy per nucleon at saturation. The references after the name of the parameter set indicate the source of the published values.}
	\label{tab:satprop1}
\end{table}
\section{Magnetised nuclear matter properties}
Table \ref{tab:satprop1} lists the saturation density and binding energy at saturation for unmagnetised matter calculated with the three parameter sets. Although the observed saturation density of nuclear matter is $0.153$ fm$^{-3}$ \cite{csg}, whenever any reference to $\rho_0$ is made it will be taken as 
\begin{eqnarray}
	\rho_0=0.148\mbox{ fm}^{-3}
\end{eqnarray}
since that is the theoretical saturation density of nuclear matter across the board for all the QHD parameter sets.\\
\\
We calculated the 
nuclear properties for magnetised matter 
by varying both the magnetic field and the coupling constant $g_b$ (although not simultaneously). As motivated in section \ref{sec:amu}, for all calculations of nuclear matter properties, (\ref{Amu}) was used for the gauge field and consequently the electric field was always taken to be zero.
\subsection{Note on magnetic field strengths}
In the calculations made during this study, the magnetic field was considered in units of fm$^{-2}$, but when plotted it was converted to gauss using (\ref{b2g}). Note that the conversion factor (\ref{b2g}) is very large. For small values of the magnetic field the occupied Landau levels will number in the ten thousands. Thus studying the system for very small values of the magnetic field was computationally very intense and hence very selectively done.
\section{Saturation density and binding energy}\label{sec:reb}
\begin{figure}
	\centering
		\includegraphics[width=1.\textwidth]{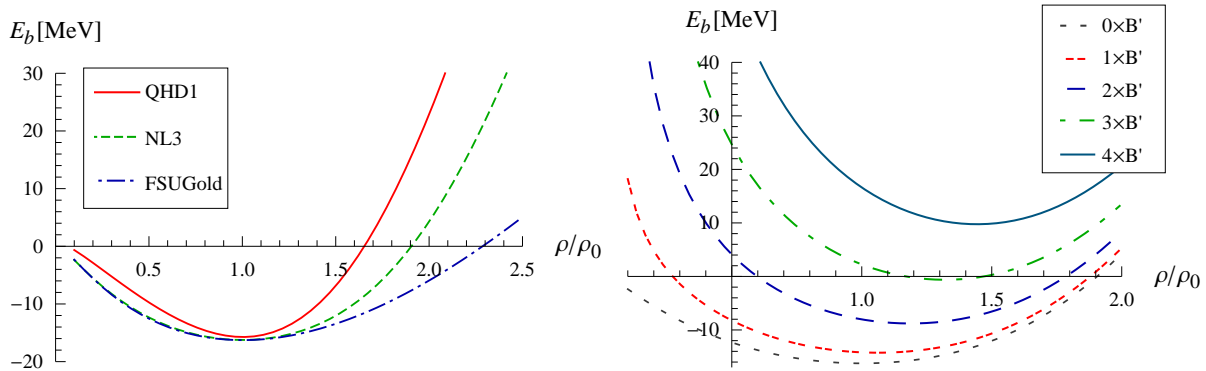}
	\caption[Binding energy of magnetised and unmagnetised nuclear matter.]{On the left is the binding energy for unmagnetised matter for the different parameter sets. On the right we have the binding energy calculated in the NL3 parameter set as a function of density for different magnetic field strengths in units of $B'$, where $B'=1.0\times 10^{17}$ G and $g_b=g_b^{(0)}$.}
	\label{fig:BE(B)NL3}
\end{figure}

\begin{figure}
	\centering
		\includegraphics[width=1.\textwidth]{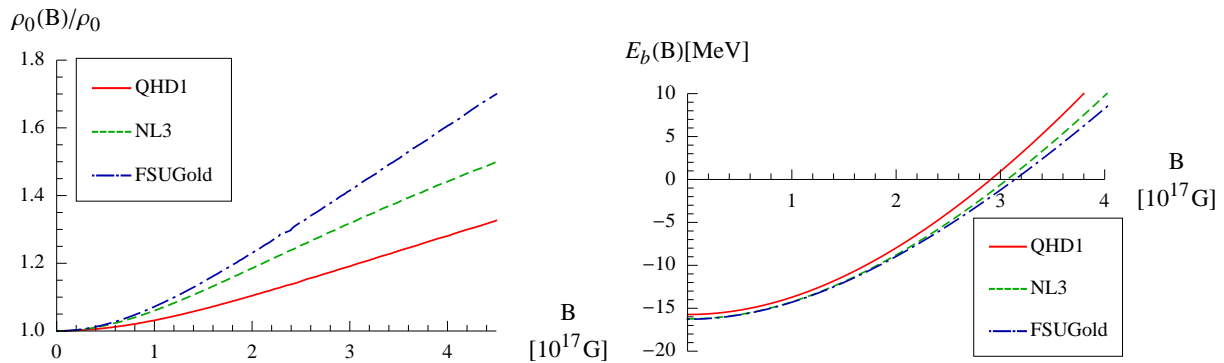}
	\caption[Binding energy at saturation and saturation density of magnetised matter in QHD.]{Combination plot of the saturation density, which is normalised with regards the $B=0$ values, and the binding energy at saturation. Plotted for the different QHD parameter sets with $g_b=g_b^{(0)}$ throughout.}
	\label{fig:BE(B)}
\end{figure}

\begin{figure}
	\centering
		\includegraphics[width=1.\textwidth]{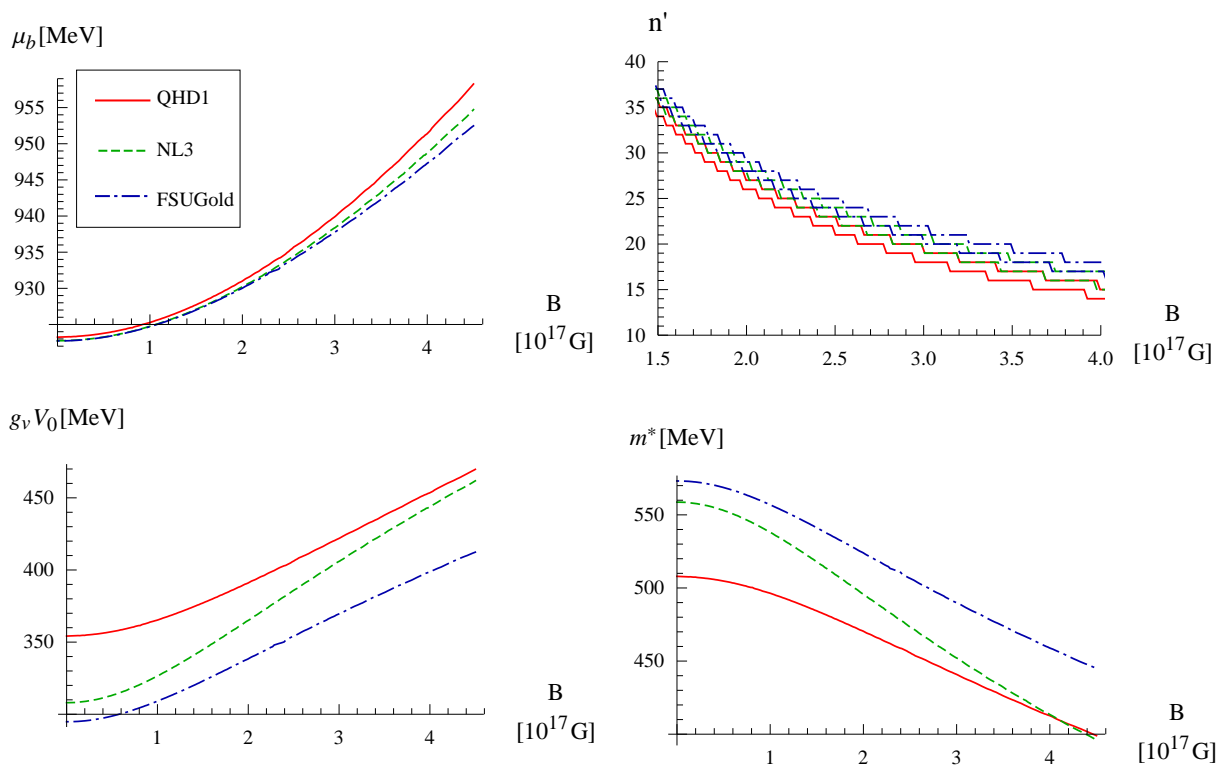}
	\caption[Plots of nuclear matter observables at saturation as a function of the magnetic field. ]{Plots of nuclear matter observables, at the saturation density, as a function of the magnetic field. Starting top left, and continuing clockwise, we have the Fermi energy of the protons and neutrons (only the neutron Fermi energy is plotted, but they do not differ noticeably). Then $n'$, from (\ref{llcon}), labels the Landau level occupation of the two proton species where the $\lambda = 1$ species occupies the most levels (for all parameter sets). This is followed by the plot of the reduced mass $m^*$, and finally the value of the product of the expectation value of the vector meson field times its coupling constant $g_v$.}
	\label{fig:BEanal}
\end{figure}

The binding energy per particle $E_b$ was calculated with 
\begin{eqnarray}
	E_b=\frac{\epsilon}{\rho}-m
\end{eqnarray}
from \cite{walecka} for a range of densities. For the corresponding energy densities we used the terms appropriate to symmetric nuclear matter from (\ref{epsLL2}). The saturation density is defined as the density at which $E_b$ is minimum.\\
\\
Figure \ref{fig:BE(B)NL3} shows the symmetric nuclear matter binding energy curves. On the left in figure \ref{fig:BE(B)NL3} $E_b$ was plotted for unmagnetised matter for all three the QHD parameter sets. On the right we show $E_b(B)$, calculated in the NL3 parameter set\footnote{An arbitrary choice: the behaviour for all parameter sets is very similar in this regard.}, for different values of an external magnetic field. For these plots the coupling between the nucleon's dipole moment and the magnetic field was kept at $g_b^{(0)}$ (the value that would reproduce the normal observed values of the magnetic dipole moments of the baryons). 
For magnetised matter the most noticeable effect is shown in the second plot in figure \ref{fig:BE(B)NL3}: at low densities the contribution of the magnetic field dominates the energy density and forces the minimum of the binding energy curve to higher densities. The minimum of the binding energy curve 
rises as $B$ increases, due to the nucleon contribution being overpowered by the magnetic field.\\
\\
Taking the above into account, the plots of the behaviour of the saturation density, $\rho_0(B)$, and $E_b(B)$ for magnetised matter holds no surprises. Figure \ref{fig:BE(B)} shows these properties, as calculated in the three parameter sets with $g_b=g_b^{(0)}$. 
From these plots we note that $E_b(B)$ behaves very similar for all parameter sets: as the magnetic field increases the system becomes less bound in a similar fashion for all the parameter sets. The system is bound up to about $3\times 10^{17}$ G. As the magnetic field increases further, the binding energy will become positive and as consequently the system will become unstable. \\
\\
As the magnetic field increases the saturation density, $\rho_0(B)$, increases correspondingly. Thus, albeit more weakly bound, the system can accommodate much denser matter as it becomes magnetised. This tendency of the saturation density was first noted by Chakrabarty et al. in \cite{chakPRL}, although they found that $E_b(B)$ increases with $B$ and thus the system becomes more strongly bound (with larger binding energy)\footnote{In this paper magnetised nuclear matter with only the $q\bar{\psi}\gamma_\mu A^\mu\psi$-coupling between the charged particles and the magnetic field was studied.}. However, the behaviour of the saturation density is distinct for the three parameter sets. Most significantly, the FSUGold $\rho_0(B)$ increases proportionally the most. Note that although the FSUGold parameter set has an additional self-coupling in the vector meson field $V_0$, as well as a coupling between $V_0$ and the isovector field $b_0$, the latter has no influence on this calculation, since we are dealing with symmetric nuclear matter and thus $b_0=0$ for all parameter sets.\\
\\
To investigate the behaviour of the $\rho_0(B)$, various properties were plotted in figure \ref{fig:BEanal}. Clockwise, from top left, we have the baryon Fermi energies, the number of occupied Landau levels, the reduced mass and finally the vector meson field. 
Since we are dealing with symmetric nuclear matter with $g_b=g_b^{(0)}$, there is almost no reason for the Fermi energies of protons and neutrons to differ and furthermore the Fermi energies from the different parameter sets show little difference. However, there is a subtle interplay between the different meson contributions to the energy density that influences the baryon densities that the various parameter sets can accommodate.\\
\\
Referring to the top right panel in figure \ref{fig:BEanal} we see that for the QHD1 parameter set the least number of Landau levels are occupied, which corresponds to it having the lowest densities. 
The values of the NL3 parameter set are somewhat obscured, but lies between that of the QHD1 and FSUGold sets. Since $g_b=g_b^{(0)}$, the spacing between the number of occupied levels pertaining to the same parameter set (representing the different particle species with $\lambda =1$ and $\lambda =-1$) are on average one. This is to be expected, since for $B>0$ only the $\lambda=1$ protons can occupy the $n=0$ levels (in the convention established by (\ref{llcon})). Hence the difference of one in the number of occupied levels. \\
\\
The presence of the vector and scalar mesons increases the energy density. As the reduced mass (\ref{mstar}) gives us a direct indication of the magnitude of the scalar meson expectation value we can, from the two bottom row plots in figure \ref{fig:BEanal}, estimate the meson contributions to the energy density. For QHD1 the values of $m^*$ are the lowest and $g_v V_0$ the highest. This means that for QHD1 the meson expectation values ($g_s\phi_0$ and $g_v V_0$)\footnote{The meson field ground state expectation values are shown in plots, or otherwise compared, multiplied by their coupling constant, since these are the meson contributions to the energy.} are the largest and therefore make the largest contribution, of all the parameter sets, to the energy density. Consequently at the saturation density, the baryon contribution of QHD1 to the energy density is the smallest. Therefore the QHD1 saturation densities should be the lowest. Applying this logic to the other parameter sets, it is apparent that the FSUGold parameter set should have the highest saturation densities since it has the lowest values of $g_s\phi_0$ and $g_v V_0$.  \\
\\
In \cite{diener}, and references therein, it has been well established that the FSUGold has a softer equation of state and can, in general terms, accommodate higher (energy) densities with a comparatively smaller increase in the pressure of the system\footnote{Whether a system has a stiff of soft equation of state is, in this case, due to the reaction of the meson expectation values to any increase in density, which is directly related to the various coupling terms and their coupling strengths. }. 
Consequently the fact that FSUGold accommodates higher densities at saturation is 
another reflection of its tendency towards a softer equations of state. The same goes for the QHD1 and NL3 parameter sets that are known to have stiffer equations of state and as a result have lower saturation densities. 
\subsection{Adjusted values of the baryon dipole moment}
\begin{figure}[ttt]
	\centering
		\includegraphics[width=1.\textwidth]{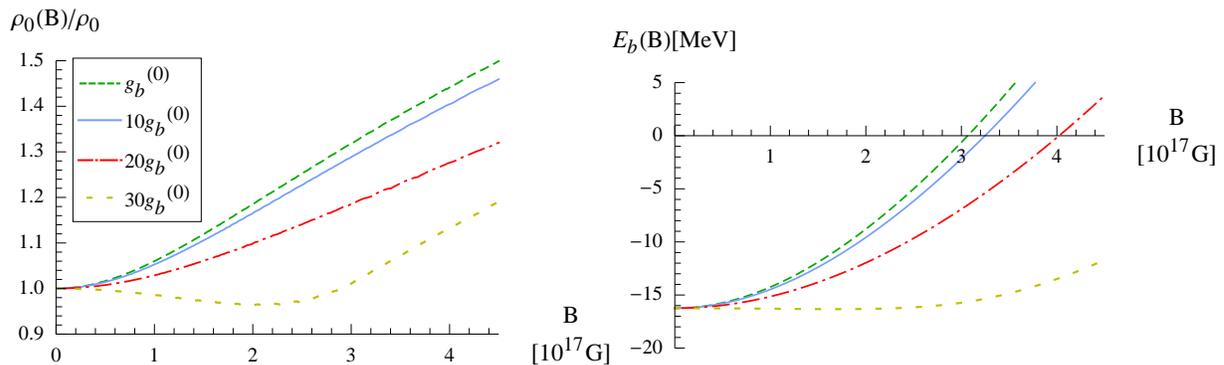}
	\caption[Saturation density and binding energy of magnetised matter as a function of $g_b$.]{Saturation density and binding energy at saturation for various values of the baryon-magnetic field coupling $g_b$. As before, $g_b^{(0)}$ indicates the value of $g_b$ that reproduces the normal values of the baryon magnetic dipole moments (\ref{normaldip}). The values of the $g_b$ for which the calculations were performed are thus indicated as a multiple of $g_b^{(0)}$.}
	\label{fig:BEgn}
\end{figure}
As a prelude to exploring the ferromagnetic phase boundary of neutron star matter, we also calculated the magnetised nuclear matter properties by adjusting $g_b$ through (\ref{gnAdj}) and (\ref{gpAdj}).  
By adjusting the coupling strength $g_b$ of the dipole coupling, we are effectively increasing the strength of the baryon dipole moment. The value of $g_b$ is adjusted in such a way that the proton and neutron dipole moments are increased by the same factor, for example $g_b=10g_b^{(0)}$ means that the dipole moments of protons and neutrons are increased by a factor of 10. We observe that the individual parameter sets respond in a very similar way to the adjustment of $g_b$ and hence we will only show the results for the NL3 parameter set (once again an arbitrary choice).\\
\\
In figure \ref{fig:BEgn} we show the NL3 values of the $\rho_0(B)$ and $E_b(B)$ for $g_b$ equal to $g_b^{(0)}$, $10g_b^{(0)}$, $20g_b^{(0)}$ and $30g_b^{(0)}$. We do not at all claim that these values of $g_b$ are necessarily feasible or attainable for the range of densities and magnetic fields we used to calculate the properties, but we show them in order to illustrate the full spectrum of the possible behaviour of magnetised matter under extreme conditions.
\begin{figure}[ttb]
	\centering
		\includegraphics[width=1.\textwidth]{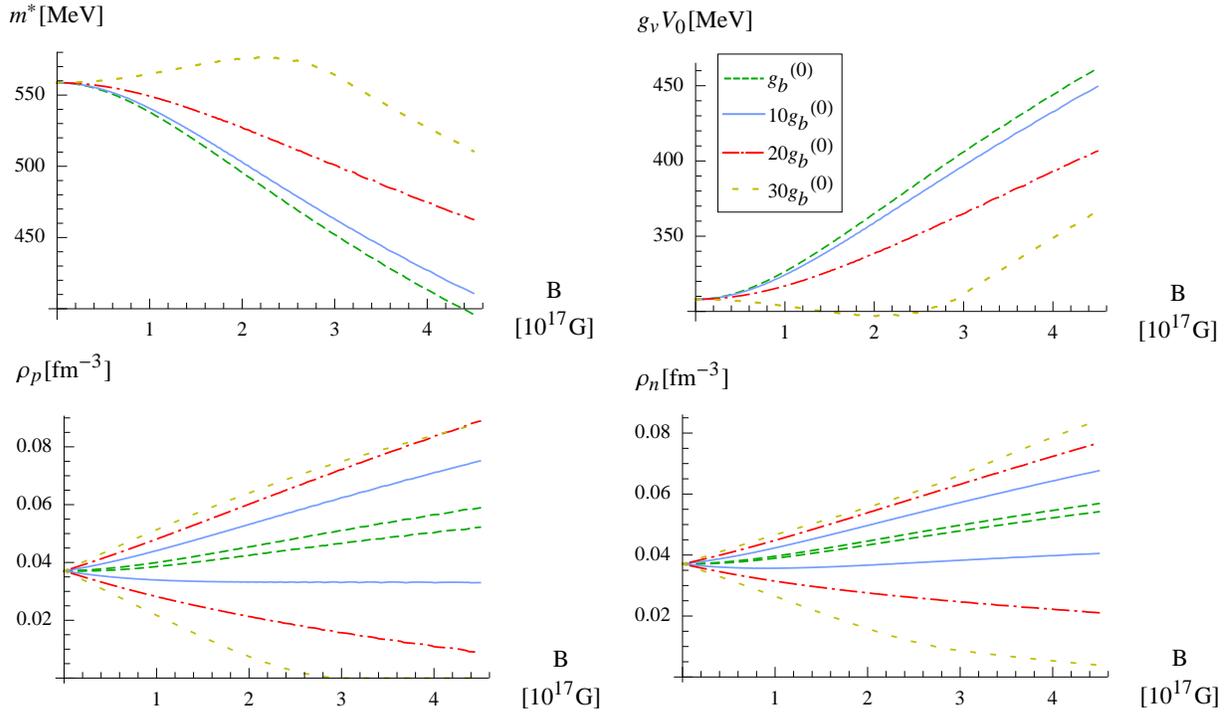}
	\caption[NL3 reduced mass, $\omega$ meson and the $\lambda=\pm 1$ species contribution to the proton and neutron densities at the saturation point for different values of $g_b$ as a function of magnetic field.]{NL3 reduced mass, $\omega$ meson expectation value and the $\lambda=\pm 1$ species contribution to the proton and neutron densities at the saturation point for different values of $g_b$ as a function of magnetic field. In the proton density plot the higher density values for different $g_b$, refers to $\lambda=1$, while for the neutrons $\lambda=-1$ would have the higher densities. }
	\label{fig:BEgnd}
\end{figure}\\
\\
From figure \ref{fig:BEgn} it is clear that as $g_b$ increases the system becomes less dense but more bound. Referring to the charged particle spectrum in a magnetic field (equation (\ref{Fp}) and illustrated in figure \ref{fig:LLap2}), we notice that by increasing $g_b$ a relative shift between the two species of protons ($\lambda=\pm1$) is produced (same for the neutrons). 
From (\ref{Fp}), for $B>0$, it is the $\lambda=1$ proton species whose minimum energy is lowered, while the $\lambda=-1$ species' minimum energy is raised. Since the neutron's dipole moment has the opposite sign to the proton's, for neutrons it will be the 
$\lambda =-1$ neutron species whose minimum energy is lowered, while $\lambda =1$ is raised. Consequently the $\lambda=1$ proton and $\lambda=-1$ neutron energy levels would be the ones preferentially filled, since populating them will lower the energy of the system. On the other hand, as $B$ increases it is clear from (\ref{Fp}) that the spacing between the levels will also increase. Therefore, if $B$ increases while the density is kept constant, the binding energy of the system will increase since the system is forced to also populate the now even higher lying $\lambda =-1$ proton energy levels. Since we are considering a saturated system, as $B$ increases the system accommodates progressively lower densities as $g_b$ increases, which will keep the energy density at a minimum. For very large values of $g_b$ the relative shift between the species energy levels is so large that only the species with the lowest energy levels is populated, while the occupation of the other is suppressed.\\
\\
In figure \ref{fig:BEgnd} the contributions of the individual species are shown for both protons and neutrons. Since we are dealing with the Landau problem for protons (and consequently their energy levels are discrete and thus more spread out than the continuous energy levels of neutrons), these baryons' $\lambda=-1$ species would be suppressed long before that of the neutrons' $\lambda=1$ species. In the case of $g_b=30 g_b^{(0)}$ the system can accommodate less protons (and consequently less neutrons as we are considering symmetric matter) since the $\lambda=-1$ proton species is almost completely suppressed for the whole range of $B$. This significantly reduces the saturation density.\\ 
\\
In figure \ref{fig:BEgnd} we also show the values of the reduced mass as well as $g_v V_0$, the omega meson ground state expectation value. The behaviour of the meson expectation values is directly correlated to the behaviour of the baryon densities through their equations of motion. 
Since the mesons are responsible for the interactions between the baryons, their behaviour primarily influences the value of the binding energy. Thus as $g_b$ increases the $\rho_0(B)$'s trends are fed back to the binding energy through the meson expectation values. 
\section{Compression modulus}
Another property of interest, as the magnetic field varies, is the compression modulus. This will give an indication of the compressibility of the magnetised nuclear matter. The compression modulus at saturation density $K$ was calculated using (\ref{K}). \\ 
\\
Firstly $K$ of symmetric nuclear matter, with the normal values of $g_b$, was calculated for the different QHD parameter sets. 
\begin{figure}[tbb]
	\centering
		\includegraphics[width=1.0\textwidth]{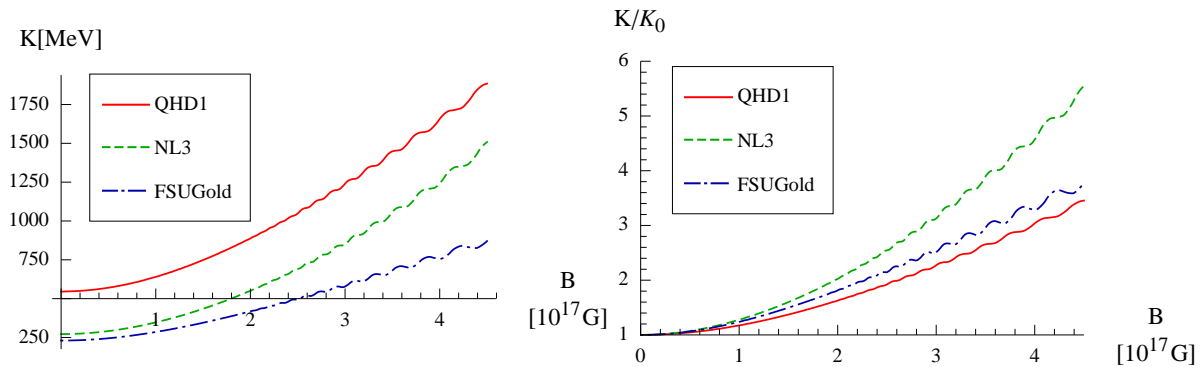}
	\caption[Compression modulus of magnetised nuclear matter for $g_b=g_b^{(0)}$ for various QHD parameter sets.]{Compression modulus of magnetised nuclear matter for $g_b=g_b^{(0)}$ for various QHD parameter sets. On the right we show the same plot, but the values are normalised with regard to their unmagnetised values.}
	\label{fig:K(B)}
\end{figure}
As the magnetic field increases $K$ also increases since it is calculated at an ever increasing density: as we have already established for increasing $B$, $\rho_0(B)$ also increases. This increase in $K$ is smooth up to a point, after which it becomes oscillatory with an ever increasing amplitude. Although not visible due to the scale of figure \ref{fig:K(B)}, this behaviour also manifests itself at smaller values of $B$. We will show that the oscillation of $K$ follows the depopulation of Landau levels\footnote{Since we start with a small magnetic field, a huge number of Landau levels (almost a thousand) will be occupied. As the field gets larger, this number will decrease since each level can accommodate more particles. }. This is not surprising, since in a Quantum Hall system one observes even more dramatic variation of the conductivity, which correlates to the population of the Landau levels \cite{yobi}. Thus we can reasonably suspect that such variational behaviour would also manifest itself in a magnetised nuclear matter system. \\
\\
Throughout the system the filling configuration (see section \ref{sec:nom}), and in particular the proton filling configuration, plays an important role. 
From (\ref{kLL}) and (\ref{rhoLL}) the magnetic field influences the proton filling configuration in two ways: through the Fermi momentum, $k_z^F(\lambda,n)$, as well as the degeneracy factor $\frac{|q_p B|}{4\pi^2}$. Depending on these, even if the density only increases by a small amount, the system may depopulate one or more Landau levels, if the levels below it in energy can accommodate the particles (by adjusting their $k_z^F(\lambda,n)$ values). These changes in the filling configuration are abrupt and not reversed as long as $B$ increases.
\begin{figure}
	\centering
		\includegraphics[width=1.0\textwidth]{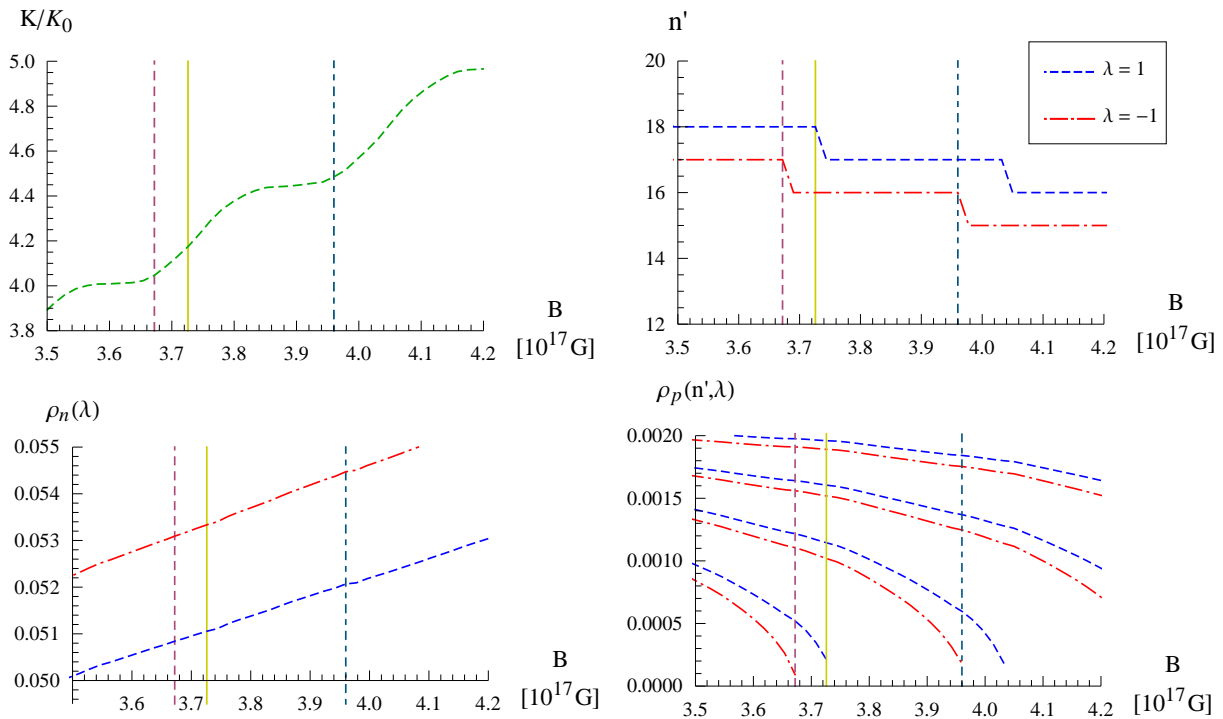}
	\caption[Analysis of NL3 K(B).]{Analysis of NL3 $K/K_0$ plot. Top right we have the Landau level occupation of 
	the two proton species. In the bottom row we show the distribution of the baryons densities among the baryon species. In the plot for $\rho_p$, the closely spaced graphs
	represent a specific value of $n'$. The vertical guidelines in all the plots denote the same values of the magnetic field in each plot. The legend applies to all panels. }
	\label{fig:Kanal}
\end{figure}\\
\\
Figure \ref{fig:Kanal} relates the points at which the depopulation takes place to the behaviour of $K$. Since $B>0$ the protons with $\lambda=-1$ has the higher energy and therefore the lower occupancy. We observe that as the occupation of the highest energy proton species' highest Landau level becomes smaller, $K$ increase only marginally.  Just before this proton species depopulates the compressibility starts to increase and continues to increase as this level depopulates. The inflection point of $K$ is where the $\lambda =1$ protons' highest Landau level depopulates. From this point onwards the changes in $K$ decrease, as the contribution of the $\lambda=-1$ protons' highest Landau level decreases, till the magnetic field reaches the point half way between $\lambda=-1$ decays. From here on $K$ changes marginally and the system contends itself by only moving particles between species (by flipping the particle's magnetic dipole moments) as well as moving particles between species' Landau levels, 
before the cycle repeats itself. Thus, as the Landau level depopulates, it is as if the system breathes: becoming more and less compressible as the levels depopulate. It would therefore also be interesting to see what happens when $g_b$ is adjusted.\\
\begin{figure}
	\centering
		\includegraphics[width=1.0\textwidth]{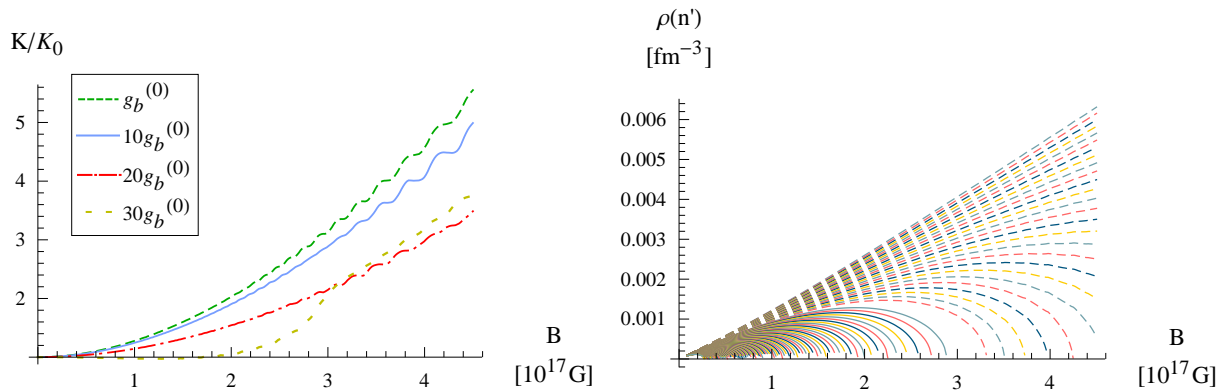}
	\caption[The compression modulus as a function of $g_b$.]{$K/K_0$ as $g_b$ varies. On the right we have the contribution to the proton density of the various Landau levels for both proton species when $g_b=30g_b^{(0)}$. The solid lines refers to the $\lambda =-1$ protons, while the dashed lines to the $\lambda=1$ protons. Note that in the plot a number of the higher $\lambda=1$ levels were omitted in order to make the plot clearer (for small $B$ a host of levels will be occupied, but quickly depopulated as $B$ grows).}
	\label{fig:K(B)gn}
\end{figure}
\\
Figure \ref{fig:K(B)gn} shows $K$ as $g_b$ varies for the NL3 parameter set. As $g_b$ increases the oscillatory behaviour of $K$ persists but $K$ increases all the more slowly, which mimics the behaviour of $\rho_0(B)$ for larger values of $g_b$. 
Correspondingly, for $g_b=30g_b^{(0)}$ $K$ behaves rather differently. It has already been established that for this value of $g_b$ the $\lambda=-1$ protons will be significantly suppressed. Initially, while some $\lambda=-1$ levels are still populated, the system is comparatively compressible since the pressure can be lowered by increasing the occupation of the $\lambda=1$ levels, albeit that the oscillatory behaviour is significantly suppressed. However, as the last $\lambda = -1$ level depopulates $K$ increases rapidly as the Landau configuration is essentially fixed since no exchange between proton species can take place anymore (which would have made the system more compressible). \\
\\
Figure \ref{fig:K(B)gn} also shows the contribution of the individual ($n'$) Landau levels to the proton density for $g_b=30g_b^{(0)}$: at a fixed $B$, summing all these contributions will give the total proton density $\rho_p$. In this plot we can clearly see the behaviour of the levels as they depopulate as well as the point where the $\lambda=-1$ levels are completely suppressed.
\section{Symmetry energy}
\begin{figure}
	\centering
		\includegraphics[width=1.0\textwidth]{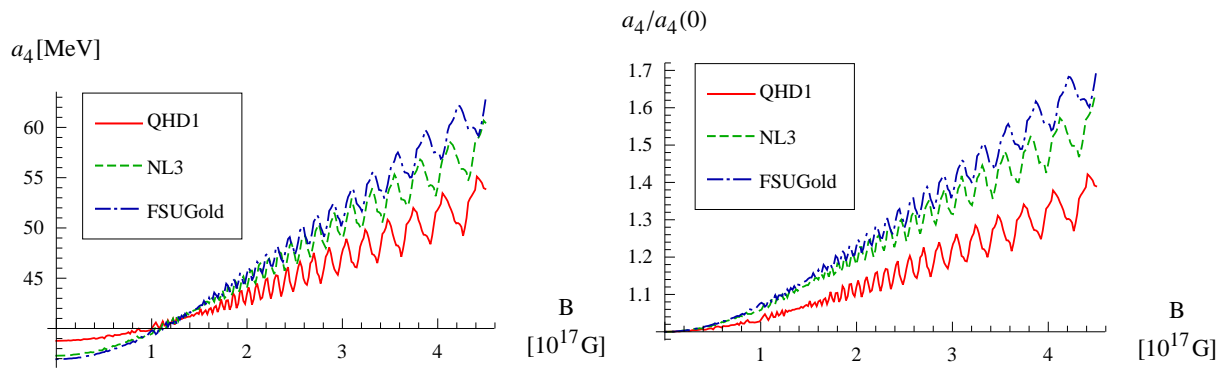}
	\caption{The unnormalised and normalised symmetry energy coefficients for magnetised nuclear matter for different QHD parameter sets.}
	\label{fig:a4(B)}
\end{figure}

The final quantity of interest, as far as nuclear matter properties are concerned, is the symmetry energy. We calculate the symmetric energy coefficient from (\ref{a4mag}) at the saturation densities obtained for different values of the magnetic field. 
Figure \ref{fig:a4(B)} shows the results.\\
\\
From (\ref{a4mag}) it is clear that the symmetry energy is concerned with the proton and neutron Fermi energies, in particular the nucleonic part $\mu'_b$. For unmagnetised symmetric nuclear matter there is no distinction between $\mu'_n$ and $\mu'_p$. Therefore $\mu'_n(t)$ and $\mu'_p(t)$ intersects at $t=0$ as the system prefers symmetric to asymmetric matter.\\
\\
Consequently the first term in (\ref{a4mag}) can effectively be taken, up to a minus sign that stems from the definition of $t$ (\ref{a4t}), as twice the value of one of the derivatives
\begin{eqnarray}
	 \frac{1}{4}\left(\left.\frac{d \mu'_n}{dt}\right|_{t=0}-\left.\frac{d \mu'_p}{dt}\right|_{t=0}\right)=\frac{1}{2}\left.\frac{d \mu'_n}{dt}\right|_{t=0}
	 =-\frac{1}{2}\left.\frac{d \mu'_p}{dt}\right|_{t=0}.
\end{eqnarray}
However, if this is not the case, from (\ref{a4mag}) we deduce that if either $\left.\frac{d \mu'_n}{dt}\right|_{t=0}$ or $-\left.\frac{d \mu'_p}{dt}\right|_{t=0}$ is larger than the other, then the system would prefer to shift from symmetric matter to favour the baryon with the smaller value.  For instance if 
\begin{eqnarray}
	\left.\frac{d \mu'_n}{dt}\right|_{t=0}>-\left.\frac{d \mu'_p}{dt}\right|_{t=0}
\end{eqnarray}
then the system would prefer to deviate from $t=0$ to $t<0$ (larger fraction of protons) and vice versa. From this point onwards we will use the short hand notation $\left.d_t \mu'_b\right|_{t=0}$, where
\begin{eqnarray}
	\left.d_t \mu'_b\right|_{t=0}=\left.\frac{d \mu'_b}{dt}\right|_{t=0}.
\end{eqnarray}
In this sense the oscillatory behaviour of $a_4$ in figure \ref{fig:a4(B)} is interesting. The overall trend for $a_4$ is to increase as the magnetic field increases, since the baryon saturation density, and therefore the Fermi energies, increases with the magnetic field. But this increase happens in a cyclic manner where $a_4$ increases sharply and then decreases almost just as sharp, although for a shorter range of $B$ in order for the nett effect to be a marginal increase from the initial point. After which the cycle repeats itself with larger amplitude. Once again, we relate this behaviour to the depopulation of the Landau levels, albeit in a more subtle way than was the case for $K$.  As far as comparing the parameter sets goes, the FSUGold parameter set has the largest values corresponding to this particular parameter set having the largest values of $\rho_0(B)$. 
\begin{figure}
	\centering
		\includegraphics[width=1.0\textwidth]{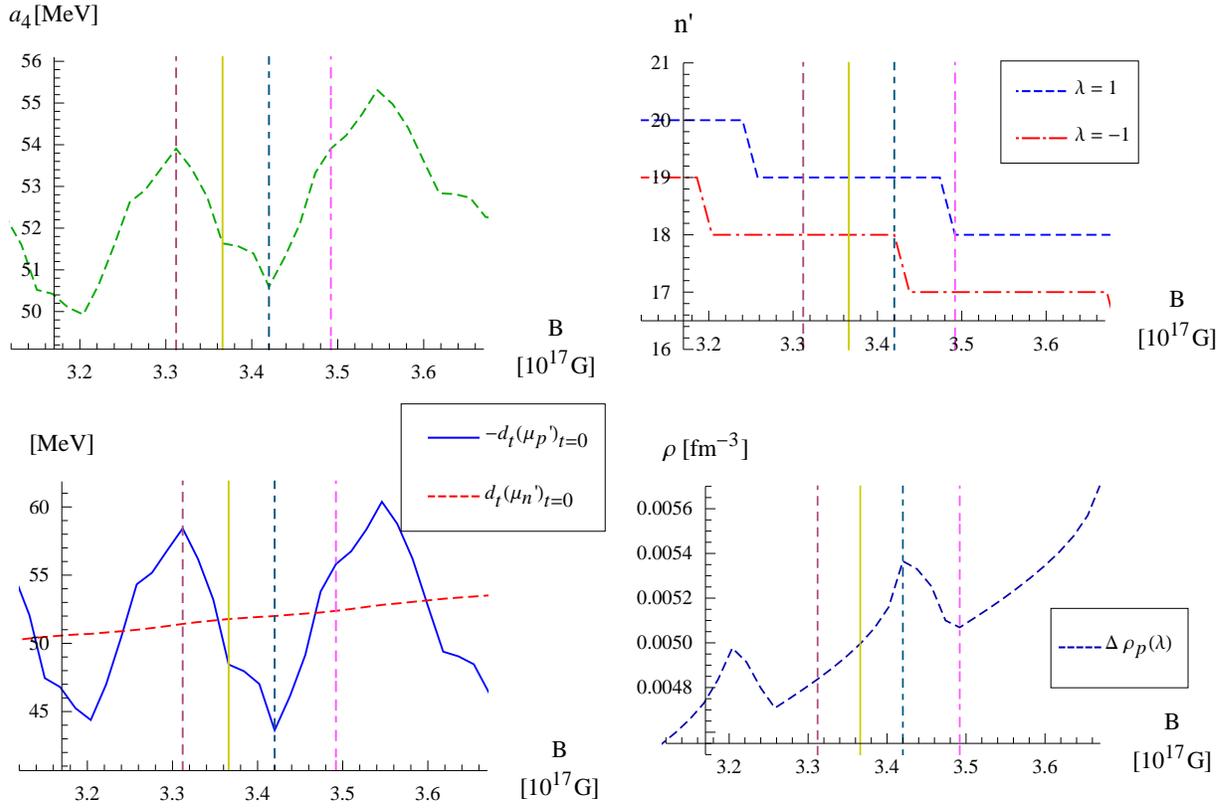}
	\caption[Analysis of the magnetised NL3 symmetry energy coefficients. ]{Analysis of the NL3 $a_4$ (top left). Clockwise right we have the Landau configuration for the proton species. Then we have the difference in density of these species. Finally the graphs of $\left.d_t \mu'_n\right|_{t=0}$ and $-\left.d_t \mu'_p\right|_{t=0}$. The vertical guidelines in all the plots denote the same values of the magnetic field in each plot.}
	\label{fig:a4(B)a}
\end{figure}
\begin{figure}
	\centering
		\includegraphics[width=1.0\textwidth]{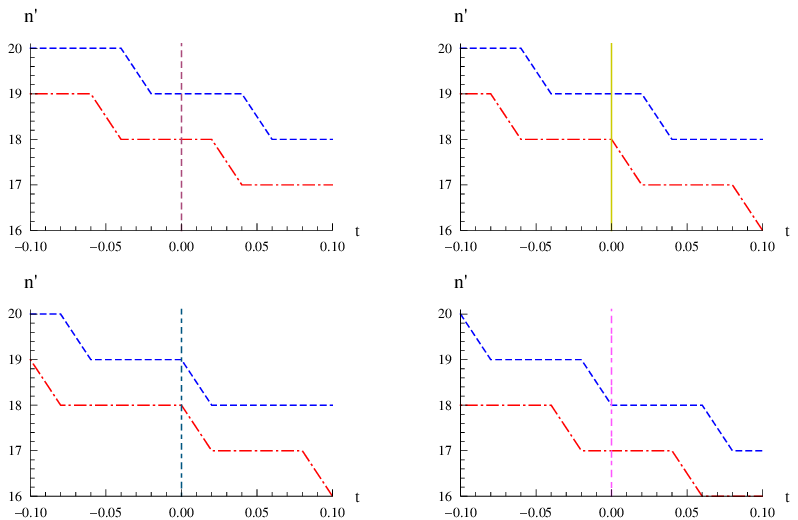}
	\caption[Plots of the Landau configuration as a function of $t$.]{Plots of the Landau configuration as a function of $t$. Each plot corresponds to a cut through the $n'$-$B$ plane along one of the vertical guidelines in the top right plot in figure \ref{fig:a4(B)a}. The lines at (0,16) in each plot corresponds to the corresponding vertical guideline in \ref{fig:a4(B)a}.}
	\label{fig:a4(B)t}
\end{figure}\\
\\
Zooming in on the NL3 $a_4$, we see in figure \ref{fig:a4(B)a} that the oscillatory behaviour is entirely due to the protons: $\left.d_t \mu'_n\right|_{t=0}$ is well behaved and contributes an almost constant amount to $a_4$. From the right hand side plots of figure \ref{fig:a4(B)a}, with the aid of figure \ref{fig:a4(B)t}, we gain an idea of the origin of the behaviour of $\left.d_t \mu'_p\right|_{t=0}$. \\
\\
Bottom right in figure \ref{fig:a4(B)a} we have the difference in density of the proton species. Top left we have the familiar Landau occupation of the protons which is closely related to the plots in figure \ref{fig:a4(B)t}. There we have the Landau occupation as a function of $t$ plotted for the values of the magnetic field indicated by the vertical guidelines in figure \ref{fig:a4(B)a}. Thus the four plots in figure \ref{fig:a4(B)t} represent perpendicular cuts through the $n'$-$B$ plane of top right in figure \ref{fig:a4(B)a}.  To indicate which plot in \ref{fig:a4(B)t} belongs to which vertical guideline in \ref{fig:a4(B)a}, the intersection line of the $n'$-$t$ plane with the $n'$-$B$ one is marked with the corresponding line.\\
\\
For the most part we will consider the plot showing the graphs of $\left.d_t \mu'_b\right|_{t=0}$ in figure \ref{fig:a4(B)a} (bottom left). Starting from the lowest valued guideline the Fermi energy of protons increases (the negative of $\left.d_t \mu'_p\right|_{t=0}$ is plotted) more steeply, since at this point the system tends to favour neutrons as $\left.d_t \mu'_n\right|_{t=0}<-\left.d_t \mu'_p\right|_{t=0}$. The next guideline indicates the point where $k_z^F$ (\ref{kLL}) of the highest $\lambda=-1$ proton energy level 
is essentially zero, but it is not yet energetically favourable to depopulate this level. From here the increase in $\mu'_p$ tapers off and reaches a minimum after the same point is reached for the highest energy $\lambda=1$ proton level. In particular, the increase in $\mu'_p$ tapers off because the $\lambda=1$ levels increase their density fraction 
by both absorbing the increase in $\rho_0(B)$ as $B$ gets larger, as well as some $\lambda=-1$ protons (in figure \ref{fig:delrho} we see that the increase in $\rho_0(B)$ is about five times smaller than the increase in difference between the densities of the proton species). The system also begins to favour $t<0$, as $\left.d_t \mu'_n\right|_{t=0}$ becomes larger than $-\left.d_t \mu'_p\right|_{t=0}$.
\begin{figure}
	\centering
		\includegraphics[width=1.0\textwidth]{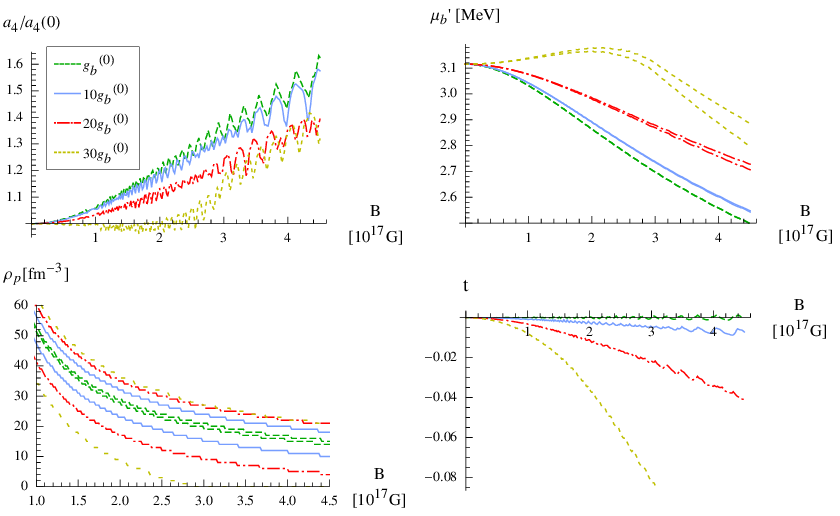}
	\caption[Normalised symmetry energy coefficients of NL3 for various values of $g_b$.]{Normalised $a_4$ of NL3 for different values of $g_b$ in the panel top left. Top right are the plots for $\mu'_b$ at the saturation densities, corresponding to the graphs in the plot on the left. For $g_b=g_b^{(0)}$ and $g_b= 10g_b^{(0)}$ the difference between $\mu'_p$ and $\mu'_n$ cannot be discerned. For the other two values of $g_b$: $\mu'_p<\mu'_n$. Bottom left we have the intersection points of $\mu'_p$ and $\mu'_n$ in $t$-space, from (\ref{a4}), indicating that the matter tend to be more proton rich as $g_b$ increases. Bottom left we show the Landau configuration at the saturation for the different values of $g_b$.  }
	\label{fig:a4(B)gn}
\end{figure}\\
\\
The steep increase in $\mu'_p$ comes to a halt when $k_z^F$ of the highest $\lambda=1$ proton level is also essentially zero. Then the highest energy $\lambda=-1$ level depopulates and the system starts to dump $\lambda=-1$ protons into $\lambda=1$ levels (which slows the increase in $\mu'_p$ considerably) till it is energetically favourable to depopulate the highest energy $\lambda=-1$ level. After which the system normalises in the sense that the difference between the density fraction of the proton species, as well as $\mu'_p$, increases steadily. Then the cycle starts over again. Thus, as the magnetic field increases, the system tends to oscillate between preferring proton rich or neutron rich matter, depending on the Landau configuration. \\
\\
Considering other values of $g_b$, which are shown in figure \ref{fig:a4(B)gn}, we see that in general the behaviour of $a_4$ is the same, but at lower values than for $a_4$ with $g_b=g_b^{(0)}$. This indicates that the matter tends to be more asymmetric, tending to be more proton-rich as $g_b$ increases.
On the right in figure \ref{fig:a4(B)gn} we have plotted $\mu'_b$ for the different values of $g_b$. Here we see that as $g_b$ increases $\mu'_p$ becomes distinctly smaller than $\mu'_n$. Thus the system starts to favour protons, since by increasing $g_b$ the energy of the (degenerate) proton levels are lowered. These degenerate proton levels can accommodate more particles at lower energies than the neutron species and hence $\mu'_p<\mu'_n$. 
Consequently $\mu'_n(t)$ and $\mu'_p(t)$ do not intersect at $t=0$ anymore, but at $t<0$. We plot the intersection point of $\mu'_n(t)$ and $\mu'_p(t)$ in the bottom right panel of figure \ref{fig:a4(B)gn}. We note that for $g_b=30g_b^{(0)}$, as the $\lambda=-1$ level becomes completely suppressed, $a_4$ increases noticeably and the system tends to favour less asymmetric matter.\\
\\
We conclude that the tendency of magnetised matter to be more proton-rich matter stems from the ability of the system to lower its energy more easily by filling the degenerate proton Landau levels than the magnetised neutron energy levels. 
\section{Electromagnetic emission}
\begin{figure}
	\centering
		\includegraphics[width=.75\textwidth]{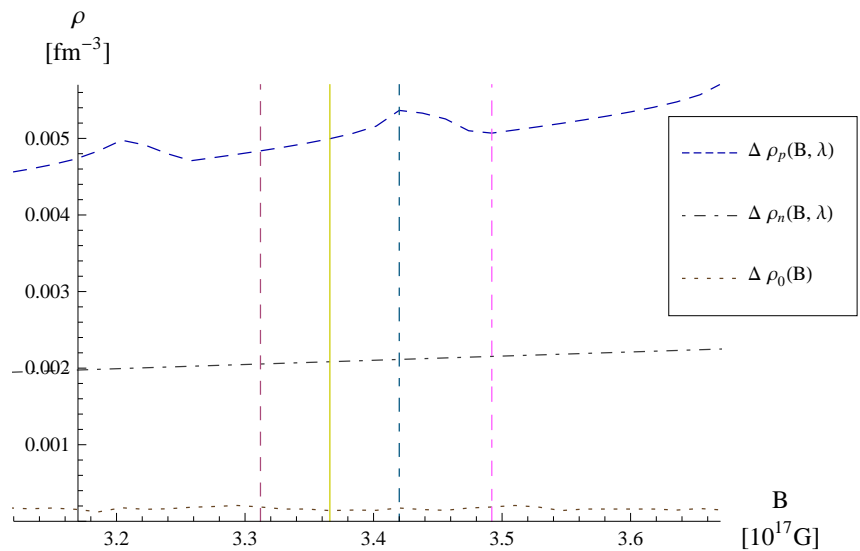}
	\caption[Saturation density for increasing $B$, as well as the difference between the densities of the proton and neutron species.]{The differences between the NL3 baryon saturation density for increasing $B$ ($g_b=g_b^{(0)}$), $\Delta\rho_0(B)$, as well as the difference between the densities of the proton and neutron species, $\Delta\rho_p(B)=\rho_p(B,+1)-\rho_p(B,-1)$ and $\Delta\rho_n(B)=\rho_n(B,-1)-\rho_n(B,+1)$ respectively. Note that the increase in $\Delta\rho_0(B)$ is considerably smaller than the relative increase in the densities of the baryon species. The plot is the expanded view of the bottom left panel in figure \ref{fig:a4(B)a}.}
	\label{fig:delrho}
\end{figure}

We note that as the magnetic field in magnetised nuclear matter increases, significant amounts of electromagnetic radiation might be released. In figure \ref{fig:delrho} we again show the bottom right panel of figure \ref{fig:a4(B)a}, but this time we include the difference between the neutron species density, $\Delta\rho_n(B)$, as well as the increase in the total baryon saturation density as $B$ changes, $\Delta\rho_0(B)$. \\
\\
From these graphs we deduce that both the lower energy proton and neutron species, 
$\rho_p(B,+1)$ and 
$\rho_n(B,-1)$, are constantly absorbing more particles 
than just the increase in $\rho_0(B)$. Therefore, in addition to the change in 
$\rho_0(B)$, a significant number of baryons must also flip their dipole moments as $B$ increases. By flipping their dipoles ($\lambda$) the particle will radiate energy, since their energies drop to a lower energy state.
\\
\\
Of course the fields considered here cannot be produced in a laboratory. However, in a stellar environment, where there are massive amounts of magnetised particles, this emission might become significant. We continue in the next chapter investigating the behaviour of magnetised neutron star matter.
\section{Summary}
We discussed the properties of magnetised nuclear matter for different QHD parameter sets and found that even for the normal values of $g_b$ the magnetic field has a profound influence. In a magnetised environment more dense matter can be bound. As $g_b$ is increased less dense matter can be more tightly bound. The compressibility depends on the filling configuration of the 
Landau levels and varies as the configuration changes. Magnetising nuclear matter tends to favour the production of more proton-rich matter, but the influence of the Coulomb interaction was not considered. A more realistic approach would be to consider charge neutral, beta-equilibrated matter, which we will do in the next chapter.

\chapter{Results: Ferromagnetism}\label{chap:resferro}
In this chapter we will focus on the results of our research pertaining to the ferromagnetic phase transition in neutron and neutron star matter. The influence of the ferromagnetic phase on neutron star properties will also be discussed. \\
\\
The manifestation of the ferromagnetic phase is essentially due to the fact that the baryon magnetic dipole moments couple to any magnetic field in the system. However, the transition to the ferromagnetic phase is only made if this coupling conspires to lower the total energy of the system. Due to the particle's spin there are two possible orientations of the magnetic dipole: parallel or anti-parallel to the magnetic field (which we have already denoted by $\lambda$). 
In the single particle baryon energies, (\ref{Fn}) and (\ref{Fp}), the different values of $\lambda$ 
reflect a split in the single particle energy spectrum, since one orientation will have a lower energy than the other. Thus, in the presence of a magnetic field, there will be an asymmetric filling of energy levels, as the lower energy levels will be preferentially filled (since we assume the system to be in its ground state). Due to this asymmetric filling the system will develop a nett magnetic dipole moment (magnetisation), as all the dipole moments are not paired with a counterpart which has the opposite value of $\lambda$.\\
\\
The magnetic field generated by the system's magnetic dipole moment is the ferromagnetic field. We assume that the system will be stable if the induced magnetic field matches the magnetic field necessary to generate the appropriate splitting between the energy levels of particles with opposite dipole moment orientations. This, in essence, is the calculation that we performed, the results of which 
are reported here. 
%
%
%
\section{Phase diagram: neutron matter}
\begin{figure}
	\centering
		\includegraphics[width=.70\textwidth]{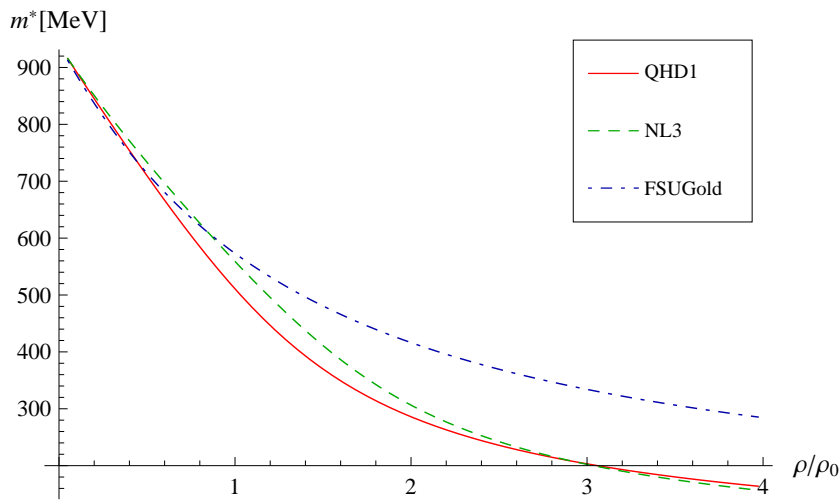}
	\caption{The reduced mass, $m^*$, at the phase boundary for QHD parameters sets in neutron matter.}
	\label{fig:neumstar}
\end{figure}

The ferromagnetic phase boundary in neutron matter is defined by the values of $g_n$ for which the energy density of unmagnetised neutron matter matches that of neutron matter magnetised by a very small constant magnetic field.
Using the formalism developed in chapter \ref{chap:ferroneu} we first calculated this phase boundary as a function of particle density. 
\begin{figure}
	\centering
		\includegraphics[width=.750\textwidth]{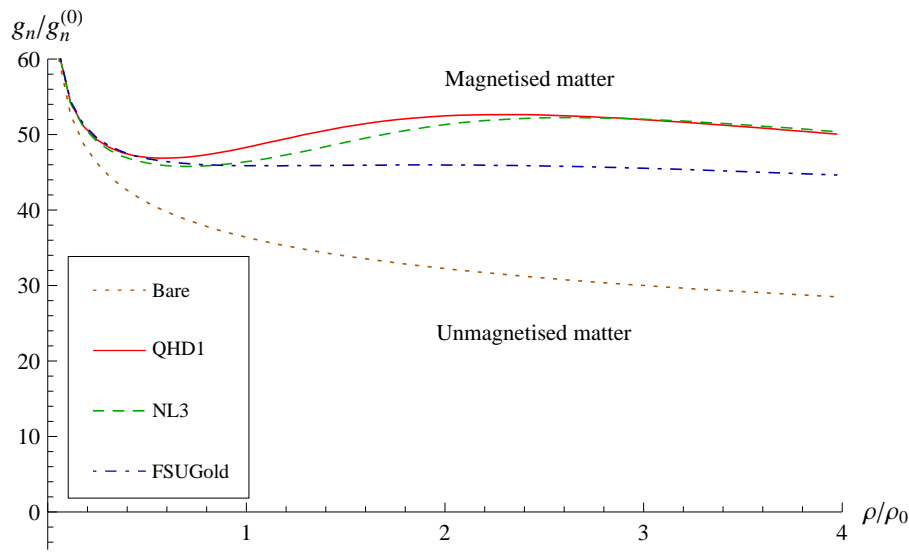}
	\caption[The ferromagnetic phase diagram as a function of $g_n$ for neutron matter. ]{The ferromagnetic phase diagram as a function of the coupling constant $g_n$ for neutron matter. Plotted on the x-axis is the neutron density 
	normalised to the density of saturated nuclear matter, $\rho_0=0.153$ fm$^{-3}$. On the y-axis the $g_n$, normalised to the normal observed magnetic dipole moment 
	of the neutron, $g_n^{(0)}$ (\ref{gnnormal}), is plotted. The value of $g_n$ observed at normal densities will thus be located at the point $(1,1)$ on the graph.}
	\label{fig:gbNeu}
\end{figure}\\
\\
Figure \ref{fig:gbNeu} shows the results 
for a non-interacting neutron gas (denoted by {\em Bare} on the plot) as well as the QHD1, NL3 and FSUGold parameter sets. From the plot it is clear that at normal densities neutron matter will only spontaneously magnetise at a value of $g_n$ that is around 45 to 50 times stronger than the one observed under normal conditions. For a non-interacting gas the energy of the system is lower compared to the interacting case due to the absence of forces between the neutrons. Consequently the phase boundary is reached at lower values of $g_n$.\\
\\
Next we consider the behaviour of the phase boundary for QHD1. This parameter set includes only the scalar (sigma) and vector (omega) mesons to the description of nuclear matter. The omega meson is responsible for the short range repulsion in the nucleon-nucleon (NN) interaction, while the scalar meson generates the long range interaction \cite{walecka}. From the neutron energy spectrum (\ref{singlepatE}) it is clear that the omega meson will not influence the phase boundary, since it shifts the energy for both choices of $\lambda$ by the same amount. Hence the difference between the neutron gas and QHD1 phase boundaries must be due to the presence of the scalar mesons. These mesons effectively modify the baryon mass, which is captured 
in the reduced mass $m^*$ (\ref{mstar}). 
Considering equation (\ref{Fn}), 
\begin{eqnarray}
	e_n({\bm k},\lambda)
	= \sqrt{\left(\sqrt{k_{\bot}^2+{m^*}^2}+\lambda g_n B\right)^2+k_{z}^2} + g_vV^0-\frac{g_\rho b_0}{2},\nonumber
\end{eqnarray}
we deduce 
that the dipole coupling to the magnetic field influences the energy in a manner similar to that of $m^*$. 
At low densities the difference between $m^*$ and $m$ is small, but as the density increases $m^*$ becomes much smaller than $m$. Correspondingly the value for $g_n$ must increase to achieve the same effect as in the $m^*= m$ case. Hence the increase in $g_n$ at the phase boundary for QHD1. The increase in the difference between the Bare and QHD1 phase boundaries are due to the change in the slope of $m^*$ as the density increases. See figure \ref{fig:neumstar} for graphs of the neutron matter $m^*$ at the phase boundary for the different parameter sets.
\\
\\
The QHD1 and NL3 parameter sets' phase boundaries are very similar, although NL3 also includes the vector rho meson. As was the case for the omega meson, the rho meson will not influence the phase boundary, since it's source is the isospin density (\ref{isorho}). For neutron matter the isospin density is a bit of a misnomer since it simply is the negative of the neutron density, which is independent of $\lambda$ and thus cannot influence the magnetic field. 
However, the small difference in behaviour of the NL3 and QHD1 phase boundaries is due to the inclusion of self-coupling terms in the scalar field of the NL3 parameter set. These were originally included to better describe the compressibility of nuclear matter (see \cite{diener} and references therein). Including these self-coupling terms modifies the reduced mass slightly and this behaviour is mirrored in the phase boundary.\\
\\
The FSUGold parameter set was developed to better constrain the density dependence of the nuclear symmetry energy without influencing the nuclear matter properties \cite{FSU}. To achieve this it includes a coupling between the omega and rho meson fields and was constrained by considering the binding energies and charge radii of magic nuclei. Therefore the FSUGold parameter set has different values of the coupling constants of the self-coupling terms in the scalar field compared to the NL3 parameter set. This leads to the reduced mass having a more gradual behaviour as the density increases than in the NL3 case. 
This is also reflected in the phase boundary.
\section{Phase diagram: neutron star matter}
\begin{figure}
	\centering
		\includegraphics[width=.750\textwidth]{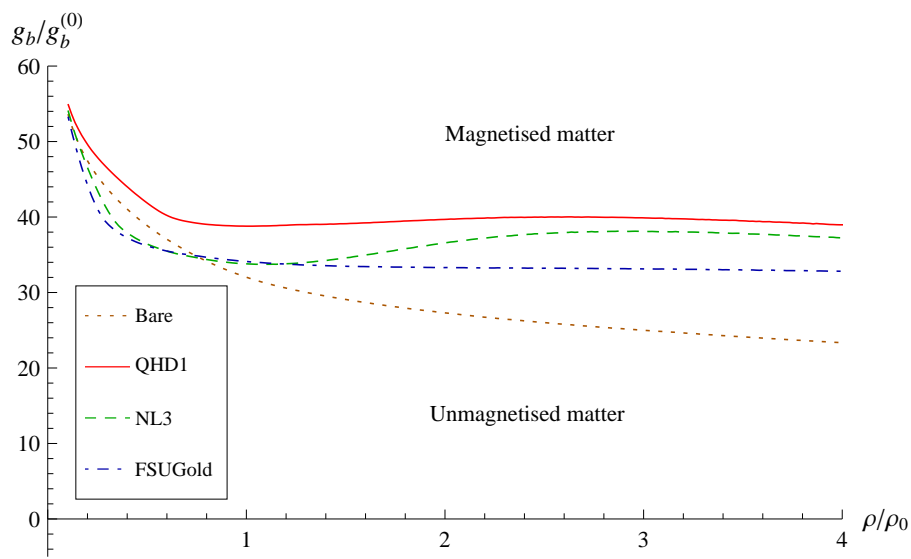}
		\caption[Ferromagnetic phase boundary for neutron star matter as a function of $g_b$.]{Ferromagnetic phase boundary for neutron star matter as a function of normalised baryon density and coupling constant $g_b$. Once again the different graphs refers to a charge neutral, beta-equilibrated gas of baryons and leptons ({\em Bare}), as well as a charge neutral, beta-equilibrated neutron star matter, calculated in the different QHD parameter sets. }
		\label{fig:gbsym}
\end{figure}

We continue by investigating the ferromagnetic phase in neutron star matter. As mentioned in section \ref{sec:nom}, we consider neutron star matter to be a charge neutral, beta-equilibrated mix of baryons and leptons, in particular neutrons, protons, electrons and muons.\\
\\
For neutron star matter we also adjusted the strength of the general baryon dipole coupling $g_b$ to the point where the energy density of magnetised matter was less or equal to that of unmagnetised matter. The values of $g_n$ and $g_p$ was adjusted to be $x$ times $g_p^{(0)}$ and $g_n^{(0)}$ according to (\ref{gnAdj}) and (\ref{gpAdj}) respectively. The results are plotted in figure \ref{fig:gbsym}.
\\
\\
When comparing the phase boundaries for neutron star matter to the ones of neutron matter, we note that in general the phase boundary is found at lower values of $g_b$. Adding any particles to the system, as well as the presence of the magnetic field, increases the energy density.  
This means that the ferromagnetic phase boundary, where the increase in the energy density brought about by the presence of the magnetic field is offset by the decrease in the energy contribution of the now magnetised baryons, is reached at lower values of $g_b$ than for neutron matter. \\
\\
Before we discuss the individual behaviour of the QHD parameter sets it is helpful to remind ourselves that the strength of the proton and neutron dipole moments do not differ substantially, but that the neutron, as well as the lepton, dipole moments have the opposite sign to the proton dipole moment. Thus the magnetisation of neutron star matter (\ref{magNSmag}) will be non-zero only if the contribution of one of the baryons dominates the other. As discussed in \cite{diener}, and references therein, neutron star matter is preferably neutron rich (although in section \ref{sec:magmat} we shall see that for magnetised matter the fractions will differ). Thus the source of the ferromagnetic field, if present, will primarily be the magnetisation of neutrons.\\
\\
From figure \ref{fig:gbsym} we notice that excluding the mesons from our model lowers the phase boundary. If we ignore the mesons, we remove the interaction between the baryons and are effectively left with a gas of non-interacting particles.  Since inverse beta-decay (\ref{invbetadec}) is the only way to lower the energy of such a non-interacting system, the proton abundance (plotted in the bottom left panel of figure \ref{fig:gbsym}) is significantly lower than for the interacting systems. 
Therefore, since the neutron fraction is higher, the phase boundary is reached at much lower values of $g_b$ than for a non-interacting system. Not surprisingly 
we observe that the proton fraction at the phase boundary increases if the interactions between the baryons are included, corresponding to the increase in $g_b$.
\begin{figure}
	\centering
		\includegraphics[width=1.0\textwidth]{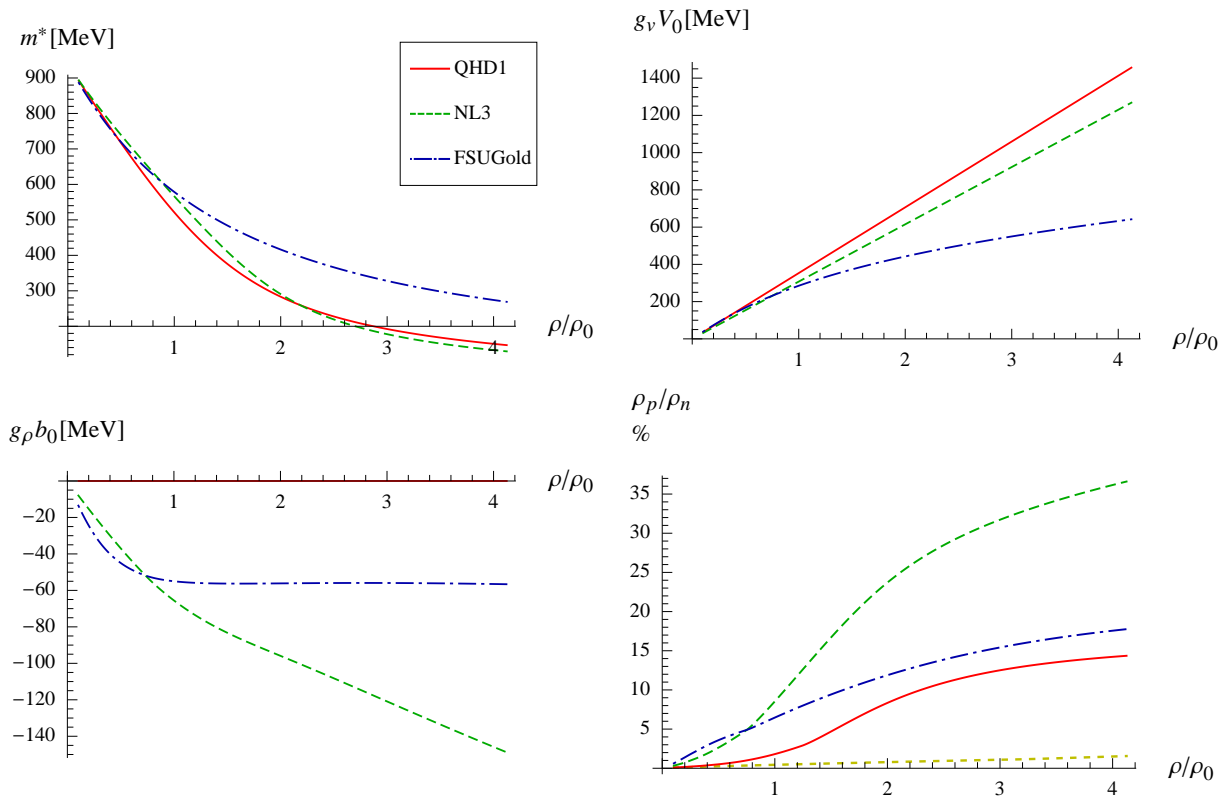}		
	\caption[Various properties of charge neutral, beta-equilibrated neutron star matter at the ferromagnetic phase boundary. ]{Various properties of charge neutral, beta-equilibrated neutron star matter at the ferromagnetic phase boundary. Clockwise from top left we have the reduced mass, the expectation values of the omega meson field, the proton density as a fraction of the neutron density and finally the expectation values of the rho meson field.  }
	\label{fig:gbanal}
\end{figure}\\
\\
For the QHD1 parameter set the behaviour of the proton fraction correlates with that of the QHD1 reduced mass (shown top left in figure \ref{fig:gbanal}). Up to a difference of $g_b B$ for magnetised matter, $m^*$ is the lower bound of the baryon single particle energies. At low values of the baryon density, $m^*$ decreases rapidly as the density increases while the proton fraction increases marginally. However, as the change in $m^*$ tapers off, the proton fraction rises since the lower bound on the baryon energies becomes more or less constant. Therefore higher and higher energy levels have to be filled, the burden of which is shared between the baryons, thereby increasing the proton fraction. This is the reason we do not see the rise of the QHD1 neutron matter phase boundary as $m^*$ tapers off in the neutron star matter version. \\
\\
For the NL3 and FSUGold parameter sets the rho meson also comes into play, but the behaviour of the expectation value of the rho meson field $g_\rho b_0$, at the phase boundary, is distinctly different for the two parameter sets. We show graphs of these in the bottom left panel in figure \ref{fig:gbanal}. From the equation of motion of $g_\rho b_0$ (\ref{FSUrhoEQMB}) we note that for NL3 $g_\rho b_0$ is directly coupled to the isospin density $\rho_p-\rho_n$. Thus, despite the increase in the NL3 proton fraction, the difference between the proton and neutron densities must also increase to yield larger values of $g_\rho b_0$. On the other hand, the FSUGold isospin density tapers off as a function of density. The difference is due to the additional couplings of the FSUGold parameter set: from the equation of motion of $g_\rho b_0$ (\ref{FSUrhoEQMB}) we see that for FSUGold $g_\rho b_0$ also couples to $g_v V_0$.\\
\\
Not surprisingly the expectation values of the omega meson field $g_v V_0$ at the boundary also differs for these two parameter sets, as we show in the top right panel of figure \ref{fig:gbanal}. From (\ref{FSUomegaEQMB}) we know that $g_v V_0$ goes like the baryon density and thus if both $\rho_p/\rho_n$ and $g_\rho b_0$ increases, $g_v V_0$ must also increase. However, for the FSUGold parameter sets the behaviour of both $g_v V_0$ and $g_\rho b_0$ is damped due to the inclusion of the $\Lambda_v$ coupling term between $g_v V_0$ and $g_\rho b_0$ in $\cal L$.\\
\\
Compared to the QHD1 phase boundary the NL3 and FSUGold phase boundaries are reached at even smaller values of $g_b$ due to the inclusion of the rho mesons. At low densities the values of $m^*$, $g_\rho b_0$ and $g_v V_0$ are very similar for both NL3 and FSUGold and correspondingly the phase boundaries are similar. 
However, at around saturation density $m^*$, $g_\rho b_0$ and $g_v V_0$ of these two parameter sets starts to differ. This behaviour is also reflected in the phase boundary, since at this point the NL3 proton fraction increases substantially. 
Accordingly, the total magnetisation of the NL3 system is suppressed, since the magnetisation of the neutrons is in part cancelled by the proton magnetisation. The nett result is that for the NL3 parameter set the phase boundary occurs at 
higher values of $g_b$ than for FSUGold parameter set. \\
\\
Once again the general trend of the parameter sets emerges: as we noted in chapter \ref{chap:resnuc} the FSUGold parameter set has the softest equation of state of the QHD parameter sets used. This entails that it can accommodate a larger number of fermions (protons, neutrons and leptons) than the other parameter sets. Accordingly, the system will magnetise at the smallest values of $g_b$ for FSUGold and at the highest values for QHD1, since the energy gain in magnetising the system benefits the densest system the most.\\
\\
We continue investigating the ferromagnetic phase of neutron star matter by considering the behaviour of the system as the baryon density increases. 
\section{Magnetised neutron star matter}\label{sec:magmat}
\begin{figure}
	\centering
		\includegraphics[height=7.5cm]{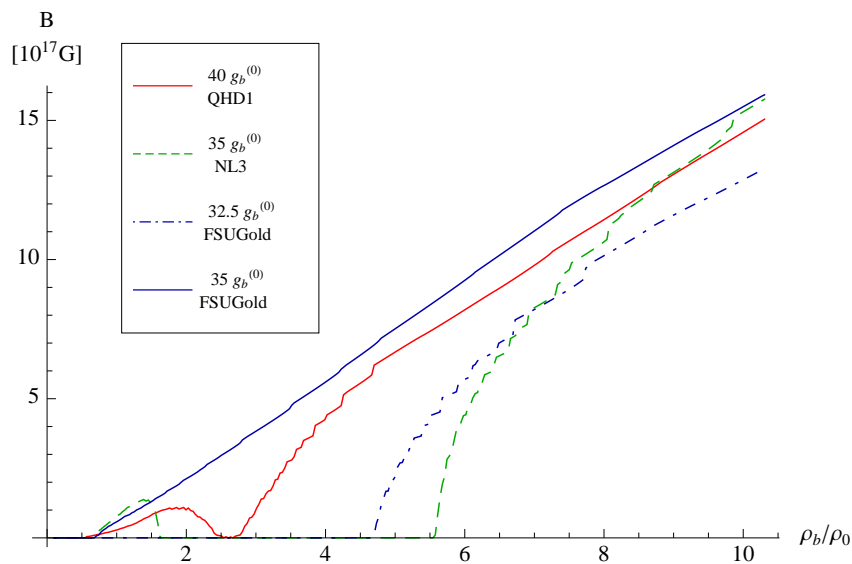}
	\caption[The ferromagnetic field as a function of baryon density and for fixed values of $g_b$.]{The ferromagnetic field as a function of baryon density and for fixed values of $g_b$.}
	\label{fig:Bs}
\end{figure}

To investigate the behaviour of the system in the ferromagnetic phase we choose fixed values of $g_n$ and $g_p$ in such a way that the magnetic dipole moments of protons and neutrons are increased by the same factor. We do this since the density dependence of the baryon magnetic dipole moments are unknown.\\ 
\\
Our choice of $g_b$ differs for the various parameter sets and are made to ensure that, 
as the densities increases, the phase boundary will be crossed for each parameter set. In figure \ref{fig:Bs} we show the ferromagnetic field (\ref{bpfull}) as a function of density for the various QHD parameter sets. In the figure we plot the ferromagnetic field for QHD1 with $g_b=40g_b^{(0)}$, NL3 with $g_b=35g_b^{(0)}$ and FSUGold with $g_b=32.5g_b^{(0)}$ as well as $g_b=35g_b^{(0)}$.
\\
\\
Taking $g_b$ to be constant is the simplest approach, but it has consequences: referring back to the phase diagram 
in figure \ref{fig:gbsym} we note that for some values of $g_b$ the QHD1 and NL3 phase boundaries can be crossed on more than one occasion\footnote{It is not entirely clear from figure \ref{fig:gbsym} for QHD1, since we only show the phase diagram up to $4\rho_0$. We do not show the rest of the phase diagram since the behaviour of the phase boundary does not differ from the established trends as the density increase. We did calculate the phase boundary to higher densities, but to get it to the point where the numerical errors are not visible in the graph requires a lot of computer time, which we did not have available.}. As was emphasised in the previous section, this behaviour of the phase boundaries is the direct result of the different meson coupling strengths and thus, as far as we can deduce, dependent on the choice of the parameter set. Although we would not rule out the possibility, it seems highly unlikely that the system would become magnetised, then switch off the magnetic field and then become magnetised again at a later stage. \\
\\
Let us now compare the graphs of the two FSUGold ferromagnetic fields in figure \ref{fig:Bs}. In figure \ref{fig:gbsym} a line representing $g_b =32.5 g_b^{(0)}$ stays relatively close to the phase boundary, but for $g_b =35 g_b^{(0)}$ it crosses the phase boundary at a point where it moves away from the boundary as the density increases. The $g_b =35 g_b^{(0)}$ graph exhibits almost linear growth with density, while the $g_b =32.5 g_b^{(0)}$ starts very steeply and then tapers off as the density increases. We consider the choice of $g_b =32.5 g_b^{(0)}$ to represent matter close to phase boundary, while $g_b =35 g_b^{(0)}$ describes matter well into the ferromagnetic phase. \\
\\
Considering all the graphs in figure \ref{fig:Bs} we deduce that the one representing the most feasible combination of $g_b$ and parameter set is probably $g_b =32.5 g_b^{(0)}$ FSUGold, since the magnetic field only switches on at a density of $5\rho_0$. However, we emphasise that this is only an approximation to the physical system and we are dealing with all variables in the simplest manner. That being said, the most interesting phenomenon is the discontinuous behaviour of the ferromagnetic field. It is exhibited by all parameter sets, while the size and frequency of the steps seem to decrease as $g_b$ gets larger. We proceed to investigate this behaviour using FSUGold with $g_b = 32.5 g_b^{(0)}$, but the discussion applies to all parameter sets and values of $g_b$.
\subsection{Ferromagnetic field}
\begin{figure}[thb]
	\centering
		\includegraphics[width=1.0\textwidth]{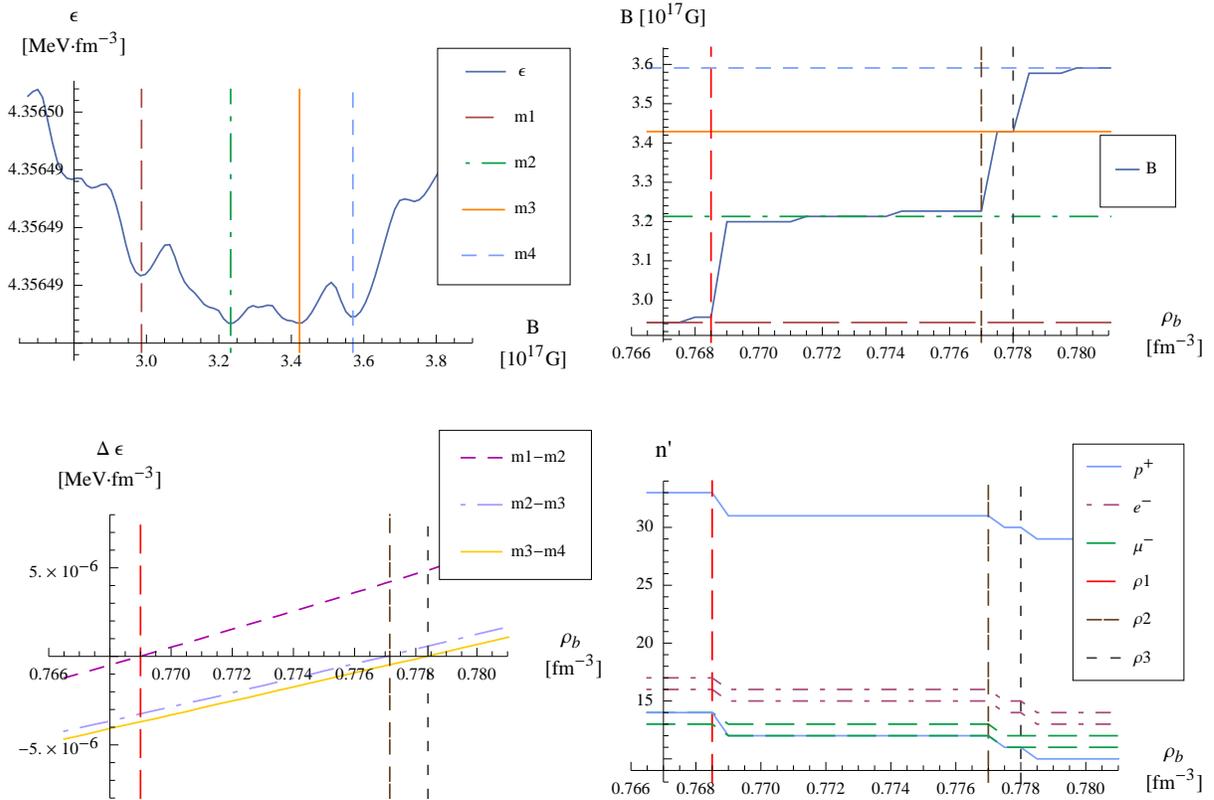}
	\caption[Various plots illustrating the behaviour of the ferromagnetic field, using $g_b =32.5 g_b^{(0)}$ in FSUGold.]{Various plots illustrating the behaviour of the ferromagnetic field, using $g_b =32.5 g_b^{(0)}$ in FSUGold. Clockwise, from top left, we have the energy density as function of magnetic field, then we zoom in on the discontinuities in the ferromagnetic field, next we show the charged particles Landau configurations and finally the energy density difference of the minima shown in the top left panel as the density increases. Please note that the small discontinuities in $B$ in the sections marked by horizontal lines in panel top right are due to numerical accuracy used to generate the plots\footnotemark. We emphasise the discontinuities in $n'$ (bottom right) and $B$ (top right) are sharp with no intermediate points: these graphs are only plotted as a solid lines to make them more visible. The guidelines apply to all plots. }
	\label{fig:moe}
\end{figure}
Our discussion will primarily refer to figure \ref{fig:moe}. To generate these plots we zoomed in on a section of the FSUGold ferromagnetic field with $g_b = 32.5 g_b^{(0)}$ graph in figure \ref{fig:Bs} and calculated the energy density surface as a function of magnetic field and baryon density.\\
\\
Since the ferromagnetic field is assigned to the values of the magnetic field, $B$, that minimises the energy density (\ref{epsLL2}), $\epsilon$, at a specific baryon density, $\rho_b$, it will follow the minimum in the energy density surface, $\epsilon(B,\rho_b)$. The top row in figure \ref{fig:moe} contains plots representing $\epsilon(B,\rho_b)$: top left we have a cut across $\epsilon(B,\rho_b)$ for a fixed density and top right is a projection of the minimum of $\epsilon(B,\rho_b)$ onto the $B$-$\rho_b$ plane (which is also plotted in figure \ref{fig:Bs}). From the cut across $\epsilon(B,\rho_b)$ we see that $\epsilon$ has various local minima, which corresponds to different values of $B$. We denote these values of $B$ as $m1,m2,m3$ and $m4$ using vertical guidelines in this plot. The vertical guidelines in the top left panel of figure \ref{fig:moe} becomes horizontal guidelines in the projection of $\epsilon(B,\rho_b)$ onto the $B$-$\rho_b$ plane (top right panel) since in this case we have $B$ on the $y$-axis. The vertical guidelines in this plot 
refer to the values of $\rho_b$ at which the discontinuities in $B$ appear. These vertical guidelines, $\rho1,\rho2$ and $\rho3$, correspond to those in the panels on the bottom row: bottom right we have the Landau level occupation of the protons and leptons as a function of $\rho_b$, while bottom left we have the difference in energy density of the various minima as a function of $\rho_b$.\\
\\
Although not shown in figure \ref{fig:moe}, we can report that all structure in $\epsilon$, shown in the top left panel, relate to either the Landau configuration of the charged particles or the behaviour of the different particles' Fermi energies. In particular, the global minima in $\epsilon$ correspond to a specific configuration of the various Landau levels and as the density increases the system jumps between these similar configurations as the different local minima becomes the global minimum. This means that the energy density increases smoothly but, as the system jumps between different Landau configurations, the magnetic field changes abruptly and significantly.  From figure \ref{fig:moe} this behaviour is clear: 
\footnotetext{To generate these graphs we used a grid to calculate $\epsilon(B,\rho_b)$ and the small discontinuities in $B$ in the panel top right reflects the step size of the grid. The larger discontinuities in $B$ are however not the results of numerical inaccuracies or choices. }
the densities at which the system switches smoothly from one energy density minima to the next (shown bottom left panel) the system also switches to different occupations of the Landau levels (bottom right). These densities corresponds exactly to the densities at which the discontinuities in $B$ appear.
\subsection{Particle densities}
Next we consider the overall behaviour of the ferromagnetised system. In particular we investigate the contribution of the different particle densities to the total density, which we will compare to the unmagnetised trends. 
\begin{figure}
	\centering
		\includegraphics[width=1.0\textwidth]{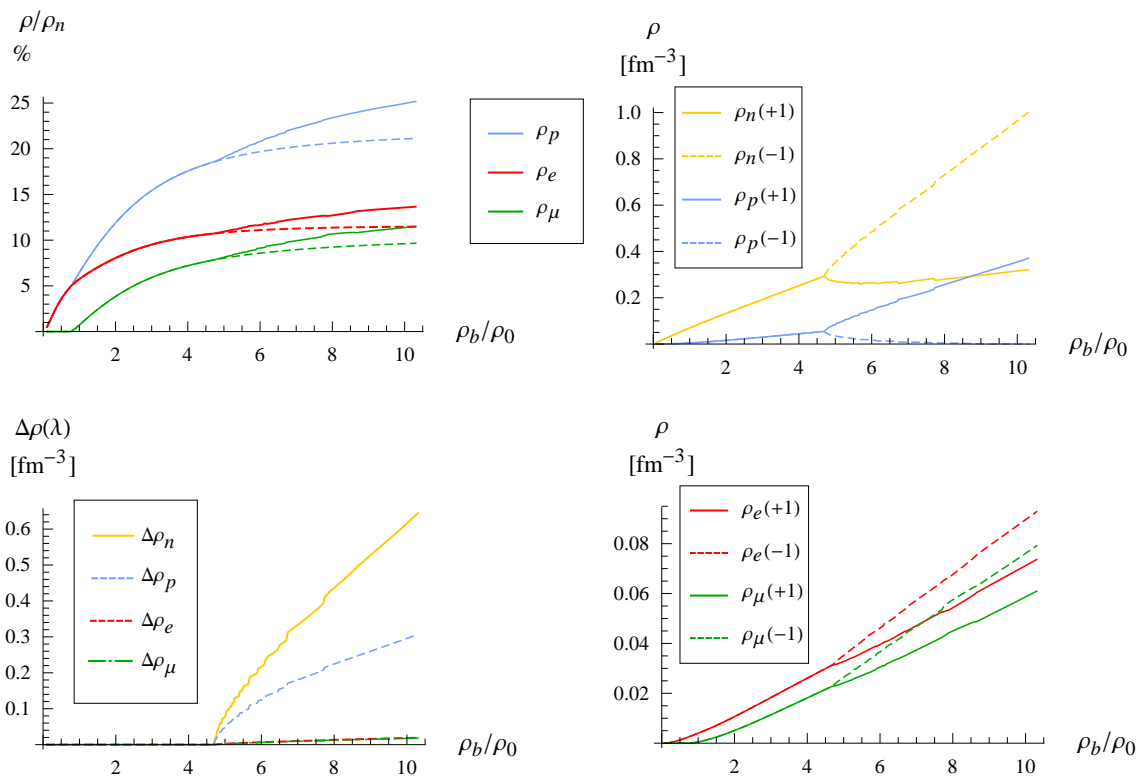}
	\caption[Different particle fractions for ferromagnetised neutron star matter in FSUGold parameter set with $g_b = 32.5 g_b^{(0)}$. ]{Different particle fractions for ferromagnetised neutron star matter in FSUGold parameter set with $g_b = 32.5 g_b^{(0)}$. Clockwise, from top left, we show the particle fractions of ferromagnetised (solid lines) and unmagnetised (dotted lines) neutron star matter. In the next two plots we have the different particle species' densities, while in the last plot we show difference in the respective particle species' densities, all as a function of total baryon density.}
	\label{fig:fbanal}
\end{figure}\\
\\
In figure \ref{fig:fbanal} we show these different contributions. Top left we have the proton, electron and muon densities as fractions of the neutron particle density for FSUGold with $g_b = 32.5 g_b^{(0)}$. The dotted lines in this plot refer to the unmagnetised values. We see that the moment the phase boundary is crossed all the charged particle fractions are relatively increased. We deduce that the advantage in magnetising the system lays therein 
that the degenerate Landau levels can accommodate particles at lower energies than the unmagnetised energy levels. Hence we conclude that the system lowers its energy (density) in the ferromagnetised state through the beta-decay of neutrons. Consequently ferromagnetised charge neutral beta-equilibrated matter tends to be more symmetric. \\
\\
The plots on the right of figure \ref{fig:fbanal} show the origin of the magnetic field: there the densities occupied by the baryon (top) and lepton (bottom) species are plotted. In all these plots the dotted graphs refer to the $\lambda=1$ species, while the solid lines are the $\lambda =-1$ species. For the neutrons and the leptons the $\lambda=1$ species have the lower densities, while for protons it is the other way around, since the proton dipole moment has the opposite sign to the neutron and lepton dipole moments. We have already noted that the neutrons are the source of the magnetic field and here we have it again: the difference between the neutron species is the largest for all the particles, as we can see in the panel on bottom left of figure \ref{fig:fbanal}. This means that neutrons have the most unpaired dipole moments and make the biggest contribution to the total magnetic dipole moment of the system. Although the sum of all the lepton densities equals the total proton density, the leptons contributes little to the ferromagnetic field, as the difference between the lepton species' densities is barely visible in figure \ref{fig:fbanal}. This is because the lepton dipole moments are 
taken at their observed values under normal densities: since they are fundamental particles, we do not assume their dipole moments to change.\\
\\
From the plots on the right we deduce that as the density increases particles are transferred between species, by flipping their magnetic dipole moments: for neutrons the density of $\lambda=1$ stays almost constant while the density of $\lambda=-1$ increases steadily. At the densities corresponding to the discontinuities in the ferromagnetic field we also see that the system rapidly undergoes beta-decay as $\lambda=1$ neutrons are converted into protons and leptons, which is clearly seen in the discontinuities in the $\Delta\rho(\lambda)$ graphs in figure \ref{fig:fbanal}.
\section{Magnetars}
We will now employ our investigation of neutron star matter to study highly magnetised neutron stars, also known as magnetars. \\
\\
Since the magnitude of the ferromagnetic field is orders of magnitude larger than the inferred pulsar magnetic field we are doubtful if the ferromagnetic phase in the pulsar interior could be the source of the pulsar magnetic field. Note the inferred magnetic fields of magnetars are about two orders of magnitude smaller than the ferromagnetic field. However, the magnetar magnetic field is inferred to be the surface field and thus the magnetic field in the interior of the star could be much higher. 
\subsection{Equation of state and mass-radius relationship}
\begin{figure}[tbt]
	\centering
		\includegraphics[width=1.0\textwidth]{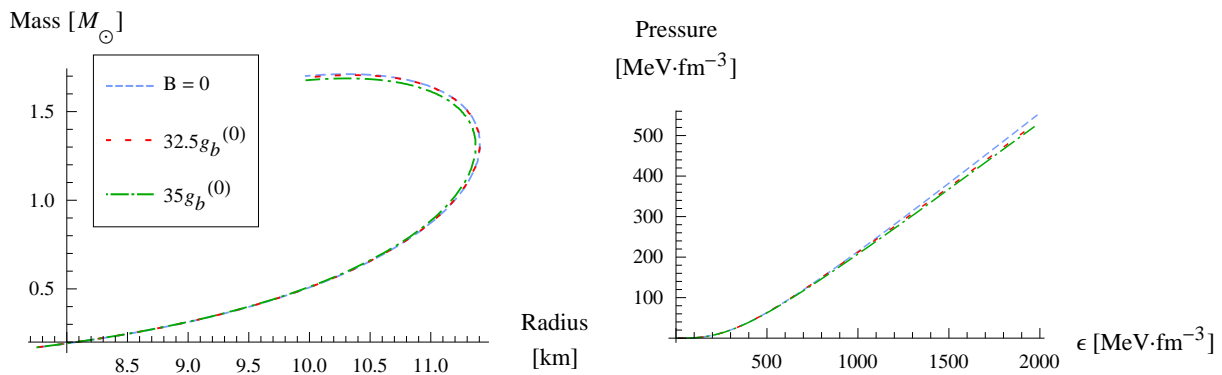}
	\caption[Equation of state and solutions of the TOV equation of ferromagnetised FSUGold neutron star matter.]{Series of TOV equation solutions, this time for neutron star matter.
	}
	\label{fig:tovf}
\end{figure}

We use the Tolman-Oppenheimer-Volkov (TOV) equation to calculate the mass-radius relationship of a neutron star. In natural units the TOV equation is \cite{csg}
\begin{eqnarray}
P'(r) = \frac{dP}{dr} = -\frac{{\cal G}\left[\epsilon(r)+P(r)\right]
\left[M(r)+4\pi r^{3}P(r)\right]}
{r\left[1-2\,{\cal G}M(r)\right]}\label{TOV}
\end{eqnarray}
with
\begin{eqnarray}
	M'(r)=\frac{dM(r)}{dr} = 4\pi \epsilon(r)r^{2}\label{TOVM},
\end{eqnarray}
and where
\begin{itemize}
	\item ${\cal G}$ is the gravitational constant,
	\item $M(r)$ is the enclosed mass of the star,
	\item $\epsilon(r)$ is the energy density, and
	\item $P(r)$ is the internal pressure of the star.
\end{itemize}
 This implies that we assume a neutron star to be a static, spherical symmetric relativistic fluid in hydrostatic equilibrium. Hence we do not consider rotational effects on the star's interior and metric.\\
\\
Since (\ref{TOV}) and (\ref{TOVM}) are two coupled differential equations we used the Runge-Kutta method to approximate the solution (see \cite{diener} and references therein for more details). This entails that we choose a central density $\rho_c$ for the star. This density correlates to the pressure at the centre of the star at which point the distance from the centre of the star ($r$), as well as the mass ($M$) enclosed by $r$, is zero. The Runge-Kutta method calculates the change in the pressure, $P(\Delta r)$, and the enclosed mass, $M(\Delta r)$, for a step from $r$ to $r+\Delta r$. However, in order to calculate $M(\Delta r)$ from (\ref{TOVM}), we need to know the equation of state, $\epsilon(P(r))$, of the matter contained in the specific step $\Delta r$ we are dealing with. For each step of $\Delta r$ we assume that the matter contained within the step is infinite nuclear matter (since $\Delta r$ is much larger than the scale of the nuclear interactions). Thus the equation of states derived in the previous chapters can be applied here. The edge of the star is defined by a chosen pressure\footnote{The pressure at the edge of the star can be zero, but in general chosen to be non-zero corresponding to the density of iron or some other heavy metal.}, $P_R$, at which point the value of $r$ is defined to be the radius, $R$, of the star and mass enclosed by $R$, $M(R)$, the mass of the star. By solving for a range of $\rho_c$, using the same equation of state, a series of solutions of the TOV equation are produced. For the canonical neutron star, $R$ is of the order of 10 km and $M$ is about 1.5 $M_\odot\,$\footnote{$M_\odot$ is referred to as a {\em solar mass} and equal to the mass of our sun. Its value is given in table \ref{tab:constants}.}\cite{csg}.\\    
\\
In our model we do not consider the neutron star to be a giant atomic nucleus. Rather it is an object consisting of layers (with thickness $\Delta r$) of charge neutral, beta-equilibrated nuclear matter. Since $\Delta r$ is of the order of meters, we can approximate the matter contained in $\Delta r$ as infinite nuclear matter and thus can use the models described in chapter \ref{chap:beta} to calculate the properties of such matter.\\
\\
As we established in chapters \ref{chap:ferroneu} and \ref{chap:beta}, the spherical symmetry of the ground state of magnetised matter is broken due to the magnetic field establishing a definite direction in the ground state. 
It is reported that this leads to a differentiation between the pressure in the longitudinal and parallel directions with regards to the magnetic field, see  \cite{isa} and references therein. Isayev and Yang state that, when the effect of the magnetisation in pure neutron matter is excluded, an externally magnetised pure neutron neutron star would be more compressed in the longitudinal direction for magnetic fields of $\approx 10^{18}$ G \cite{isa}. They also report that as the magnetic field increases to $\approx 10^{19}$ G the longitudinal component to the pressure vanishes, leading to instabilities in this direction. However, in using the TOV equation (\ref{TOV}) to calculate the mass-radius relationship of the star, we ignore these possible anisotropies in the pressure and continue to apply the magnetised equation of state to a static spherical symmetric star. We also assume that for each point in the star the magnetic field points in the same direction. We are aware that this is a very simple and na\"ive approximation and treat it as such.\\
\\
A further simplification is how we treat the magnetic field in the neutron star interior: it comes down to that for each $\Delta r$ layer in our TOV calculations we assume it to be a solenoid with edge current such that the ferromagnetic field within that layer is screened from other layers. However, we will ignore the discrepancies due to the difference in the currents flowing at the boundary of adjacent $\Delta r$ layers (since each of these layers are at a different density and thus have a different value of the ferromagnetic field). Another consequence is that matter at densities that lies below the ferromagnetic phase boundary is considered to be unmagnetised and to exist in a vanishing magnetic field. Thus neutron star matter at the lower densities, as well as matter outside of the star, is unaware of the ferromagnetised core matter at higher densities, which is of course not what we observe in nature. These simplifications are all related to the boundary conditions imposed on (\ref{rmfbeqm}) as was discussed then. More realistically the boundary conditions imposed on (\ref{rmfbeqm}) must take account of more realistic edge currents that will not lead to complete screening of the magnetic field outside the ferromagnetised core of the star. Here we ignore these complications to get a simple qualitative idea of the effect of the ferromagnetic equation of state on the neutron star.\\
\\
\begin{figure}
	\centering
		\includegraphics[width=1.0\textwidth]{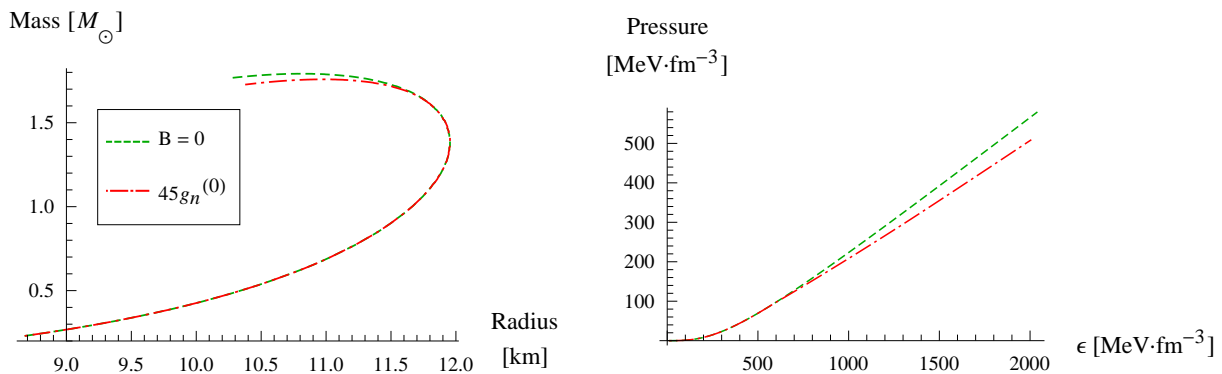}
	\caption[Equation of state and solutions of the TOV equation of ferromagnetised FSUGold neutron matter.]{Series of TOV equation solutions for unmagnetised as well as ferromagnetised FSUGold pure neutron matter, on the left. The corresponding equations of states are shown on the right. Note that we did not include a crust for the star in the equation of state\footnotemark.}
	\label{fig:tovfn}
\end{figure}
In figure \ref{fig:tovfn} we show the mass-radius relationship for a neutron star with a FSUGold pure neutron matter equation of state. For the neutron matter equation of state we chose $g_n =45g_n^{(0)}$. Contrary to what was reported in \cite{astron} we observe that the equation of state is visibly softer for ferromagnetised matter. This is the result of lower energy states being available to ferromagnetised neutrons and consequently the lower values of the energy density. Despite the softer character of the equation of state there is only a slight difference in the mass-radius relationships of unmagnetised and ferromagnetised equations of state.
\footnotetext{Usually the equation of state for neutron-rich nuclei confined to a crystalline lattice, representing the crust of the neutron star, is preferred over the nuclear matter equation of state at low densities for the neutron star equation of state. See for instance \cite{diener} and references therein. Including a crust stiffens the low densities equation of state and the low mass solutions of the TOV equation consequently tend to have increasingly large radii.}\\
\\
In figure \ref{fig:tovf} we show the mass-radius relationship for a neutron star with a FSUGold equation of state. The three graphs corresponds to unmagnetised matter and ferromagnetised matter with $g_b = 32.5g_b^{(0)}$ and $g_b =35g_b^{(0)}$ respectively. We also show the corresponding equation of states in the panel on the right of figure \ref{fig:tovf}. We did not include a crust for the star as was done in \cite{diener} and thus just considered the star to be made of nuclear matter without any crystalline structure at low densities. From both plots we observe that the transition into the ferromagnetic state softens the equation of state, which results in stars with smaller radii and lower maximum masses. However, the softening is a rather small effect, with the difference in the maximum masses less than one tenth of a solar mass.
\\
\\
\begin{figure}[tbh]
	\centering
		\includegraphics[width=1.0\textwidth]{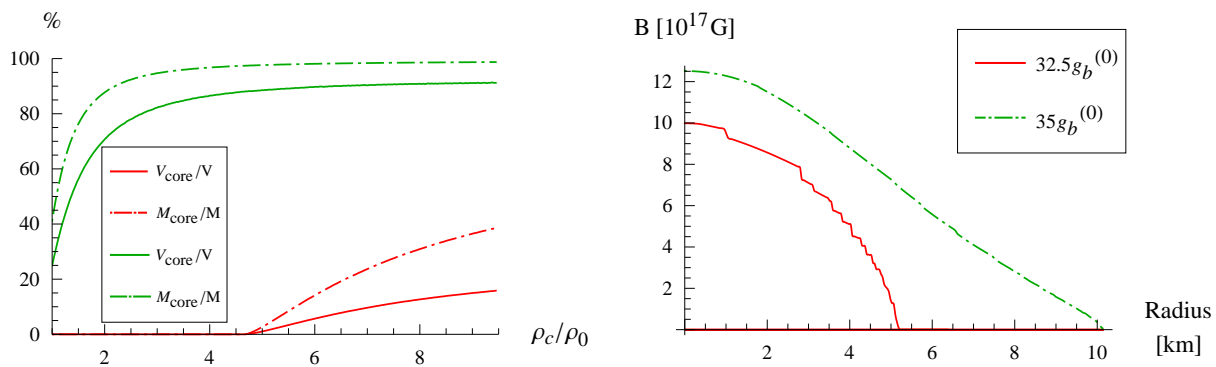}
	\caption[Ratios of the volume and mass of the ferromagnetic core as well as the profile of the magnetic field in the ferromagnetic neutron star interior.]{On the left the ratio of volume of the ferromagnetic to the total volume of the star is plotted (solid lines) for $g_b = 32.5g_b^{(0)}$ and $g_b =35g_b^{(0)}$ in the FSUGold parameter set. The ratio of the mass of the ferromagnetic core to the total mass is the dot-dash graph. The profile of the magnetic field in the maximum mass star for the two values of $g_b$ is shown on the right. The colour of the graphs correspond to the same value of $g_b$ in all the plots.}
	\label{fig:tovfa}
\end{figure}
\begin{figure}[tbh]
	\centering
		\includegraphics[width=1.0\textwidth]{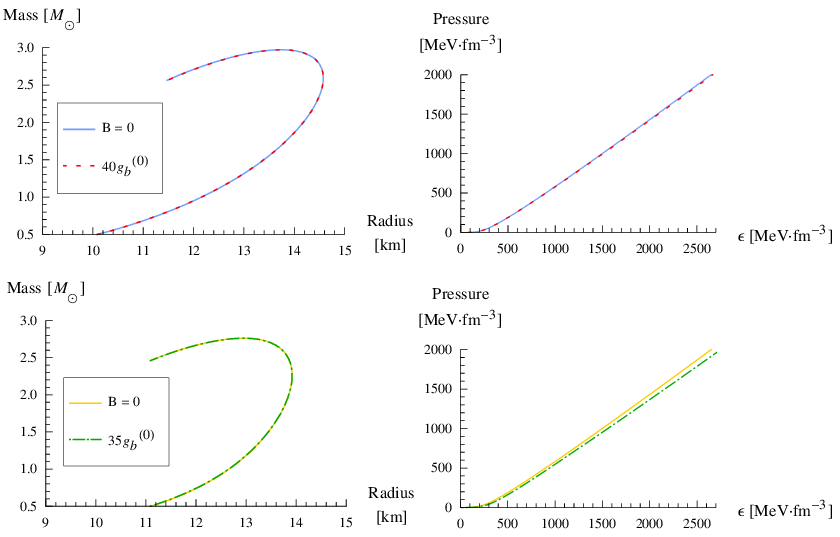}
	\caption[Ferromagnetised and unmagnetised equations of state and solutions of the TOV equation for NL3 and QHD1.]{Ferromagnetised and unmagnetised equations of state and solutions of the TOV equation for NL3 and QHD1. In the top row we show the QHD1 results for $g_b = g_b^{(0)}$ and $g_b = 40g_b^{(0)}$. In the bottom row the NL3 results for $g_b = g_b^{(0)}$ and $g_b = 35g_b^{(0)}$.}
	\label{fig:tovnq}
\end{figure}
For the two ferromagnetised solutions of the TOV equation in figure \ref{fig:tovf} the ferromagnetic phase boundary is crossed at different densities. Since the density in the star is the highest at the centre and lowest at the surface, the ferromagnetic phase, if present, will always form in the core of the star.  In figure \ref{fig:tovfa} we show the ratios of the volume and mass of the ferromagnetic core to the total volume and mass of the star. For $g_b = 32.5g_b^{(0)}$ the phase boundary is only crossed deep in the interior of the star, at about $5\rho_0$. The ferromagnetic core is therefore relatively small. For $g_b = 35g_b^{(0)}$ the phase boundary is crossed almost immediately thus the ferromagnetic phase is present throughout most of the star.\\
\\
In the plot on the right of figure \ref{fig:tovfa} we show the profile of $B$ in the interior of the maximum mass star of each ferromagnetised TOV solution series. 
For the $35g_b^{(0)}$ graph the field is present for the biggest part of the star. For the $32.5g_b^{(0)}$ graph the field switches off when we cross the ferromagnetic phase boundary into the unmagnetised phase.
 Thus we assume the magnetic field to be totally contained in the interior of the star with no influence on the matter 
 outside of the ferromagnetic phase. As mentioned earlier, this is of course an oversimplification since we are effectively ignoring the effect of the magnetic field on the crust of the star, which lies outside the ferromagnetised densities. \\
 \\
However, by comparing the two ferromagnetised FSUGold TOV solution series we would not expect the mass-radius relationship to differ substantially if we were to include a magnetised matter equation of state for matter outside of the ferromagnetic core: for $35g_b^{(0)}$ almost the whole star is magnetised without an significant change in the mass-radius relationship. However, considering the results from chapter \ref{chap:resnuc}, magnetised nuclear matter in the crust of the star might significantly alter other properties of the star.\\
\\
The softening of the equation of state due to the ferromagnetised matter is also observed for the other parameter sets, although the softening is significantly smaller for these generic stiffer equations of state. We show their solutions in figure \ref{fig:tovnq}.
\subsection{Emission}
\begin{figure}
	\centering
		\includegraphics[width=.75\textwidth]{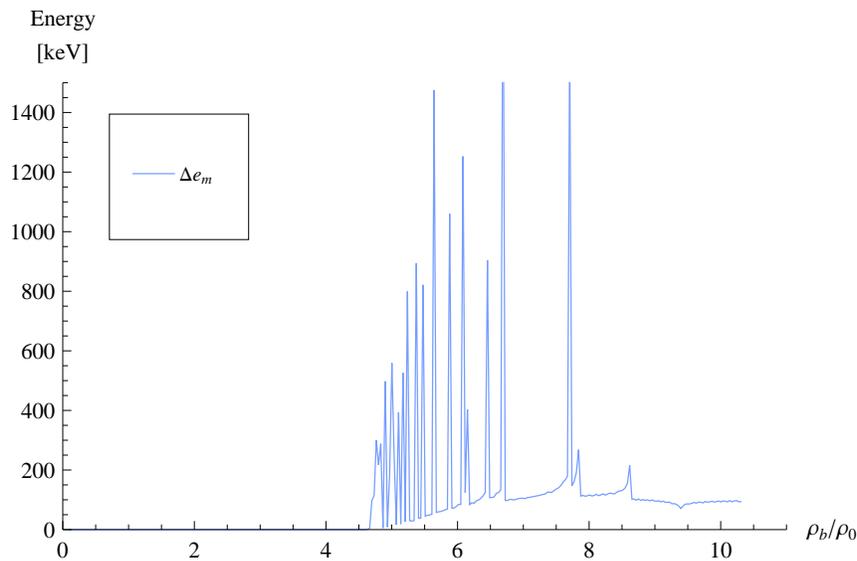}
	\caption[Energy emitted by particles as function of magnetic field.]{Energy emitted by particles as the magnetic field increases.}
	\label{fig:EdB}
\end{figure}

This section is more speculative. We wish to establish a ball park estimate of the type of emission we can expect from a star containing ferromagnetised nuclear matter which is cooling down (and consequently increasing the star's density).\\
\\
As we have seen from figures \ref{fig:tovf} and \ref{fig:tovnq} the equation of state of a ferromagnetised system does not differ much from the unmagnetised one. However, in the ferromagnetised system there is a distinct difference between, what we will call, the matter energy density $\epsilon_m$ and the magnetic energy density $\epsilon_B$. The total energy density of the ferromagnetised system is $\epsilon$, from (\ref{epsLL}), and we define
\begin{align}
	\epsilon=\epsilon_m+\epsilon_B
\end{align}
with
\begin{align}
	\epsilon_B=\frac{1}{2}B^2.
\end{align}
Therefore $\epsilon_B$ is the contribution of the ferromagnetic field $B$ to the energy density. For 
ferromagnetised matter there is some balance between $\epsilon_m$ and $\epsilon_B$, while for unmagnetised matter $\epsilon=\epsilon_m$.\\
\\
In figure \ref{fig:Bs} we showed the different ferromagnetic fields for different parameter sets and values of $g_b$. We observed that the fields do not increase smoothly as a function of density, but have a discontinuous behaviour. At these discontinuities $B$ increases rapidly, which signals a preference of the system to increase $\epsilon_B$ as one or more charged particle Landau levels depopulate. However, in the interim between these discontinuities, $B$ also increases steadily. From figure \ref{fig:fbanal} we can relate this more continuous increase to the system moving particles between particle species: on the right hand panel of figure \ref{fig:fbanal} we see that as $\rho_b$ increases there is a definite preference to increase the density of only one species of individual particle species, namely the one which has the lowest energy states. Thus the nett dipole moment of the system increases, which strengthens the magnetic field.\\
\\
The magnetic field couples to the photons in the system that provides a mechanism to radiate energy 
out of the system. We are interested at what frequencies both the continuous and discontinuous increase in $B$ can radiate energies. As a ball park estimate, we divided $\epsilon_B$ by the total density of particles in the system, which is the sum of the baryon and lepton densities,
\begin{align}
	\rho_b+\rho_l,
\end{align}
to establish the average magnetic energy per particle at a given density\footnote{Of course this energy is not stored in the particles, but in the magnetic field, but since we are interested in the change in $\epsilon_B$ as the density changes, we normalised it to a hypothetical quantity per particle.}. We then calculated the difference in the (ferro)magnetic energy per particle for a small change in density, which we shall call $\Delta \mbox{e}_m$. This corresponds to the energy transferred to the ferromagnetic field as the density changes. $\Delta \mbox{e}_m$ is plotted for FSUGold with $g_b = 32.5g_b^{(0)}$ in figure \ref{fig:EdB}.
\\
\\
Since we are interested in the change in magnetic energy per particle, the derivative should actually have been taken. However, due to the discontinuities in $B(\rho_b)$ this is difficult to calculate and thus we opted for a more simple calculation. 
Since this is a rough estimate of the derivative, it will be an overestimation of the true change in the magnetic energy per particle.\\
\\
From figure \ref{fig:EdB} we see that $\Delta \mbox{e}_m$ is obviously zero when the magnetic field is zero. When the ferromagnetic field switches on, the spikes in $\Delta \mbox{e}_m$ corresponds to the abrupt changes in the magnetic field at low densities, due to the number of occupied Landau levels decreasing rapidly as the magnetic field increases. At these depopulation of Landau levels $\Delta \mbox{e}_m$ has values between 300 and 1000 keV. \\
\\
In the interim, between the spikes, $\Delta \mbox{e}_m$ has significant lower values of about 10 to a couple of 100 keV. These interim periods between spikes corresponds to small, stable (smooth) increases in the magnetic field, shown in figure \ref{fig:Bs}. In this case $\Delta \mbox{e}_m$ stems from the conversion between individual particle species, i.e. flipping of particles' dipole moment.\\
\\
Since X-rays have photon energies of 1 - 100 keV, while $\gamma$-rays have energies between 100 and 1200 keV, based on the observations above, we would not be surprised if a ferromagnetic star radiates X- and $\gamma$-rays. These are consistent with the energies detected from magnetars \cite{chap14}. In terms of luminosities, the change in $\epsilon_B$ corresponding to the spikes in \ref{fig:EdB} are of the order of 0.1 to 1 MeV/fm$^{3}$, which corresponds to a whopping $10^{53}$ erg/m$^{3}$. Thus only a small amount of ferromagnetic material is needed 
to achieve the observed luminosities. \\
\\
Since Woods et al. observed that a giant flare in SGR 1900+14 correspondeds to a reconfiguration of the star's magnetic field \cite{Brecon}, we would like to speculate that the flare might have been caused by the depopulation of various Landau levels in the ferromagnetic interior of the star. Since it has been reported that outbursts in AXPs may be accompanied by glitches \cite{AXPprogress}, it would be interesting to note the behaviour of the star's moment of inertia as the ferromagnetic phase boundary is crossed.
\section{Summary}
We showed the ferromagnetic phase diagram for neutron and neutron star matter as a function of the strength of the dipole coupling 
and baryon density. We correlated the behaviour of the phase boundary to the contributions of the meson field expectation values and different particle fractions.\\
\\
We also investigated the effect of the ferromagnetic phase on the neutron star equation of state and speculated on possible observational consequences of the presence of such a phase.

\chapter{Conclusions and future prospects}\label{chap:con}
In this dissertation we presented our study of ferromagnetism in neutron and neutron star matter. \\
\\
We investigated the ferromagnetic phase in these types of matter by including the coupling
\begin{eqnarray}
	-\frac{g_b}{2}\bar{\psi}_b{ (x)}\sigma^{\mu\nu}F_{\mu\nu}\psi_b{ (x)}
\end{eqnarray}
in the QHD Lagrangian density. We included this coupling only for the baryons, in order 
to incorporate the composite nature of the baryons in the Lagrangian density since otherwise the baryons are considered to be point particles. It couples the magnetic field to the baryon dipole moment and for a non-zero magnetic field it induces a splitting in the single particle baryon energy spectrum. We assume that the magnetised lower energy levels will be preferentially filled. This asymmetric filling of energy levels produces a nett dipole moment in the system: the source of the ferromagnetic field.\\
\\
The ferromagnetic phase boundary was calculated by adjusting $g_b$ till the energy density of the magnetised system, including the contribution from the magnetic field, was found to be at a lower energy than the unmagnetised system. The rationale behind adjusting $g_b$ is that the baryon magnetic dipole moment is, at least in part, due to the baryon's internal charged structure. As the  
density increases we expect the influence of the internal baryon (quark) degrees of freedom in the interaction with the magnetic field to grow, which will effectively increase the strength of baryon dipole moment. 
Of course the density dependence of the baryon dipole moment is unknown, so we kept the values within the range that we think might be reasonable. \\
\\
We also considered the behaviour of the ferromagnetised system. The ferromagnetic field was calculated by minimising the total energy density (at a fixed baryon density) as a function of magnetic field. In the ferromagnetic phase the energy density increases smoothly, but the magnetic field exhibits erratic discontinuities as it increases. We matched these discontinuities to the depopulation of Landau levels, which has a dramatic effect on the system in terms of how the energy is distributed between the individual particle densities and the magnetic field.\\
\\
The equation of state of ferromagnetised matter was derived and employed to solve the TOV equation, calculating the mass and radius relationship of a neutron star. We found that although the ferromagnetic phase does not significantly alter the mass-radius relation of the star its presence in the neutron star interior will generate magnetic field strengths, which are of the expected order of magnitude assumed to be present in the interior of magnetars. We speculate whether the behaviour of the ferromagnetised system could account for the observed behaviour of the SGRs and AXPs and present some evidence to support this notion. It is generally assumed that the observed behaviour of SGRs and AXPs are driven by the decay of their superstrong magnetic field. However, if the ferromagnetic phase is present, this behaviour could be driven by an increase in the magnetic field. \\
\\
We also considered the properties of magnetised symmetric nuclear matter. The presence of the Landau levels distinctly affects the behaviour of the saturated system: as the matter becomes more magnetised the saturation density increases and nuclear matter tend to be more proton-rich. However, the system is less strongly bound as $B$ increases. 
The compressibility of the system exhibits an oscillatory character 
as the Landau level filling configuration changes, giving some clues as to what might be the externally magnetised behaviour of symmetric nuclear matter. This type of matter might be found in the crust of a neutron star, but we did not include this in our equation of state. However, calculating the moment of inertia of a magnetised and ferromagnetised system is high on the future agenda.\\
\\
Finally we presented an almost complete derivation of the magnetised baryon Dirac spinors and eigenvectors as well as the calculation of the magnetic dipole moment in the non-relativistic limit. These are included as addenda. \\
\\
Although we cannot comment on the true density dependence of the baryon magnetic dipole moment, we conclude that if the ferromagnetic phase boundary is reached the effect would be dramatic and spectacular. We believe that the description presented here can contribute to the understanding of the behaviour of highly magnetised matter at extreme densities, as are assumed to be found in the interior of magnetars. We will continue this work by calculating possible changes in the moment of inertia of ferromagnetised stars, since we believe that the depopulation of Landau levels may also have an influence on the rotational behaviour of magnetars. If the moment of inertia changes abruptly while the angular momentum stays conserved then the angular velocity of the star must change. If the moment of inertia of the star displays the same abrupt behaviour as the ferromagnetic field, then the depopulation of Landau levels will most probably be accompanied by a glitch in the star's rotation. Thus correlating flares in magnetars with glitches may provide us with very strong evidence of the presence of a ferromagnetic phase in the star's interior. \\
\\
Note that the dynamics of the Landau levels is relevant regardless of the source of the magnetic field. Therefore, even if it turns out that the origin of the magnetic field is not ferromagnetism, this study of the dynamics of Landau levels is still important and this work provides a basis for doing this.

\appendix
\chapter{Relativistic mean-field approximation}\label{ap:RMF}
Here we will further illustrate the relativistic mean-field approximation. We will focus on how to implement the RMF approximation regarding the calculation of various observables, as well as providing insights into the interpreting the effect of the approximation on the system.
\section{Minimisation of energy density}
In the RMF approximation we essentially only consider the field potentials and assume that there is no kinetic contribution to the fields. Therefore applying the RMF approximation to the Lagrangian density, $\cal L$, rids it of all the (meson) field derivatives.  Thus deriving the equation of motion for any the meson fields using the Euler-Lagrange equation (\ref{EL}) reduces to 
\begin{align}\label{rmfmot}
	\begin{split}
		\partial_\nu\left(\frac{\partial {\cal L}}{\partial (\partial_\nu \phi_\alpha)}\right) - 
		\frac{\partial {\cal L}}{\partial \phi_\alpha} &= 0\\
		&=-\frac{\partial {\cal L}}{\partial \phi_\alpha}.
	\end{split}
\end{align}
Correspondingly, the contributions to energy-momentum tensor (\ref{EMTi}) by these fields reduces to
\begin{align}
	\begin{split}
		 T^{\mu\nu} &=  \frac{\partial\cal{L}}{\partial (\partial_\mu \phi_\alpha)}\,
		\partial^\nu \phi_\alpha - \cal{L}\eta^{\mu\nu}\\
		&=- \cal{L}\eta^{\mu\nu}, 
	\end{split}
\end{align}
and thus the energy density $T^{00}$ is
\begin{align}\label{epsmot}
	\begin{split}
		  T^{00} =\epsilon = - \cal{L}.
	\end{split}
\end{align}
Therefore, (\ref{rmfmot}) reduces to
\begin{align}\label{eqmmot}
	\begin{split}
		-\frac{\partial {\cal L}}{\partial \phi_\alpha}=0=\frac{\partial \epsilon}{\partial \phi_\alpha}
	\end{split}
\end{align}
and the equation of motion of the meson fields are equivalent to the minimisation of the energy density.\\
\\
Of course the baryons still contribute to $T^{\mu\nu}$ through the first term in (\ref{EMTi}), thus (\ref{epsmot}) does not technically hold. However, what we want to do here is to construct an effective energy density, consisting of the potential terms of the meson fields and including the baryon densities to which the meson fields couples. With this effective energy density, (\ref{eqmmot}) will hold in terms of the minimisation of the energy density. Thus minimising this effective energy density will result in equations equal to the equation of motion of fields in the RMF approximation calculated with the Euler-Lagrange equation.
\section{Calculating densities}
Any product of the Dirac wave function, $\psi$, in the RMF approximation with its Dirac adjoint, $\bar{\psi}$, can be calculated by firstly constructing $\psi$ and then calculating the product in question. Here we will illustrate another method to calculate such products. This method taken from \cite{csg} and this particular illustration is heavily based on the one given in \cite{diener}. This method is based on the assumption that the RMF approximation is such that it minimises the energy density of the system (we will illustrate this point in the next section) as well as the assumption that by minimising the single particle energies the energy density will also be a minimum.\\
\\
It states that the expectation value of an operator ($\Gamma$) in the ground state can be given in terms of the expectation value of the single particle state:
\begin{eqnarray}
	\left(\bar{\psi}\Gamma\psi\right)_{{\bm k},s},
\end{eqnarray}
where
\begin{itemize}
	\item ${\bm k}$ denotes the momentum, and
	\item $s$ the spin of the single-particle state.
\end{itemize}
The expectation value in the many-nucleon system is then \cite{csg}
\begin{eqnarray}
	\left\langle \bar{\psi}\Gamma\psi\right\rangle  = 
	\sum_s\int\frac{d{\bm k}}{(2\pi)^3}\left(\bar{\psi}\Gamma\psi\right)_{{\bm k},s}\,
	\Theta\big[\,\mu-e({\bm k},s)\big]\,\label{NKGexpect},
\end{eqnarray}
where
\begin{itemize}
	\item $e({\bm k},s)$ 
	are the positive single-particle energies, 
	\item $\mu$ is the chemical potential/Fermi energy, and
	\item $\Theta[\,\mu-e({\bm k},s)]$ is a step function with 
	\begin{eqnarray}
		\Theta[\,\mu-e({\bm k},s)] = \left\{
		\begin{array}{cc}
			1&\mbox{if}\  e({\bm k},s)
			\leq \mu\\
			0&\mbox{if}\ e({\bm k},s)> \mu
		\end{array}\right..
	\end{eqnarray}
\end{itemize}
All the possibilities for $\Gamma$ will in general appear in the Dirac Hamiltonian \cite{csg}. The Dirac Hamiltonian $H_D$ can be constructed from the equation of motion of $\psi$ (Dirac equation) as
\begin{eqnarray}
	H_D\psi =  i \frac{\partial}{\partial t}\psi.
\end{eqnarray}
Taking the single-particle expectation value of $H_D$
\begin{eqnarray}
	\left(\psi^\dagger H_D\psi\right)_{{\bm k},s} &=& e({\bm k})\left({\psi}^\dagger\psi\right)_{{\bm k},s}\nonumber\\
	&=& e({\bm k})\label{singexp},
\end{eqnarray}
since 
\begin{eqnarray}
	\left({\psi}^\dagger\psi\right)_{{\bm k},s} =1\,\forall\, {\bm k},s
\end{eqnarray}
as the spinors are normalised and $\psi$ is of the form of (\ref{Kexpand}). Taking the derivative of the left-hand side of equation (\ref{singexp}) with respect to any variable ($\zeta$) yields
\begin{eqnarray}
	\frac{\partial}{\partial\zeta}\,\left(\psi^\dagger H_D\psi\right)_{{\bm k},s} &=& 
	\left(\psi^\dagger\,\frac{\partial H_D}{\partial\zeta}\,\psi\right)_{{\bm k},s}
	+ e({\bm k})\,\frac{\partial}{\partial\zeta}\,\left(\psi^\dagger\,\psi\right)_{{\bm k},s}\nonumber\\
	& = & \left(\psi^\dagger\,\frac{\partial H_D}{\partial\zeta}\,\psi\right)_{{\bm k},s}\label{derivHD},
\end{eqnarray}
since $\psi$ is an eigenfunction of $H_D$ and consequently the second term on the right vanishes \cite{csg}. Thus equation (\ref{derivHD}), after considering (\ref{singexp}), yields
\begin{eqnarray}
	\frac{\partial}{\partial\zeta}\,\left(\psi^\dagger H_D\psi\right)_{{\bm k},s} 
	&=&\frac{\partial}{\partial\zeta}\,e({\bm k})\nonumber\\
	 & = & \left(\psi^\dagger\,\frac{\partial H_D}{\partial\zeta}\,\psi\right)_{{\bm k},s}.\label{singlepatexp}
\end{eqnarray}
Any expectation value can therefore be obtained by using the general expression (\ref{NKGexpect}): taking the derivative of $H_D$ for the appropriate choice of $\zeta$ to yield $\Gamma$ and obtaining the expression for the single particle expectation value of $\Gamma$ from equation (\ref{singlepatexp}).
\section{Symmetry energy}\label{ap:a4}
Here we will illustrate the derivation of (\ref{a4mag}). In general $a_4$ is given by (\ref{a4})
\begin{eqnarray}
	a_{4} = \frac{1}{2}\left(\frac{\partial^{2}}{\partial t^{2}}\frac{\epsilon}{\rho_b}\right)_{t=0}\mbox{ with }\left(t\equiv\frac{\rho_{n} -
		 \rho_{p}}{\rho_b}\right).\label{apa4}
\end{eqnarray}
When the $t$-dependence of the baryon densities can be expressed in terms of the magnitude of the Fermi momentum, $k^F$, (\ref{apa4}) holds completely. Then (\ref{a4unmag}) can be derived by writing the energy density and all the baryon densities in terms of $k^F$ and simply taking the second derivative. If however, you choose to express $a_4$ in terms of the Fermi energies of the baryons then, instead of the partial derivative, the total derivative needs to be taken. Since the $t$ dependence of the baryon densities implies that the Fermi momentum is also $t$ dependent, but this is not obvious from the expression. \\
\\
To derive (\ref{a4mag}), we will make use of the fact that since
$\rho_b=\rho_p+\rho_n$ is constant as $t$ varies,
	from (\ref{apa4}) we have that 
	\begin{eqnarray}
		\rho_p-\rho_n=-t \rho_b,
	\end{eqnarray}
therefore
		\begin{subequations}
			\begin{eqnarray}
				\rho_n &=& \frac{t+1}{2}\rho_b,\mbox{ and}\\
				\rho_p &=& \frac{1-t}{2}\rho_b,
			\end{eqnarray}
		\end{subequations}
and consequently
		\begin{subequations}
		\begin{eqnarray}
				\frac{d}{dt}\rho_n &=& \frac{1}{2}\rho_b,\mbox{ and}\\
				\frac{d}{dt}\rho_p &=& -\frac{1}{2}\rho_b.
			\end{eqnarray}
		\end{subequations}
Furthermore defining $\chi$ as
	\begin{eqnarray}
		\chi = \frac{m_\rho^2}{2\left(m_\rho^2+2g_\rho^2\Lambda_v(g_vV_0)^2\right)},
	\end{eqnarray}
we can express $\frac{d}{dt}b_0$, using (\ref{FSUrhoEQM}), as
	\begin{eqnarray}
		\frac{d}{dt}b_0= -\frac{g_\rho}{m_\rho^2}\chi\rho_b.
	\end{eqnarray}
We also define $\mu'_n$ and $\mu'_p$ as the nucleon contribution to the baryon Fermi energies, where
	\begin{subequations}
		\begin{eqnarray}
			\mu'_p &=& \mu_p - g_v V_0 - \frac{1}{2}g_\rho\,b_0,\mbox{ and}\\
			\mu'_n &=& \mu_n - g_v V_0 + \frac{1}{2}g_\rho\,b_0.
		\end{eqnarray}
	\end{subequations}
Considering the energy density of symmetric nuclear matter consists of contributions from the baryons ($\epsilon_b$) and mesons ($\epsilon_{meson}$), where $\epsilon_b$ refers to the contribution of (\ref{epsFSUexp}) to $\epsilon$ and $\epsilon_{meson}$ to the rest, then
\begin{eqnarray}
	\frac{d^2}{dt^2}\epsilon=\frac{d^2}{dt^2}\epsilon_b+\frac{d^2}{dt^2}\epsilon_{meson}.
\end{eqnarray}
If we again use the label $\alpha$ to distinguish the baryons, then we have from (\ref{rhoB})
\begin{eqnarray}
	\frac{d\rho_\alpha}{dt}=\frac{\partial\rho_\alpha}{\partial \mu_\alpha}\frac{d\mu_\alpha}{dt}=
	\sum_s\int\frac{d{\bm k}}{(2\pi)^3}\,\delta\!\left(\mu_\alpha-e_\alpha({\bm k},s)\right)
\end{eqnarray}
and correspondingly that, from (\ref{epsFSUexp}) as well as keeping (\ref{enerconven}) in mind,
	\begin{align}
		\begin{split}
			\frac{d}{dt}\sum_s\int\frac{d{\bm k}}{(2\pi)^3}\,E_\alpha({\bm k},s)\,\Theta[\,\mu_\alpha-e_\alpha({\bm k},s)]
			&=\sum_s\int\frac{d{\bm k}}{(2\pi)^3}\,E_\alpha({\bm k},s)\delta\!\left(\mu_\alpha-e_\alpha({\bm k},s)\right)\\
			&=\mu'_\alpha\frac{\partial\rho_\alpha}{\partial \mu_\alpha}\frac{d\mu_\alpha}{dt}.
		\end{split}
\end{align}
After some algebra we can than show that
\begin{eqnarray}
	\frac{d^2\epsilon_b}{dt^2}=\frac{1}{2}\left(\frac{d}{dt}\big[\mu'_n-\mu'_p\big]\right)+\frac{g_\rho^2}{m_\rho^2}\chi\rho_b^2,
\end{eqnarray}
as well as
\begin{eqnarray}
	\frac{d^2}{dt^2}\epsilon_{meson}=-\frac{1}{2}\frac{g_\rho}{m_\rho^2}\chi\rho_b^2.
\end{eqnarray}
Thus, substituting back the expression for $\chi$, we have
\begin{eqnarray}\label{a4gn}
	\frac{1}{2}\left(\frac{d^{2}}{dt^{2}}\frac{\epsilon}{\rho_b}\right)_{t=0}= \frac{1}{4}\left(\left.\frac{d \mu'_n}{dt}\right|_{t=0}-\left.\frac{d \mu'_p}{dt}\right|_{t=0}\right)
	+\frac{1}{8}\left(\frac{g_\rho^2\, \rho_b}{m_\rho^2 + 2\Lambda_vg_\rho^2(g_vV_0)^2}\right).
\end{eqnarray}

\chapter{Magnetic dipole moment}\label{ap:dipole}
Here the non-relativistic limit of the modified Dirac equation will be investigated by using the {\sl Foldy-Wouthuysen} transformation. This will be done to identify the various terms in the Dirac equation's contribution to the magnetic dipole moment of the proton and the neutron. This derivation is strongly based on the one given in \cite{gross}.\\
\\
To simplify the calculation the meson contributions will be ignored, resulting in the equation of motions for the protons and neutrons being
	\begin{eqnarray}
			\Big[\gamma^{\mu}\left(i\partial_{\mu}-q_b A_\mu\right)
			-\frac{g_b}{2}\sigma^{\mu\nu}F_{\mu\nu}-m_b\Big]\psi{ (x)} = 0\label{DEap2},
	\end{eqnarray}
where the index $b$ refers to the different particle species.\\
\\
In the non-relativistic limit the mass $m$ is the dominant contribution to the particle's energy. All other energies (potentials) or momenta are assumed to be much smaller than $m$ and thus the positive energy solution has an energy close to $m$. To emphasise the non-mass additions to the energy ($E$) these contributions will be labelled by $T$, where
\begin{eqnarray}
	 T=E-m.
\end{eqnarray}
To investigate the non-relativistic limit the positive energy solution will be expanded and terms of the order $(v/c)^2\sim(p/m)^2\times\mbox{(leading terms)}$ will be considered. It will also be assumed that all potentials ($A^\mu$) are of the same order as the kinetic energy ($\propto m v^2 = p^2/m$, and therefore the order of the leadings terms) so that the expansion will be up to the order $(v/c)^2\sim p^4/m^3$ \cite{gross}.\\
\\
As in \cite{gross} a positive energy solution of the form
\begin{eqnarray}
	\psi(r,t)=\left[\begin{array}{c}\chi(r)\\\eta(r)\end{array}\right]e^{-iEt}
\end{eqnarray}
will be assumed where $E=T+m$. Reverting to the notation of the Dirac matrices, equation (\ref{DEap2}) is rewritten as
\begin{subequations}
	\begin{eqnarray}\label{Hap}
			T\left[\begin{array}{c}\chi\\\eta\end{array}\right]
			&=&\Big[{\bm \alpha}\!\cdot\!\big({\bm p}-q_b{\bm A}\big)+ q_b A^0+
			g_b\beta\left(i\bm \alpha \!\cdot\!\bm E- \bm\Sigma \!\cdot\!\bm B\right)+\beta m -m\Big]\left[\begin{array}{c}\chi\\\eta\end{array}\right]\label{H1}\\
			&=&
			\begin{tabular}{rl}
				$\left[
				\begin{array}{cc}
						q_bA^0-g_b\bm\sigma\!\cdot\!\bm B												  	&   {\bm \sigma}\!\cdot\!\left({\bm p}-q_b{\bm A}+i g_b \bm E\right)\\
					{\bm \sigma}\!\cdot\!\left({\bm p}-q_b{\bm A}-i g_b \bm E\right)		&   	q_bA^0+g_b\bm\sigma\!\cdot\!\bm B-2m		 
				\end{array}
				\right]$
				&
				$\left[\begin{array}{c}\chi\\\eta\end{array}\right]$
		\end{tabular}\\
		&=&H\left[\begin{array}{c}\chi\\\eta\end{array}\right],
	\end{eqnarray}
\end{subequations}
with $\bm E \mbox{ and }\bm B$ the electric and magnetic fields. Thus obtaining coupled equations for $\chi$ and $\eta$:
\begin{subequations}
	\begin{eqnarray}\label{coupled}
		T\chi &=&  {\bm \sigma}\!\cdot\!\left({\bm p}+i g_b \bm E\right)\eta+(q_bA^0-g_b\bm\sigma\!\cdot\!\bm B)\chi,\mbox{ and}\label{eerste}\\
		(2m+T)\eta &=& {\bm \sigma}\!\cdot\!\left({\bm p}-i g_b \bm E\right)\chi+(q_bA^0+g_b\bm\sigma\!\cdot\!\bm B)\eta.\label{tweede}
	\end{eqnarray}
\end{subequations}
As shown in \cite{gross} using equation (\ref{tweede}) to obtain an expression for $\eta$ and then solving for $T$ using equation (\ref{eerste}) will yield an energy ($T$) dependent Hamiltonian, due to the explicit energy dependence of $\eta$. This is undesirable and thus the equations must be transformed so that the off-diagonal elements of the (transformed) Hamiltonian are small so that the leading order estimate for $\eta$ will be energy independent and yield the effective Hamiltonian to the desired accuracy. This transformation is known as the Foldy-Wouthuysen transformation \cite{gross}. The off-diagonal terms only need to be calculated up to $\mathcal{O}(m^{-1})$.\\
\\
For convenience equation (\ref{H1}) is re-written as
	\begin{eqnarray}
		T\left[\begin{array}{c}\chi\\\eta\end{array}\right]
		=\Big[{\bm \alpha}\!\cdot\!\left(\bm p-\bm V\right)+ V^0-
			g_b\beta\bm\Sigma \!\cdot\!\bm B+\beta m -m\Big]\left[\begin{array}{c}\chi\\\eta\end{array}\right],
	\end{eqnarray}
where $\bm V=(q_b{\bm A}-i g_b \beta \bm E)$ and $V^0=q_bA^0$.
Since the off-diagonal terms are all dependent on $\bm\alpha$, the transformation should also depend on $\bm\alpha$, since any even product of $\bm \alpha$s will be diagonal. A simple unitary transformation
\begin{eqnarray}
	U=U^{\dagger}=A\beta+\frac{\Lambda}{m}\bm\alpha\!\cdot\!\bm p\ \ \mbox{with}\ \ A=\sqrt{1-\frac{\Lambda^2p^2}{m^2}}
\end{eqnarray}
will be used ($\Lambda$ is parameter which will be determined later ).
Considering the unitarity of $U$ and thus the fact that the norm of the transformed wave functions will not change, equation (\ref{Hap}) transforms to
\begin{eqnarray}
		T\left[\begin{array}{c}\chi'\\\eta'\end{array}\right]
		=U H U^\dagger\left[\begin{array}{c}\chi'\\\eta'\end{array}\right]
		= H'\left[\begin{array}{c}\chi'\\\eta'\end{array}\right].
\end{eqnarray}
Expanding $A$ to
\begin{eqnarray}
	A\cong 1-\frac{\Lambda^2p^2}{2m^2}
\end{eqnarray}
and calculating the off-diagonal terms of $H'$ up to $\mathcal{O}(m^{-1})$ (thus $A\sim 1$ is sufficient) it can be shown that the off-diagonal terms are
\begin{eqnarray}
	H'_{\mbox{\scriptsize{off-diag}}}=2\Lambda{\bm \alpha}\!\cdot\!\bm p-{\bm \alpha}\!\cdot\!\left(\bm p-\bm V\right)+\mathcal{O}(m^{-2}).\label{H'}
\end{eqnarray}
Choosing 
\begin{eqnarray}
	\Lambda=\frac{1}{2}
\end{eqnarray}
 $H'_{\mbox{\scriptsize{off-diag}}}$ is approximated to be 
\begin{eqnarray}
		H'_{\mbox{\scriptsize{off-diag}}}\cong{\bm \alpha}\!\cdot\!\bm V
\end{eqnarray}
and since
\begin{eqnarray}
		{\bm \alpha}\!\cdot\!\bm V  = 
		\left[
			\begin{array}{cc}
				0 & {\bm \sigma}\!\cdot\!\bm V\\
				{\bm \sigma}\!\cdot\!\bm V & 0
			\end{array}	
		\right],
\end{eqnarray}
the coupled equations (\ref{coupled}) become
\begin{subequations}
	\begin{eqnarray}\label{coupled2}
		T'\chi' &=&  H'_{11}\chi'+{\bm \sigma}\!\cdot\!\bm V\eta',\mbox{ and}\label{eerste2}\\
		T'\eta &=& {\bm \sigma}\!\cdot\!\bm V\chi'-2m\,\eta'\label{tweede2},
	\end{eqnarray}
\end{subequations}
where only the leading order contributions have been retained in all terms except for $H'_{11}$. Since $T$ is of $\mathcal{O}(m^{-1})$ the lower component of the Dirac spinor ($\eta$) is dominated by the $2m$ the $T\eta'$ term and be ignored in equation (\ref{tweede2}) and so equation (\ref{tweede2}) can be solved as
\begin{eqnarray}
	T'\chi' =  \left(H'_{11}+\frac{{\bm \sigma}\!\cdot\!\bm V{\bm \sigma}\!\cdot\!\bm V}{2m}\right)\chi'=\left(H'_{11}+\frac{V^2}{2m}\right)\chi'.\label{T'}
\end{eqnarray}
$H'_{11}$ is calculated from $H'$ (\ref{H'}) for $\Lambda=\frac{1}{2}$ and can be shown to be
\begin{align}
	\begin{split}
		H'_{11}\cong&-\frac{p^2}{2m^2}+ V^0-\frac{p^2}{8M^2}V^0-V^0\frac{p^2}{8M^2}+\frac{\bm \sigma\!\cdot\!\bm p\, V^0\bm \sigma\!\cdot\!\bm p}{4m^2}\\
		&+\,\frac{\bm \sigma\!\cdot\!(\bm p-\bm V^{(u)})\,\bm\sigma\!\cdot\!\bm p+\bm \sigma\!\cdot\!\bm p\,\bm \sigma\!\cdot\!(\bm p-\bm V^{(u)})}{2m}-\frac{p^4}{8m^3}\\
		&+\,g_b\,\bm\sigma\!\cdot\!\bm B-g_b\frac{p^2}{4m^2}\bm\sigma\!\cdot\!\bm B+\frac{g_b}{4m^2}\bm\sigma\!\cdot\!\bm p\,\bm\sigma\!\cdot\!\bm B\,\bm\sigma\!\cdot\!\bm p,
	\end{split}
\end{align}
where $\bm V^{(u)}=(q_b{\bm A}-i g_b\bm E)$. Using the properties of the Pauli-matrices and replacing $\bm p$, $V^0$ and $\bm V^{(u)}$ with
\begin{subequations}
	\begin{eqnarray}
		\bm p&=&-i\bm\nabla,\\
		 V^0&=&q_b A^0,\mbox{ and}\\
		 \bm V^{(u)}&=&(q_b{\bm A}-i g_b\bm E),
	\end{eqnarray}
\end{subequations}
$T'$ is calculated using equation (\ref{T'}) to be
\begin{align}
	\begin{split}
		H'_{11}+\frac{V^2}{2m}=&
		\frac{(\bm p-q_b{\bm A}+i g_b\bm E)^2}{2m}+q_b A^0-\frac{p^4}{8M^3}+\frac{q_b\nabla^2A^0}{8M^2}
		+\frac{q_b}{4M^2}\bm\sigma\!\cdot\!\left([\nabla A^0]\times\bm p\right)\\
	&-\,\frac{p^4}{8m^2}+\frac{g_b}{4m^2}\bm p\!\cdot\!\bm B\,\bm\sigma\!\cdot\!\bm p
		-\frac{q_b}{2m}\bm\sigma\!\cdot\!\bm B+g_b\bm\sigma\!\cdot\!\bm B.\label{Heff}
	\end{split}
\end{align}
This is the effective Hamiltonian in the non-relativistic limit up to order $p^4/m^3$. The terms of interest with regards to the magnetic dipole moment are the last two in equation (\ref{Heff}). The nuclear magneton is defined as \cite{PDGconst}
\begin{eqnarray}
	\mu_N=\frac{e \hbar}{2 m_p}=\frac{q_p}{2 m}=\mu_N^{(dip)},
\end{eqnarray}
where $e$ the charge of the proton ($q_p$ in the notation used here)\footnote{Also remember natural units, where $\hbar=c=1$, is used}. Noting that the spin-operator is 
\begin{eqnarray}
	\bm S=\frac{\bm\sigma}{2},
\end{eqnarray}
the last two terms in equation (\ref{Heff}) becomes
\begin{eqnarray}
	-\frac{q_b}{2m}\bm\sigma\!\cdot\!\bm B+g_b\bm\sigma\!\cdot\!\bm B=-\left(2-\frac{2\,g_b}{\mu_N^{(dip)}}\right)\mu_N^{(dip)}\bm S\!\cdot\!\bm B.
\end{eqnarray}
Thus the inclusion of the $\sigma^{\mu\nu}F_{\mu\nu}$-term in the Lagrangian modifies the magnetic dipole moment by adding the amount $-2g_b/\mu_N^{(dip)}$ to the bare value of $2$, which is due to the charge of the particle coupling to the electromagnetic field. Defining $g_b$ to be
\begin{eqnarray}
	g_b=-\frac{\kappa_b\mu_N^{(dip)}}{2}
\end{eqnarray}
this will lead to the particle's magnetic moment being altered by
\begin{eqnarray}
	-\left(2-\frac{2\,g_b}{\mu_N^{(dip)}}\right)\mu_N^{(dip)}\bm S\!\cdot\!\bm B=-\left(2+\kappa_b\right)\mu_N^{(dip)}\bm S\!\cdot\!\bm B.\label{magdip}
\end{eqnarray}
Since the observed magnetic dipole moment of protons and neutrons are 
\begin{subequations}\label{normaldip}
	\begin{align}
		2.793\,\mu_N^{(dip)}&\mbox{, and}\\
		-1.913\,\mu_N^{(dip)}&
		\end{align}
respectively \cite{PDGmuon}, $\kappa_p$ and $\kappa_n$ must be
	\begin{eqnarray}
		\kappa_p &=&  0.793\label{kapp}\mbox{, and}\\
		\kappa_n &=& -1.913\label{kapn}
	\end{eqnarray}
\end{subequations}
to reproduce the observed values of the baryon magnetic dipole moment. Note that since neutrons have no charge, $\kappa_n$ has to carry the full value of the neutron magnetic dipole moment. 
\section{Adjusting magnetic dipole moments} 
In this work we are interested the response of the system when the value of the magnetic dipole units are changed. These adjustments will in general the magnetic dipole will be adjusted in units of $\mu_N^{(dip)}$. This will ease the comparison of the adjusted values to those measured under normal laboratory conditions (\ref{normaldip}).
\subsection{Neutrons}
If the magnetic dipole moment is to be increased by a factor of $x$ times the normal value, $g_n$ has to be adjusted to 
\begin{eqnarray}\label{gnadj}
	g_n=-x\frac{\kappa_n\mu_N^{(dip)}}{2}.
\end{eqnarray}
\subsection{Proton}
In the case of protons, this will differ slightly since the $\bar{\psi}q_b A_\mu\psi$-coupling automatically introduces the factor of 2 in equation (\ref{magdip}). If a dipole moment of $x$ times its normal value is required, $g_p$ has to be adjusted in the following way:
 \begin{eqnarray}\label{gpadj}
	g_p=-\frac{x(2+\kappa_p)-2}{2}\mu_N^{(dip)}.
\end{eqnarray}

\chapter{Relativistic description of a charged particle in a magnetic field}\label{ap:landau}
In this appendix the spectrum and wave functions of neutral and charged relativistic particles in a magnetic field are derived. The calculation of the particle density of such a system will also be illustrated.\\
\\
Of specific interest is the influence of the $\bar{\psi}\sigma^{\mu\nu}F_{\mu\nu}\psi$ coupling (between the magnetic dipole moment of the fermions and the magnetic field) on the single particle energy spectrum. This derivation is based the ones by Broderick et al. \cite{Brod00} and Mao et al. \cite{mao}. However, notation specific to this work is used and details not explicitly dealt with in the mentioned papers are included here.
\section{Spectrum without the $\bar{\psi}\sigma^{\mu\nu}F_{\mu\nu}\psi$ coupling}\label{ap:land:spec}
\begin{figure}
	\centering
		\includegraphics[width=1.\textwidth]{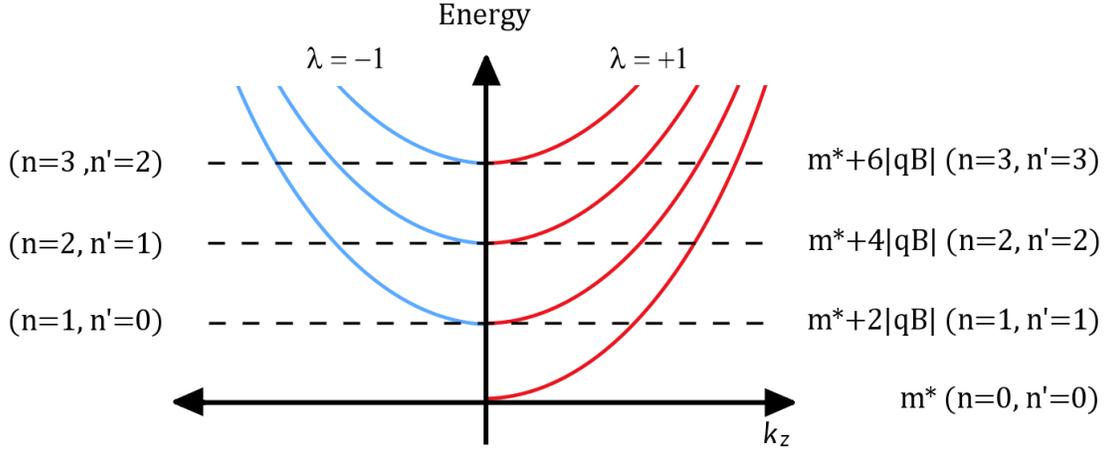}
	\caption[Illustration of the Landau levels occupied by protons in a magnetic field.]{Illustration of the Landau levels (in one half of the Fermi surface) occupied by protons in a magnetic field pointing in the positive $z$-direction. Labels in brackets pertains to the specific choice of $\lambda$. 	 }
	\label{fig:LLap}
\end{figure}

In contrast to neutral particles, which have a continuous spectrum, charged particles occupy quantised energy levels in the presence of a magnetic field. This is known as Landau quantisation and the quantised levels as the Landau levels \cite{gross,zuber}. These levels are labelled by the integer $n$.\\
\\
The energy spectrum of a relativistic fermion (thus described by the Dirac equation) with charge $q$ and mass $m$ in a magnetic field ${\bf B}=B\hat{z}$ is \cite{zuber}
\begin{subequations}\label{LLenercon}
	\begin{eqnarray}
		E\,(k_z,n,\lambda)&=&\sqrt{k_z^2+m^2+2|q B|n}\\
		&=&\sqrt{k_z^2+m^2+2|q B|\left(n'+\frac{1}{2}-\alpha\,\frac{\lambda}{2}\right)}\label{LL2},
	\end{eqnarray}
\end{subequations}
where
\begin{itemize}
	\item $k_z$ is the momentum component along ${\bf B}$,
	\item $n=0,1,2,...$ labels the different Landau levels,
	\item $\alpha=\frac{q B}{|q B|}=\pm 1$, and
	\item $\lambda=\pm 1$ is the eigenvalue of $\sigma_z$.
\end{itemize}
In (\ref{LL2}) $n$ is expanded as $n=n'+\frac{1}{2}-\alpha\,\frac{\lambda}{2}$ to show the dependence on $\lambda$. When the $\lambda$-dependence is considered it is clear that the $n=0$ (lowest) Landau level can only be occupied by particles with either $\lambda=1$ or $\lambda=-1$, depending on the value of $\alpha$. The $n=1$ level can however be occupied by particles with $\lambda=1$ and $\lambda=-1$, independent of the value of $\alpha$. This can lead to confusion when the number of occupied Landau levels has to be counted. For example, if $\alpha=1$ and we were to use $n$ to label the levels then particles with $\lambda=1$ will occupy levels from $n=0$, while particles with $\lambda=-1$ will occupy levels from $n=1$. However, if we use $n'$ to label the levels then the lowest level occupied by any particle (regardless of $\alpha$ or $\lambda$) will be $n'=0$. For an illustration of the different libelling schemes see Figure \ref{fig:LLap}.\\
\\
To aid the calculation of the different fermion densities we will use the $n'$ notation in this work. 
\subsection{Lepton spectrum}
Since $\alpha$ distinguishes between positive and negative charged particles for a given value of $B$, (\ref{LL2}) is the most general expression for the charged particle's energy spectrum in a magnetic field. In this work we will not include the $\bar{\psi}\sigma^{\mu\nu}F_{\mu\nu}\psi$ coupling when describing leptons and therefore (\ref{LL2}) will be used in our calculations to describe the lepton spectrum. 
\section{Including the $\bar{\psi}\sigma^{\mu\nu}F_{\mu\nu}\psi$ coupling}
However, we will include the $\bar{\psi}\sigma^{\mu\nu}F_{\mu\nu}\psi$ coupling in our description of fermions. From the Lagrangian (\ref{LBB}) the equation of motion (\ref{DEBB}) for fermions in a magnetic field is 
\begin{align}
		\begin{split}
			\left[\gamma^{\mu}\left(i\partial_{\mu}-q_p\frac{1+\tau_3}{2} A_\mu
			- g_{v}V_{\mu}
			- \frac{g_\rho}{2}{\bm \tau}\cdot{\bf b}_\mu
			\right)
			-\frac{g_b}{2}F^{\mu\nu}\sigma_{\mu\nu}-\left(m-g_{s}\phi
			\right)\right]\psi{ (x)}= 0.
		\end{split}
\end{align}
Making the RMF approximation and using the notation of the Dirac-matrices as well our choice of $A^\mu$, 
the above equation can be rewritten as\footnote{Essentially from here on we assume $\psi{ (t,\,\bm x)}$ not to be an isodoublet spinor, but just a $4\times1$ Dirac spinor. Equation (\ref{matDE}) is however valid for both protons and neutrons, since $\tau_0$ is the eigenvalue of $\tau_3$ and the charge $q_b$ will differentiate the two baryon species. }
\begin{eqnarray}
	\left(i\partial_0-g_vV_0 -\frac{g_\rho}{2}\tau_0b_0\right)\psi{ (x)}=
	\big({\bm \alpha}\!\cdot\!\left({\bm p}-q_p{\bm A}\right)-g_b\beta \bm\Sigma \!\cdot\!\bm B+\beta m^*      \big)\psi{ (x)}\label{matDE}.
\end{eqnarray}
Defining 
\begin{eqnarray}
	\bm \pi = (\bm p - q_p \bm A)\label{pie},
\end{eqnarray}
where $\bm p$ is the momentum operator ($\bm p = -i\bm\nabla$), and ignoring the meson contributions turns equation (\ref{matDE}) into 
\begin{eqnarray}
	i\partial_0\psi=\big({\bm \alpha}\!\cdot\!{\bm \pi}-g_b\beta \bm\Sigma \!\cdot\!\bm B+\beta m^*      \big)\psi.
\end{eqnarray}
Assuming $\psi{ (x)}$ to be of the form
\begin{eqnarray}
	\psi{ (t,\,\bm x)}=\left[\begin{array}{c}\chi
	\\\eta
	\end{array}\right]e^{-iEt}\label{assump}
\end{eqnarray}
the coupled equations for $\chi$ and $\eta$ can be derived:
\begin{subequations}
	\begin{eqnarray}
		\left(E-m+g_bB\sigma_z\right)\chi&=&{\bm \sigma}\!\cdot\!{\bm \pi}\,\eta,\mbox{ and}\label{chi}\\
		\left(E+m-g_bB\sigma_z\right)\eta&=&{\bm \sigma}\!\cdot\!{\bm \pi}\,\chi.\label{eta}
	\end{eqnarray}
\end{subequations}
Multiplying equation (\ref{eta}) by $\left(E+m+g_bB\sigma_z\right)$ then produces
\begin{eqnarray}
	\left(\left(E+m\right)^2-\left(g_bB\right)^2\right)\eta=\left(E+m+g_bB\sigma_z\right){\bm \sigma}\!\cdot\!{\bm \pi}\,\chi,
\end{eqnarray}
so that equation (\ref{chi}) becomes
\begin{eqnarray}
	\left(E-m+g_bB\sigma_z\right)\chi&=&
	\frac{{\bm \sigma}\!\cdot\!{\bm \pi}\left(E+m+g_bB\sigma_z\right){\bm \sigma}\!\cdot\!{\bm \pi}}
	{\left(E+m\right)^2-\left(g_bB\right)^2}
	\,\chi\nonumber\\
	&=&\frac{\left(E+m-g_bB\sigma_z\right)}
	{\left(E+m\right)^2-\left(g_bB\right)^2}
	\left({\bm \sigma}\!\cdot\!{\bm \pi}\right)^2\chi
	+\frac{2g_b B \pi_z{\bm \sigma}\!\cdot\!{\bm \pi}}{\left(E+m\right)^2-\left(g_bB\right)^2}\chi.\label{chi2}
\end{eqnarray}
Here we have used the relation
\begin{eqnarray}
	\left\{\sigma^i,\sigma^j\right\}=2\delta^{ij}\dblone_2,
\end{eqnarray}
which implies that
\begin{eqnarray}
	{\bm \sigma}\!\cdot\!{\bm \pi}\sigma_z{\bm \sigma}\!\cdot\!{\bm \pi}=
	-\sigma_z({\bm \sigma}\!\cdot\!{\bm \pi})^2+2\pi_z{\bm \sigma}\!\cdot\!{\bm \pi}.
\end{eqnarray}
Multiplying equation (\ref{chi2}) by $\left(E+m+g_bB\sigma_z\right)$ produces
\begin{eqnarray}
	\left(\left(E+g_bB\sigma_z\right)^2-m^2\right)\chi&=&
	({\bm \sigma}\!\cdot\!{\bm \pi})^2\,\chi
	+\frac{2g_b B\left(E+m+g_bB\sigma_z\right)}{\left(E+m\right)^2-\left(g_bB\right)^2}
	\pi_z({\bm \sigma}\!\cdot\!{\bm \pi})\chi,\label{Fa}
\end{eqnarray}
which is rewritten as
\begin{eqnarray}
	F_\lambda\chi&=&
	({\bm \sigma}\!\cdot\!{\bm \pi})^2\,\chi
	+a_\lambda
	\pi_z({\bm \sigma}\!\cdot\!{\bm \pi})\chi\label{chi3}
\end{eqnarray}
by introducing $F_\lambda$ and $a_\lambda$ as
\begin{subequations}
	\begin{eqnarray}
		F_\lambda&=&\left(E+g_bB\sigma_z\right)^2-m^2,\mbox{ and}\\
		a_\lambda&=&\frac{2g_b B\left(E+m+g_bB\sigma_z\right)}{\left(E+m\right)^2-\left(g_bB\right)^2}.
	\end{eqnarray}
\end{subequations}
From this point onwards we will consider protons and neutrons separately, thus $g_b$ will become $g_p$ and $g_n$ respectively.
\subsection{Protons}
For our choice of $A^\mu$ (\ref{Amu}) it can be shown that
\begin{eqnarray}
	{\bm \pi}\!\times\!{\bm \pi}=iq_p B\hat z
\end{eqnarray}
and using the commutation relations of the Dirac matrices $({\bm \sigma}\!\cdot\!{\bm \pi})^2$ can then be expanded as
\begin{eqnarray}
	({\bm \sigma}\!\cdot\!{\bm \pi})^2&=&{\bm \pi}\!\cdot\!{\bm \pi}+i\bm\sigma\cdot({\bm \pi}\!\times\!{\bm \pi})\nonumber\\
	&=&{\bm \pi}\!\cdot\!{\bm \pi}-\sigma_z q_p B.
\end{eqnarray}
Thus equation (\ref{chi3}) becomes
\begin{eqnarray}
	F_\lambda\chi&=&
	{\bm \pi}\!\cdot\!{\bm \pi}\chi- q_p B\sigma_z \chi
	+a_\lambda
	\pi_z({\bm \sigma}\!\cdot\!{\bm \pi})\chi.\label{chi4}
\end{eqnarray}
At this point an ansatz has to be made in order to proceed. Although it is not obvious from \ref{chi4} we are in fact dealing with a quantum mechanical harmonic oscillator. Thus, based on the off-diagonal components of the last term in \ref{chi4}, raising and lowering operators ($\xi_\pm$) will be defined as
\begin{eqnarray}
	\xi_\pm=\frac{1}{\sqrt{2|q_pB|}}\big( p_x\pm i\left(|q_pB| x-\alpha k_y\right) \big)\label{xi},
\end{eqnarray}
where once again $\alpha=\pm1$. Although it is probably not necessary, $\alpha$ is included to describe the most general case. Note that $p_x$ is an operator, while $k_y$ is the eigenvalue of $p_y$. Applying (\ref{xi}) to rewrite (\ref{chi4}) in matrix form yields
\begin{align}\label{chi5}
		\begin{split}
		F_\lambda\chi=&\ 
		2|q_pB|\left[
			\begin{array}{cc}
				\xi_+\xi_{{\scriptscriptstyle -}}+\frac{1}{2}-\frac{\alpha}{2} & 0\\
				0 & \xi_+\xi_{{\scriptscriptstyle -}}+\frac{1}{2}+\frac{\alpha}{2}
			\end{array}	
		\right]\!\chi\\
		&+p_z^2\,\chi
		+a_\lambda
		\left[
			\begin{array}{cc}
				p_z^2 & \sqrt{2|q_pB|}\xi_\alpha\, p_z\\
				\sqrt{2|q_pB|}\xi_{{\scriptscriptstyle (-\alpha)}}\,p_z & -p_z^2
			\end{array}	
		\right]\chi
		\end{split}
\end{align}
and we deduce that the specific form of the ansatz for $\chi$ will also depend on $\alpha$. $\chi$ for $\alpha=1$ must be 
\begin{subequations}\label{chiansatz}
\begin{align}
	\chi&=\left[
		\begin{array}{c}
			I_{n,k_y}(x)\\
			i\omega I_{n-1,k_y}(x)
		\end{array}
		\right]e^{ik_yy+ik_zz},\\
\intertext{while for $\alpha=-1$ }
	\chi&=\left[
		\begin{array}{c}
			i\omega I_{n-1,k_y}(x)\\
			I_{n,k_y}(x)
		\end{array}
		\right]e^{ik_yy+ik_zz},
\end{align}
\end{subequations}
where in both instances $\omega$ needs to be determined and $I_{n,k_y}(x)$ is
\begin{eqnarray}\label{Lwfn}
	I_{n,k_y}(x)=\frac{N}{\sqrt{2^nn!}}\, e^{-\frac{1}{2}\left(\sqrt{|q_pB|}\left(x-\alpha\frac{k_y}{|q_pB|}\right)\right)^2}H_n\left(\sqrt{|q_pB|}\left(x-\alpha\frac{k_y}{|q_pB|}\right)\right),
\end{eqnarray}
where
\begin{itemize}
	\item $H_n$ the Hermite polynomial
	\begin{eqnarray}
		H_n(x)=(-1)^ne^{x^2}\frac{d^n}{dx^n}e^{-x^2},\mbox{ and}
	\end{eqnarray}
	\item $N$ the normalisation factor so that 
	\begin{eqnarray}\label{Inorm}
		\int^\infty_{-\infty} I_{n,k_y}(x)I_{n,k_y}(x)dx=1.
	\end{eqnarray}
\end{itemize}
By applying the raising and lowering operator to $I_{n,k_y}(x)$ it can be shown that 
\begin{subequations}
	\begin{eqnarray}
		\xi_+ I_{n,k_y}(x)&=&i\sqrt{\frac{n+1}{|q_pB|}}I_{n+1,k_y}(x),\\
		\xi_- I_{n,k_y}(x)&=&-i\sqrt{n|q_pB|}I_{n-1,k_y}(x),\mbox{ and}\\
		\xi_+\xi_-&=&\frac{1}{2|q_pB|}\big( p_x^2+\left(|q_pB| x-\alpha k_y\right)^2-|q_pB| \big)
	\end{eqnarray}
\end{subequations}
while $I_{-1,k_y}(x)$ is defined to be zero \cite{mao}.\\
\\
Using these properties and substituting the expressions of $F_\lambda$ and $a_\lambda$ back into (\ref{chi5}) yields two coupled equations. For $\alpha =1$
\begin{subequations}
	\begin{align}
		\begin{split}
			\left(E+m+g_pB\right)(&E +m-g_pBz)(E-m+g_pB)\\
			&=2|q_p B|n\left(E+m-g_pB\right)+k_z^2\left(E+m+g_pB\right)-\sqrt{2n}\omega k_z 2g_p B,
		\end{split}
	\intertext{and}
		\begin{split}
			(E+m+g_pB)(&E+m-g_pB)(E-m-g_pB)\\
			&=2|q_p B|n(E+m+g_pB)+k_z^2\left(E+m-g_pB\right)-\sqrt{2n} k_z 2g_p B\frac{2|q_p B|}{\omega}.
		\end{split}
	\end{align}
\end{subequations}
For $\alpha=-1$ they are
\begin{subequations}
	\begin{align}
		\begin{split}
		(E+m+g_pB)(&E+m-g_pB)(E-m+g_pB)\\
		&=2|q_p B|n(E+m-g_pB)+k_z^2(E+m+g_pB)-\sqrt{2n}\frac{2|q_p B|}{\omega} k_z 2g_p B,
		\end{split}
	\intertext{and}
		\begin{split}
		(E+m+g_pB)(&E+m-g_pB)(E-m-g_pB)\\
		&=2|q_p B|n(E+m+g_pB)+k_z^2(E+m-g_pB)-\sqrt{2n}\omega k_z 2g_p B.
		\end{split}
	\end{align}
\end{subequations}
Using these equations we can eliminate $\omega$ and establish that
\begin{eqnarray}
	\omega=-\frac{\alpha}{k_z\sqrt{2n}}\left(2|q_p B|n+\left(E+m+\alpha g_pB\right)\left(m+\alpha\lambda\sqrt{m^2+2|q_p B|n}\right)\right).\label{omega}
\end{eqnarray}
Then we can solve for $E$ and find
\begin{eqnarray}
	E=\pm\sqrt{k_z^2+\left(\sqrt{m^2+2|q_p B| n}+\lambda g_p B\right)^2}.\label{Ell1}
\end{eqnarray}
Matching (\ref{Ell1}) to the $E$ when the $\bar{\psi}\sigma^{\mu\nu}F_{\mu\nu}\psi$ coupling is not included (\ref{LL2}), the $\alpha$- and $\lambda$-dependence of the energy spectrum can be deduced. Thus the spectrum of protons in a magnetic field is
\begin{eqnarray}
	E(k_z,\lambda,n)=\pm\sqrt{k_z^2+\left(\sqrt{m^2+2|q_p B| \left(n'+\frac{1}{2}-\alpha\,\frac{\lambda}{2}\right)}+\lambda g_p B\right)^2}.\label{Ell}
\end{eqnarray}
Upon including the meson contributions the energy spectrum for positive energy protons is 
\begin{eqnarray}
	e(k_z,\lambda,n)=\sqrt{k_z^2+\left(\sqrt{m^{*2}+2|q_p B| \left(n'+\frac{1}{2}-\alpha\,\frac{\lambda}{2}\right)}+\lambda g_p B\right)^2}+g_v V_0+\frac{g_\rho b_0}{2}.\label{ELL}
\end{eqnarray}
\subsection{Proton eigenvectors}
The proton eigenvectors can be calculated by substituting the expressions for $\omega$ (\ref{omega}) into the ansatz for $\chi$ (\ref{chiansatz}) and using equation (\ref{eta}) to calculate the lower component of $\psi{ (t,\,\bm x)}$ (\ref{assump}).
\subsection{Neutrons}
Since for neutrons $q = 0$, $\bm \pi$ (\ref{pie}) simply becomes 
\begin{eqnarray}
	\bm \pi= \bm p.
\end{eqnarray}
We will assume that the momentum-dependence of the spinor (\ref{assump}) is contained in the wave component of the spinor and therefore
\begin{eqnarray}
	\psi{ (t,\,\bm x)}=\left[\begin{array}{c}\chi
	\\\eta
	\end{array}\right]e^{-iEt+i\bm k\cdot\bm x}.\label{assump2}
\end{eqnarray}
We define $G_\lambda$ as
\begin{eqnarray}
	G_\lambda=F_\lambda-k^2-a_\lambda\sigma_z k_z,
\end{eqnarray}
where ${\bf k}^2=k_x^2+k_y^2+k_z^2$. After considering that
\begin{eqnarray}
	(\bm\sigma\!\cdot\!{\bm p})^2\,e^{-iEt+i\bm k\cdot\bm x}=(\bm\sigma\!\cdot\!{\bm k})^2\,e^{-iEt+i\bm k\cdot\bm x}=\bm k^2e^{-iEt+i\bm k\cdot\bm x},
\end{eqnarray}
equation (\ref{chi3}) becomes
\begin{eqnarray}
	G_\lambda \chi = a_\lambda\pi_z(\sigma_x k_x + \sigma_y k_y)\chi.\label{Glambda}
\end{eqnarray}
By inspection it is clear that $G_\lambda$ and $a_\lambda$ are diagonal and thus all off-diagonal elements are contained on the right-hand side of equation (\ref{Glambda}). Furthermore, we know that $\chi$ is a $2\times1$ matrix, which we assume to be
\begin{eqnarray}
	\chi{ (\bm k)}=\left[\begin{array}{c}\mu{ (\bm k)}\\
	\nu{ (\bm k)}\end{array}\right].
\end{eqnarray}
Thus (\ref{Glambda}) becomes a set of two coupled equations
\begin{subequations}
	\begin{align}
		G_{+1}\,\mu&=a_{+1}(k_x-ik_y)k_z\,\nu\label{G1},\mbox{ and}\\
		G_{-1}\,\nu&=a_{-1}(k_x+ik_y)k_z\,\mu\label{G2}.
	\end{align}
\end{subequations}
Using (\ref{G2}) to solve for (\ref{G1}) we find that
\begin{eqnarray}
	G_{+1}\,G_{-1}=a_{+1}\,a_{-1}(k_x^2+k_y^2)k_z^2.\label{G3}
\end{eqnarray}
From (\ref{Fa}) $a_\lambda$ is
\begin{eqnarray}
	a_\lambda=\frac{2 g_n B}{E+m-\lambda g_n B}
\end{eqnarray}
and $G_\lambda$ becomes
\begin{eqnarray}
	G_\lambda=(E+\lambda g_n B)^2-m^2-k^2-\frac{2 g_n B}{E+m-\lambda g_n B}.
\end{eqnarray}
Using the result above equation (\ref{G3}) can be solved to yield
\begin{eqnarray}
	E=\pm\sqrt{k_z^2+\left(\sqrt{k_\perp^2+m^2}+\lambda g_n B\right)^2},
\end{eqnarray}
where $k_\perp^2=(k_x^2+k_y^2)$.
Including the meson contributions the energy spectrum for a positive energy neutron is 
\begin{eqnarray}
	e(\bm k,\lambda)=\sqrt{k_z^2+\left(\sqrt{k_\perp^2+m^2}+\lambda g_n B\right)^2}+g_v V_0-\frac{g_\rho b_0}{2}.\label{EN}
\end{eqnarray}
\subsection{Neutron eigenvectors}
According to \cite{Brod00} the eigenvectors can be constructed by setting $\mu(\bm k)=1$ for the $\lambda = 1$ positive energy spinor. Then $\nu$ can be calculated from (\ref{G2}). For the $\lambda=-1$ positive energy spinor $\nu(\bm k)$ is set to $1$. However the explicit construction of the eigenvectors is not necessary within the scope of this work, since in the RMF approximation the fermion densities can also be derived by minimizing the energy density. 
\section{Charged particle densities in a magnetic field}\label{apsec:Lp}
Consider a quantum mechanical particle in a three-dimensional box with volume $L^3$ and a magnetic field pointing in the $z$-direction. The particle has two possible spin orientations which are labelled by $\lambda$. The Landau levels are labelled by $n=0,1,2,...$.\\
\\
For a given choice of $\lambda$ and $n$ the movement in the $z$-direction is that of a free particle. With periodic boundary conditions $k_z$ can assume values of 
\begin{eqnarray}
	k_z=\frac{2\pi l}{L}\label{Lkz}
\end{eqnarray}
where
\begin{itemize}
	\item $l=0,\pm 1,\pm 2,\pm 3,...,\pm l_{max}$, and
	\item $L$ is the length of the square box.
\end{itemize}
The value of $l$ is restricted by the Fermi energy which implies (for a given choice of $\lambda$ and $n$) a Fermi momentum $k_z^F$, where
\begin{eqnarray}
	k_z^F(\lambda,n)=\frac{2\pi l_{max}}{L}.
\end{eqnarray}
In the $xy$-plane the charged particles behaves like harmonic oscillators localised at various points in the $x$-direction. For $k_y$ the same relation (\ref{Lkz}) applies as for $k_z$. However, from the wave function (\ref{Lwfn}) that the Gaussian (harmonic oscillator) part is localised in the $x$-directions as a function of $k_y$ as\footnote{We omitted $\alpha$ without loss of generality, since the the quantity of interest is the absolute distance.}
\begin{eqnarray}
	x(k_y)=\frac{k_y}{|q_pB|}.
\end{eqnarray}
Thus the spacing between two consecutive harmonic oscillators is 
\begin{eqnarray}
	\Delta x&=&x_{l+1}-x_{l}\\
	&=& \frac{k_y(l+1)-k_y(l)}{|q_pB|}\\
	&=& \frac{2\pi}{L|q_pB|}
\end{eqnarray}
With the spacing the number of particles that can be accommodated in the $xy$-plane is limited, since
\begin{eqnarray}
	N(x,y)=\frac{L}{\Delta x}=\frac{L^2|q_pB|}{2 \pi}.
\end{eqnarray}
Re-writing the above by defining the area $A=L^2$ to
\begin{eqnarray}
	N(x,y)/A=\frac{|q_p||B|}{2 \pi}
\end{eqnarray}
a well-known quantity in the Landau problem is recovered, namely the magnetic flux density divided by the fundamental flux quantum, $\Phi_0$, where
\begin{eqnarray}
	\Phi_0=\frac{2 \pi\hbar}{e}=\frac{2 \pi}{|q_p|}
\end{eqnarray}
in the units used in this work\footnote{``$e$'' being the fundamental charge in this instance.}. 
It is well known that the degeneracy of each Landau level corresponds to the number of fundamental flux quanta penetrating the system. It is for this reason that the magnetic flux density $B$ divided by $\Phi_0$ is precisely the number of states per unit area in each Landau level \cite{yobi}. 
The magnetic length ($l_B$) is also defined in terms of $\Phi_0$ as 
\begin{eqnarray}
	B\,{l_B}^2=\Phi_0
\end{eqnarray}
and ${l_B}^2$ gives an indication of the area occupied by one state.\\
\\
Thus the total number of particles, including the contribution of the states with different $k_z$, in the box is given by
\begin{align}
	\begin{split}
		N&=\sum_{\lambda,n}\sum_{l=-l_{max}}^{l_{max}}\frac{L^2|q_pB|}{2 \pi}\\
		&=\sum_{\lambda,n} 2 l_{max}\frac{L^2|q_pB|}{2 \pi}\\
		&=\sum_{\lambda,n} 2\frac{k_z^F(\lambda,n) L}{2\pi }\frac{L^2|q_pB|}{2 \pi}\\
		&=2\frac{L^3 |q B|}{4\pi^2}\sum_{\lambda,n} k_z^F(\lambda,n).
	\end{split}
\end{align}
Thus the particle density is given by
\begin{eqnarray}\label{Lprho1}
	\rho=\frac{N}{L^3}=\frac{|q B|}{2\pi^2}\sum_{\lambda,n} k_z^F(\lambda,n).
\end{eqnarray}
In this work we are primarily considered with the RMF approximation, which is made for the system in the thermodynamic limit (where $N\rightarrow\infty$ and $V\rightarrow\infty$). 
Since $V\rightarrow\infty$ implies that $L\rightarrow\infty$, in this limit $k_z$ (\ref{Lkz}) becomes continuous. However, (\ref{Lprho1}) for $\rho_p$ still applies. In this case it is simply obtained through 
\begin{align}
	\begin{split}
		\rho_p
		&=\sum_{\lambda,n}\frac{|q_p B|}{4\pi^2}\int({\psi}^\dagger\psi)_{n,\lambda}\Theta\big[\,\mu-e(k_z,\lambda,n)\big]dk_z\\
		&=\sum_{\lambda,n}\frac{|q_p B|}{4\pi^2}\int^{k_z^F(\lambda,n)}_{-k_z^F(\lambda,n)}
		({\psi}^\dagger\psi)_{n,\lambda}dk_z\\
		&=\frac{|q_p B|}{2\pi^2}\sum_{\lambda,n} k_z^F(\lambda,n).\label{Lrho}
	\end{split}
\end{align}
In calculating $\rho_p$ we have neglected the effects of the system boundary in the $x$-direction, which we assume to be negligible in the thermodynamic limit. These results therefore hold within the bulk in the limit where $N$ and $V$ tend to infinity in a fixed ratio. Although each Landau level is therefore infinitively degenerate, the finite particle density in the bulk implies that multiple Landau levels may never the less be occupied.\\
\\
Although the expression for $\rho$ (\ref{Lrho}) was derived making use of the protons Landau problem, the result applies to magnetised charged particles in general.



\end{document}